\documentclass[prx,twocolumn,showpacs,preprintnumbers,float]{revtex4-2}
\usepackage[colorlinks=true,urlcolor=blue,citecolor=blue,linkcolor=blue,breaklinks=true]{hyperref}
\usepackage{graphicx}
\usepackage{amsmath}
\usepackage{amssymb}
\usepackage{times}
\usepackage[dvipsnames]{xcolor}
\usepackage{ulem}
\usepackage{physics}
\usepackage{cleveref}
\usepackage{xr}
\usepackage{multirow}
\usepackage{comment}
\usepackage{siunitx}

\def\d{\delta}
\def\e{\epsilon}

\def\k{\kappa}

\def\r{\rho}

\def\s{\sigma}

\def\w{\omega}
\def\ua{\uparrow}
\def\da{\downarrow}
\def\Vec#1{\mathbf #1}

\def\kk{\Vec{k}}
\def\qq{\Vec{q}}

\def\rr{\Vec{r}}
\def\RR{\Vec{R}}
\def\bg{\text{bg}}

\def\AN{\text{AN}}
\def\TCFM{\text{TCFM}}
\def\sr{\s_\text{resol.}}

\newcommand{\Tc}{T_\text{c}}
\newcommand{\Snor}{\Sigma^\text{nor}}
\newcommand{\Sano}{\Sigma^\text{ano}}
\newcommand{\Stot}{\Sigma^\text{tot}}

\newcommand{\Sbg}{\Sigma_\text{bg}}
\newcommand{\Smfl}{\Sigma_\text{MFL}}

\newcommand{\Akw}{A(\Vec{k},\omega)}
\newcommand{\Gkw}{G(\Vec{k},\w)}
\newcommand{\Fkw}{F(\Vec{k},\w)}
\newcommand{\Wkw}{W(\Vec{k},\w)}

\newcommand{\rqw}{\r(\Vec{q},\w)}
\newcommand{\kAN}{\Vec{k}_\text{AN}}
\newcommand{\kN}{\Vec{k}_\text{N}}

\newcommand{\eck}{\epsilon_{c\Vec{k}}}

\newcommand{\edk}{\epsilon_{d\Vec{k}}}

\newcommand{\Dck}{D_{c\Vec{k}}}
\newcommand{\Ddk}{D_{d\Vec{k}}}
\newcommand{\wpk}{\omega_{p\Vec{k}}}

\newcommand{\Vk}{V_\Vec{k}}

\newcommand{\cks}{c_{\Vec{k}\s}^{\phantom \dagger}}
\newcommand{\cksd}{c_{\Vec{k}\s}^{\dagger}}
\newcommand{\dks}{d_{\Vec{k}\s}^{\phantom \dagger}}
\newcommand{\dksd}{d_{\Vec{k}\s}^{\dagger}}
\newcommand{\Vi}{\hat{V}_\text{imp}}
\newcommand{\Vset}{V_\text{set}}
\newcommand{\Iset}{I_\text{set}}
\newcommand{\Gloc}{\hat{G}_\text{loc}}
\newcommand{\rloc}{\rho_\text{loc}}
\newcommand{\gqw}{g(\Vec{q},\w)}
\newcommand{\Lqw}{L(\Vec{q},\w)}

\begin{document}

\title{Unified description of cuprate superconductors by fractionalized electrons emerging from \\ integrated analyses of photoemission spectra and quasiparticle interference}

\author{Shiro Sakai$^{1,2}$, Youhei Yamaji$^{3,4}$, Fumihiro Imoto$^5$, Tsuyoshi Tamegai$^6$, Adam Kaminski$^{7,8}$, Takeshi Kondo$^9$, Yuhki Kohsaka$^{2,10}$, Tetsuo Hanaguri$^2$, and Masatoshi Imada$^{1,6}$}
\affiliation{
$^1$ \mbox{Physics Division, Sophia University, Chiyoda, Tokyo 102-8554, Japan}\\
$^2$ \mbox{RIKEN Center for Emergent Matter Science, Wako, Saitama 351-0198, Japan} \\
$^3$ \mbox{Research Center for Materials Nanoarchitectonics, National Institute for Materials Science,} \\
\mbox{Tsukuba, Ibaraki 305-0044, Japan} \\
$^4$ \mbox{RIKEN Center for Computational Science, Kobe, 650-0047, Japan}\\
$^5$ \mbox{Schr\"odinger Inc., Tokyo 100-0005, Japan}\\
$^6$ \mbox{Department of Applied Physics, University of Tokyo, Tokyo 113-8656, Japan}\\
$^7$ \mbox{Ames National Laboratory, U.S. Department of Energy, Ames, IA 50011, USA}\\
$^8$ \mbox{Department of Physics and Astronomy, Iowa State University, Ames, IA 50011, USA}\\
$^9$ \mbox{ISSP, University of Tokyo, Kashiwa, Chiba 277-8581, Japan}\\
$^{10}$ \mbox{Department of Physics, Kyoto University, Kyoto 606-8502, Japan}
} 

\date{\today}
\begin{abstract}
Electronic structure of high-temperature superconducting cuprates is studied
by analyzing experimental data independently obtained from two complementary spectroscopies, one, quasiparticle interference (QPI) measured by scanning-tunneling microscopy and the other, angle-resolved photoemission spectroscopy (ARPES) and by combining these two sets of data in a unified theoretical analysis.
Through explicit calculations of experimentally measurable quantities, we show that a simple two-component fermion model (TCFM) representing electron fractionalization succeeds in reproducing various detailed features of these experimental data: 
ARPES and QPI data are concomitantly reproduced by the TCFM in full energy and momentum spaces. The measured QPI pattern reveals a signature characteristic of the TCFM, distinct from the conventional single-component prediction, supporting the validity of the electron fractionalization in the cuprates. The 
integrated analysis also solves the puzzles of ARPES and QPI data that are seemingly inconsistent with each other. The overall success of the TCFM offers a comprehensive understanding of the electronic structure of the cuprates, in particular the unoccupied side of the spectra, of which momentum-resolved structure has long been unexplored experimentally. We further predict that a characteristic QPI pattern should appear in the unoccupied 
high-energy part if the fractionalization is at work. We propose that integrated-spectroscopy analyses offer a promising way to explore challenging issues of strongly correlated electron systems.
\end{abstract}

\maketitle

\section{Introduction}\label{sec:intro}
The mechanism of high-temperature superconductivity in cuprates \cite{bednorz86}  remains a major challenge in condensed matter physics \cite{keimer15}. In nearly 40 years since its discovery, an enormous amount of experimental data has been reported and accumulated by using a variety of experimental tools. Since each experimental result has often been analyzed independently, a comprehensive and consistent understanding based on a unified picture is often lacking.
To overcome such disjointed analyses and sometimes inconsistent conclusions, it is important to combine results from different but complementary experimental tools, in particular, energy ($\omega$) and momentum ($\kk$) resolved spectroscopies, and analyze them in an integrated fashion using a unified theoretical framework to reach a consistent understanding of the electronic structure.

As one of the major experimental tools, angle-resolved photoemission spectroscopy (ARPES) \cite{damascelli03,sobota21} has been developed and improved in step with the advances in the research history of the cuprates. 
A great advantage of ARPES is its accessibility to a momentum-resolved single-particle spectral function, while its applicable scope is limited to the occupied part of the spectra.
Although the unoccupied part of the spectra may hold a key to unlocking the mechanism of the high-temperature superconductivity, as well as the anomalous metallic behavior above the critical temperature ($\Tc$), no high-resolution photoemssion technique has been established for this purpose: Inverse photoemission spectroscopy, which can in principle measure the unoccupied spectra, has not yet reached an energy resolution sufficiently high to discuss the above issues.

Another experimental technique that can access the single-particle excitation spectra is the scanning tunneling spectroscopy/microscopy (STS/STM) \cite{fischer07}.
This technique can reveal the real-space electronic structure of a sample surface in an atomic resolution, measuring a quantity proportional to the local density of states (LDOS) at each position on both sides of the Fermi energy at a high energy resolution.
While atomic-scale modulation in the LDOS occurs within a unit cell, larger-scale modulation can be observed when a defect or impurity exists on the sample surface. 
This is owing to the interference effect between electron waves before and after the scattering by the defect or impurity, and is called quasiparticle interference (QPI) effect \cite{hasegawa93,crommie93}.
The QPI pattern observed for the cuprates shows various interesting signals \cite{hoffman02,mcelroy03,hanaguri04,hanaguri07,kohsaka08,lee09,alldredge12,he14,fujita14,machida16,gu19,yu19}, while it requires careful theoretical analyses to interpret them in terms of the structure of the single-particle excitation spectrum.

These experimental techniques have indeed revealed various anomalous electronic structures in the cuprates, e.g., the momentum-dependent pseudogap \cite{ding96} and superconducting gap \cite{shen93} including the arc-like truncated Fermi surface \cite{norman98,shen05,kondo09,kondo13}, growth of the pseudogap distinct from the superconducting gap \cite{tanaka06,kondo07}, asymmetric energy dependence of the density of states (DOS) between occupied and unoccupied sides \cite{hanaguri04} in contrast to the electron-hole symmetry expected in the superconducting phase, and the dopant-induced spatial inhomogeneity~\cite{pan01} 
as well as the stripe and checkerboard type inhomogeneity \cite{hanaguri04,kohsaka07}. 
For the occupied spectra, comparisons have been made
between the QPI results and autocorrelation of the ARPES data  \cite{markiewicz04,cheng05,mcelroy06,chatterjee06,fujita20}.
These analyses have argued that most of the strong QPI signals can be attributed to scatterings between tips of banana-shaped constant-energy contours of the spectral function, namely, endpoints of arc-like conspicuous spectral intensity.
However, a discrepancy
between ARPES and QPI/STM results has also been pointed out and remains puzzling:
For instance, the dispersion of a QPI signal (dubbed $\qq_7$ below) related to the $d$-wave superconducting gap seems to cut the Fermi energy at a finite momentum \cite{kohsaka08,lee09,vishik09}, unlike the clear $d$-wave gap structure observed in ARPES.

To consider these issues, we would first point out that the experimentally measured QPI spectra do not necessarily have the same spatial dependence, or equivalently the same momentum-space structure, as the LDOS: While the tunneling spectrum is proportional to the LDOS at each spatial position, the normalization factor could depend on the position, giving a spatial dependence different from that of the LDOS. Therefore, a more faithful calculation of the experimentally measurable quantities, beyond the LDOS, is necessary to discuss the QPI pattern. 

Here, by such a faithful calculation of the QPI pattern, we show that a simple theoretical model, derived from an electron fractionalization due to a strong correlation effect, gives a consistent explanation for many essential features of the observed results of ARPES, STM/STS and QPI, resolving the puzzles and showing the power of integrated spectroscopy analysis. It renders support for the validity of the fractionalization as well.
The model, which we call the two-component fermion model (TCFM) \cite{sakai16,sakai16PRB}, materializes the fractionalization in the simplest form of a one-body Hamiltonian involving a measurable quasiparticle and a hidden fermion, which hybridize with each other.
Through the hybridization with the hidden fermion, the quasiparticle acquires a pole singularity in the self-energy, which describes nonperturbative correlation effects and resultant formation of the Mott gap in Mott insulators and the pseudogap in doped Mott insulators.

The concept of the electron fractionalization has been proposed in various contexts such as the spin-charge separation in 1D Tomonaga-Luttinger liquid~\cite{tomonaga50,luttinger63}, solitons in polyacetylene~\cite{heeger88}, fractional quantum Hall state~\cite{laughlin83} and slave-boson or slave-fermion approaches for doped Mott insulators~\cite{sachdev03,kotliar86}. The present fractionalization is distinct from any of these proposals and represents two bistable fermion excitations splintered off from one electron to describe physics around the Mott insulator such as the cuprates.

The cuprates are considered to be doped Mott insulators and often modeled by the two-dimensional (2D) Hubbard model as one of the simplest models. 
Without paying serious attention to the fractionalization, the 2D Hubbard model has been extensively studied by the cluster extension \cite{kotliar01,maier05RMP} of the dynamical mean-field theory (cDMFT)~\cite{metzner89,georges96}
and related methods, for the ARPES spectra 
\cite{huscroft01,maier02,senechal04,civelli05,maier05RMP,senechal05,kyung06,aichhorn07,kancharla08,civelli08,liebsch09,okamoto10PRB82,chen12,gull13PRL,kohno14}, electronic Raman spectroscopy \cite{lin12,sakai13,loret16,loret17}, optical conductivity \cite{lin09,ferrero10,sakai13}, and STS \cite{ferrero09,sakai10,sordi12PRL,sakai13}.
These analyses were able to capture some aspects of the cuprate physics.

On the other hand, aside from the above clarifications, cDMFT studies have also revealed the existence of self-energy poles in the 2D Hubbard model, naturally accounting for the emergence of the Mott gap in the Mott insulator and the pseudogap in the low-energy region of the underdoped normal state, both of which are beyond the scope of the perturbative treatments of the electron correlation and are manifestations of the Mott physics~\cite{stanescu06,civelli09PRL,kyung09,sakai09,sakai10,gull15,sakai18}. Furthermore, the cDMFT studies in the $d$-wave superconducting phase revealed the existence of the poles in the normal and anomalous self-energies and their unexpected cancellation in the electron Green's function~\cite{sakai16,sakai16PRB}. However, the nature and root of the self-energy poles had been highly nontrivial at the time of the discovery of the cancellation.

The TCFM was first proposed~\cite{sakai16,sakai16PRB} to offer a unique way of understanding the above mentioned emergence of the self-energy poles and their enigmatic cancellation in the superconducting state discovered in the cDMFT studies.
The self-energy poles can be interpreted as a consequence of the hybridization of a quasiparticle with a hidden fermion degree of freedom, which is correctly described by the TCFM.
It was shown~\cite{sakai16PRB,imada24} that the Hubbard model at any carrier densities can be generally and exactly mapped onto a multi-component non-interacting fermion model with a hybridization between original electrons and hidden fermions  (see Sec. II.C.1 and II.C.2 in Ref.~\cite{sakai16PRB}, see also Sec.4 in Ref.~\cite{imada24}). Based on the Hubbard model studies, it was proposed~\cite{sakai16,sakai16PRB} that the Mott gap and pseudogap can be obtained as the hybridization gaps of these multi-component fermions. In particular, in the pseudogap and high-temperature superconducting states of our interest, essential physics can be captured by the mapping to the TCFM, which has only one hidden fermion energetically well separated from the energy scale of the Mott gap and can 
therefore be treated separately by incorporating the effect of the Mott-gap formation as the renormalization of the TCFM parameters.
For a review, see also Ref.~\cite{sakai23}. 

Furthermore, without relying on the Hubbard model, the machine-learning analysis of the ARPES data has shown the cancellation of the self-energy peaks consistently with the prediction of the TCFM~\cite{yamaji21}, suggesting the dominance of the fractionalization in real cuprates. 
The TCFM with the fractionalized electrons has also predicted an enhancement of high-energy excitonic signal below $T_{\rm c}$ in resonant inelastic X-ray spectroscopy~\cite{imada21}, which has been evidenced experimentally~\cite{singh22}.

Recently, other related methods such as ghost orbital in Ref.~\cite{lanata17} and an ancilla qubit in Ref.~\cite{zhang20}, capable of describing nonperturbative effects and the emergence of spectral gaps and self-energy poles, were proposed.

It has been argued that one of the two fermions in the TCFM is interpreted as the conventional quasiparticle and the other as an electron weakly bound to a hole~\cite{sakai18} or the incoherent part of the same electron~\cite{imada19}, respectively, which emerge as a consequence of local bistability ascribed to a direct consequence of the Mottness. The bistability directly generates emergent attraction and the Cooper pairs. In {\it ab initio} calculations of the cuprates, the emergence of the bistability and resultant effective attraction were evidenced counterintuitively in the local repulsive interaction term, where the local emergent attraction has shown one-to-one correspondence to the calculated $d$-wave superconducting long-ranged order~\cite{schmid23}.

Since the electron fractionalization offers an entirely different picture from the conventional electronic structure and experimental evidence is few, it is desired to test whether the fractionalization is indeed a valid and useful concept in real cuprates by thorough and combined analyses using independent spectroscopy tools. In this paper, we explore the consistency of ARPES and QPI in their integrated analyses by employing the TCFM to further critically examine the validity of the electron fractionalization and its microscopic basis. For this purpose, we first determine the TCFM parameters so as to reproduce the ARPES experimental result \cite{palczewski10} measured for an optimally-doped Bi$_2$Sr$_2$CaCu$_2$O$_{8+\d}$ (Bi2212).
We then calculate the QPI pattern based on the same model, and finally compare it with the experimental results for the same material.
For this comparison, we faithfully calculate 
experimentally measurable QPI spectra without resorting to the LDOS, which has often been used in the literature but is not directly measurable by QPI.
We find that the TCFM can reproduce the main features of both the ARPES and QPI experimental results at the same time and solve the known puzzle of the seemingly inconsistent results between the ARPES and QPI data on the quasiparticle dispersions mentioned above.

We further find that the available experimental QPI data are not consistent with the single-component description without fractionalization but are accounted for by the TCFM.
Through these integrated spectroscopy analyses, we establish that the electron fractionalization indeed takes place in cuprate superconductors. We also predict another possible piece of evidence for the fractionalization to be observed in the future QPI measurements for the high-energy region.

All in all, our work establishes that the hypothesis of electron fractionalization proposed in Refs.~\cite{sakai16,sakai16PRB,sakai18,imada19} is indeed at work in the integrated-spectroscopy analysis of ARPES and QPI for the real cuprate compound.

The rest of the paper is organized as follows.  In Sec.~\ref{sec:model}, we introduce the TCFM Hamiltonian. Section \ref{sec:method} explains our method (details of the method are described in Appendix \ref{sec:method_detail}). In Sec.~\ref{sec:arpes}, we show the ARPES experimental data.
The calculated results are presented and their comparison with the experimental data are discussed in Sec.~\ref{sec:result}.
Section \ref{sec:conclude} is devoted to a summary and outlook.

\section{Model}\label{sec:model}
As is outlined in Sec.~\ref{sec:intro}, we consider the following TCFM Hamiltonian, 
\begin{align}
H&= \sum_{\kk,\s}\left[ \eck \cksd \cks + \edk \dksd \dks 
         + \Vk( \dksd \cks +\cksd \dks ) \right] \nonumber\\
         &-\sum_{\kk}\left[ \Dck c_{\kk\ua} c_{-\kk\da} +\Ddk d_{\kk\ua} d_{-\kk\da} +\text{h.c.}\right], \label{eq:tcfm}
\end{align}
where $c$ and $d$ represent a quasiparticle and the hidden fermion, respectively.
$\e_{c(d) \kk}$ and $D_{c(d)}$ denote the normal and anomalous dispersions of $c$ ($d$), and $\Vk$ denotes the hybridization between them.
Integrating out the $d$ degrees of freedom from $H$, we obtain an effective action for $c$ (quasiparticle) degrees of freedom, where their normal and anomalous dispersions are modified by 
\begin{align}
 \Snor_{\TCFM}(\kk,\w)&= \frac{\Vk^2(\w+\edk)}{\w^2-\edk^2-\Ddk^2},\label{eq:snor}
\end{align}
and
\begin{align}  
  \Sano_{\TCFM}(\kk,\w)&=- \frac{\Vk^2 \Ddk}{\w^2-\edk^2-\Ddk^2}, \label{eq:sano}
\end{align}
respectively \cite{sakai16PRB,sakai16}.
These modifications can be interpreted as a self-energy for $c$.

Namely, as far as there is only a single self-energy pole in the low-energy region, the correlation effects in a many-body system (such as the Hubbard model and first-principles one for the cuprates) can be mapped onto a {\it one-body} Hamiltonian of Eq.~(\ref{eq:tcfm}), where the low-energy dynamics of the quasiparticle ($c$) can be described by the ``bare'' dispersions ($\e_c$ and $D_c$) and the self-energy corrections, Eqs.~(\ref{eq:snor}) and (\ref{eq:sano}).
In other words, the presence of a self-energy pole in a many-body system manifests a fractionalization of an electron into a quasiparticle (continuously connected from the bare electron through a renormalization) and a hidden fermion (corresponding to a self-energy pole, i.e., incoherent excitation).
In the particular case relevant to high-$\Tc$ cuprates, which are doped Mott insulators, this hidden fermion may correspond to an electron trapped around doped holes, namely, a fermionic part of a weakly bound exciton \cite{imada19}, that appears as the incoherent part of the carrier in experimental measurements such as ARPES, optics and transport, because such a state can reduce the interaction energy and hence be a low-energy excitation \cite{yamaji11PRB, yamaji11PRL}.
As was introduced in Sec.~\ref{sec:intro}, the {\it ab initio} calculations of the cuprates support the emergence of the bistable excitations, one, the conventional quasiparticle represented by $c$ and the other, the hidden fermion $d$, because the local Coulomb energy has double-well type dependence on the carrier density with the negative curvature in between, which induces emergent local attraction between carriers as well \cite{schmid23}.
A more detailed physical picture is discussed in Refs.~\cite{sakai16,sakai16PRB,sakai18,imada19,schmid23}. 

In this study, we do not address the origin of $d$ but utilize the Hamiltonian, Eq.~(\ref{eq:tcfm}), to calculate the QPI pattern.
However, just in case, we note that the origin of the two components has nothing to do with the separation of Cu-$d$ and O-$p$ contributions nor the bilayer structure of Bi2212. It has been established that the single-band description by the antibonding orbital of strongly hybridized Cu-$d$ and O-$p$ orbitals is enough for the analysis of low-energy physics of cuprates~\cite{moree22}. 
Instead, the two components in the TCFM emerge from the strong correlation effect of this single-band degree of freedom.
Although a unit cell of Bi2212 contains two CuO$_2$ layers, only the bonding orbital residing near the Fermi level is relevant to our analysis thanks to the bonding-antibonding splitting, justifying the single-layer TCFM Hamiltonian.

We determine the model parameters in Eq.~(\ref{eq:tcfm}) by fitting the ARPES experimental result presented in Sec.~\ref{sec:arpes}.
As the CuO$_2$ planes, well separated by dopant layers, form a square lattice, we employ 
\begin{align}
\e_{c(d) \kk}&=t_{c(d),0}-2t_{c(d),1}(\cos k_x+\cos k_y)\nonumber\\
&-4t_{c(d),2} \cos k_x \cos k_y-2t_{c(d),3} (\cos 2k_x + \cos 2k_y)\label{eq:ecdk}
\end{align}
with $t_{c(d),i}$ ($i=0,1,2,3$) denoting the onsite potential, and the nearest-, second-, and third-neighbor hoppings of the $c$ ($d$) fermion. 
The hybridization,
\begin{align}
\Vk=V_0 +V_1(\cos k_x +\cos k_y) +2V_2 \cos k_x \cos k_y,\label{eq:vk}
\end{align}
is parametrized in a similar way. 
We assume the $d$-wave superconductivity by \begin{align}
D_{c(d)\kk}=\frac{1}{2}D_{c(d),0}(\cos k_x-\cos k_y).\label{eq:dcdk}
\end{align}
The detailed procedure for the fitting is described in Appendix \ref{ssec:fit}.

While the TCFM Hamiltonian introduced in Eq.~(\ref{eq:tcfm}) captures the most singular structure of self-energy,
interactions among $c$, $d$ and other degrees of freedom, such as excitons mentioned previously, may remain~\cite{imada19}, which cause continuum structure in the self-energy of $c$ and $d$ fermions. 
To take these effects into account, we phenomenologically add less-singular components in the self-energy of $c$ and $d$ that induce their finite lifetimes and reproduce the tail or satellite structures in the real cuprates' spectral peaks.
The candidates for the origin of these additional self-energies
have been discussed already in the literature as the marginal Fermi liquid (MFL)~\cite{varma89}, possibly caused by the Planckian dissipation~\cite{sachdev10,zaanen19} near the Fermi level and the background~\cite{damascelli03,carbotte11} at high energies. We include these contributions as formulated in Appendix \ref{ssec:sig}.

\section{Method}\label{sec:method}
We determine the parameters in the TCFM, Eq.~(\ref{eq:tcfm}), by fitting the ARPES data \cite{palczewski10}, which will be presented in Sec.~\ref{sec:arpes}.
Then, the self-energy for the $c$ fermion is obtained through Eqs.~(\ref{eq:snor}) and (\ref{eq:sano}), with additional contributions (background and MFL components) mentioned above. With this self-energy,
the Green's function for the $c$ fermions,
\begin{align}
 \hat{G}(\kk,\w)=
 \left(
    \begin{array}{cc}
     \Gkw & \Fkw \\
     \Fkw & G^\ast(\kk,-\w)
    \end{array}
   \right),   
\label{eq:Ghat}
\end{align}
is given as Eq.~(\ref{eq:nambu}) in Appendix \ref{sec:method_detail}. Then, the spectral function $A(\kk,\omega)$ and the density of states (DOS) for $c$ fermions can be derived from $G(\kk,\omega)$. See Appendices \ref{ssec:sig} and \ref{ssec:fit} for technical details.

While the ARPES and STM offer information about $A(\kk,\omega<0)$ and the DOS, the QPI measures the local ($\rr$-dependent) differential conductance, 
\begin{align} 
   g(\rr,eV) \equiv \dv{I}{V} =e\Iset \frac{\rho(\rr,eV)}{\int_0^{e\Vset}\rho(\rr,\w)d\w}, \label{eq:grw_txt}
\end{align}
modulated by the impurity scattering, where $\rho(\rr,\w)$ is the local density of states.
Here, $e$ is the electron charge and $V$ denotes the bias voltage applied to the sample. $\Iset$ and $\Vset$ refer to the set-point current and voltage, respectively, used for the constant-current scanning.
Since the denominator of Eq.~(\ref{eq:grw_txt}) depends on $\rr$ and $\Vset$, the spatial modulations in $g(\rr,eV)$ and $\rho(\rr,\w)$ may exhibit qualitative differences depending on $\Vset$ \cite{kohsaka07}.
To mitigate this so-called set-point effect, we introduce a normalized conductance $L(\rr,eV)$ defined as
\begin{align}
  L(\rr,eV)\equiv \frac{V}{I}\dv{I}{V}=\frac{V}{I}g(\rr,eV).\label{eq:Lrw_txt}
\end{align}
While the spatial modulations may still differ between $L(\rr,\w)$ and $\rho(\rr,\w)$, $L(\rr,\w)$ is independent of $\Vset$ and hence reflects the intrinsic electronic structure more explicitly than $g(\rr,\w)$.
See Appendix \ref{sssec:gandL} for more details.

$\r(\rr,\w)$ can be decomposed into the nonperturbed part $\r_0(\rr,\w)$ (directly given by the TCFM) and the impurity-induced modulation $\d\r(\rr,\w)$ as $\r=\r_0+\d\r$. The Fourier transform of $\d\r(\rr,\w)$ is calculated as
\begin{align}
  \d\r(\qq,\w)&=\frac{i}{2\pi N_{\kk}N_c} \sum_{\rr'\kk} w_{\kk}(\rr')w_{\kk+\qq}^\ast(\rr') e^{-i\qq\cdot\rr'} \nonumber\\
                &\times[\hat{G}(\kk,\w)\Vi(\qq)\hat{G}(\kk+\qq,\w) \nonumber\\
                &-\{\hat{G}(\kk+\qq,\w)\Vi(\qq)\hat{G}^\ast(\kk,\w)\}]_{11}, \label{eq:drhoqw_txt}
\end{align}
where $\Vi$ is the impurity potential,
$w_{\kk}(\rr)$ is the tail of the Wannier function at the surface, and $N_{\kk}$ ($N_c$) denotes the number of momentum points (real-space points in the unit cell) used in the calculation. See Appendix \ref{sssec:ldos} for details.
In practice, we first calculate Eq.~(\ref{eq:drhoqw_txt}) and obtain $\d\r(\rr,\w)$ through the inverse Fourier transformation. Then, we calculate $g(\rr,eV)$ and $L(\rr,eV)$ through Eqs.~(\ref{eq:grw_txt}) and (\ref{eq:Lrw_txt}). By Fourier-transforming $g(\rr,eV)$ and $L(\rr,eV)$, we obtain $g(\qq,eV)$ and $L(\qq,eV)$ of our main interest.

\section{ARPES experimental data}\label{sec:arpes}
\begin{figure}[tb]
\centering
\includegraphics[width=0.48\textwidth]{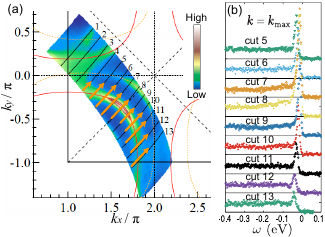}
\caption{ARPES experimental result for the optimally-doped Bi2212 \cite{palczewski10}. (a) Momentum-space map of the low-energy spectral intensity. 
The red solid (orange dashed) curves denote the normal-state Fermi surface (shadow band), black solid lines denote the momentum cuts along which the dispersion is measured, and the orange arrows on them denote the data used in the fitting in this study. (b) The EDCs at $\kk=\kk_{\rm max}$ along the cuts 5 to 13.} \label{fig:arpes}
\end{figure}

We use high-resolution laser ARPES experimental data measured for an optimally-doped Bi2212 sample ($\Tc=90$K) \cite{palczewski10}, which has a layered perovskite structure with conducting CuO$_2$ double layers separated by one Ca layer and sandwiched by SrO and BiO layers.
Figure \ref{fig:arpes}(a) shows a momentum-space map of the low-energy spectral intensity in the superconducting state ($T=12$K).
The strongest intensity in the lower-left quadrant of the Brillouin zone (denoted by red solid curves) roughly corresponds to the position of the normal-state Fermi surface.
Black lines, 1 to 13, represent the momentum cuts used in the following analysis.
We define $\kk_{\rm max}$ for each cut as the wave number giving the strongest low-energy intensity.
We note that another signal around $(2\pi,0)$ comes from a superstructure modulation specific to Bi2212 \cite{norman95}.
This superstructure modulation can also be seen in the QPI experimental data as additional spots in the direction of the modulation while such additional spots do not exist in the direction perpendicular to the modulation. In the analysis in the subsequent sections, we exclude these superstructure-induced spectra in the ARPES data by choosing the momentum sections denoted by orange arrows, and compare the calculated QPI results with the experimental data in the direction without the modulation.

Figure \ref{fig:arpes}(b) shows the energy-distribution curves (EDCs) at $\kk=\kk_{\rm max}$ for cuts 5-13.
The low-energy peak shifts to a higher binding energy as the momentum goes from the node (N, cut 8), where the superconducting gap vanishes, to the antinode (AN, cut 13), where the superconducting gap is maximized.
Note that the dispersion along the cut 8 crosses the Fermi energy at $\kk=(0.38\pi,0.38\pi)$ while the spectral peak of the minimal binding energy is located at $\kk=(\pi,0.14\pi)$ along the cut 13, as was identified in Ref.~\cite{palczewski10}. Hereafter, we denote $(0.38\pi,0.38\pi)$ as $\kN$ and $(\pi,0.14\pi)$ as $\kAN$.
Notably, a flat intensity is seen below $\sim -0.1$ eV for all the momenta. 
This feature cannot be reproduced solely by the TCFM, which applies only to a low-energy region ($|\w|\lesssim 0.1$ eV).
We discuss this feature in Appendix \ref{sssec:bg}, where we take it into account by introducing an additional self-energy.
The energy dispersions along cuts 4-13 are displayed in Fig.~\ref{fig:disp}.

\section{Results}\label{sec:result}
\subsection{Spectral function}\label{ssec:akw}
We first show the results obtained by fitting the ARPES data through the optimization procedure described in Appendix \ref{ssec:fit}.
The best parameters of the model (\ref{eq:tcfm}) are determined as $t_{c0}=0.700$, $t_{c1}=0.545$, $t_{c2}=-0.218$, $t_{c3}=0.0250$, $t_{d0}=0.0545$, $t_{d1}=-0.0161$, $t_{d2}=-0.0137$, $V_0=0.0548$, $V_1=0.0623$, $V_2=-0.0608$, $D_{c,0}=0.0668$, $D_{d,0}=0.0912$,   $\sr=0.008$, and $\eta=0.003$ in units of eV, and $f_{\bg,d}=0.222$ and $c_\bg=0.218$.
For these parameters, the cost function, Eq.~(\ref{eq:c}), is 0.452.
The present optimized model indeed reproduces the unoccupied part of the STS spectrum as well.
To confirm the quality of the fitting, we also quantify the deviation from the ARPES data and conclude that our model reproduces the inherent spectrum within the experimental noise
(see Appendix \ref{sssec:cost} for details).

\subsubsection{Dispersion}\label{sssec:disp}
\begin{figure*}[tb]
\centering
\includegraphics[width=0.96\textwidth]{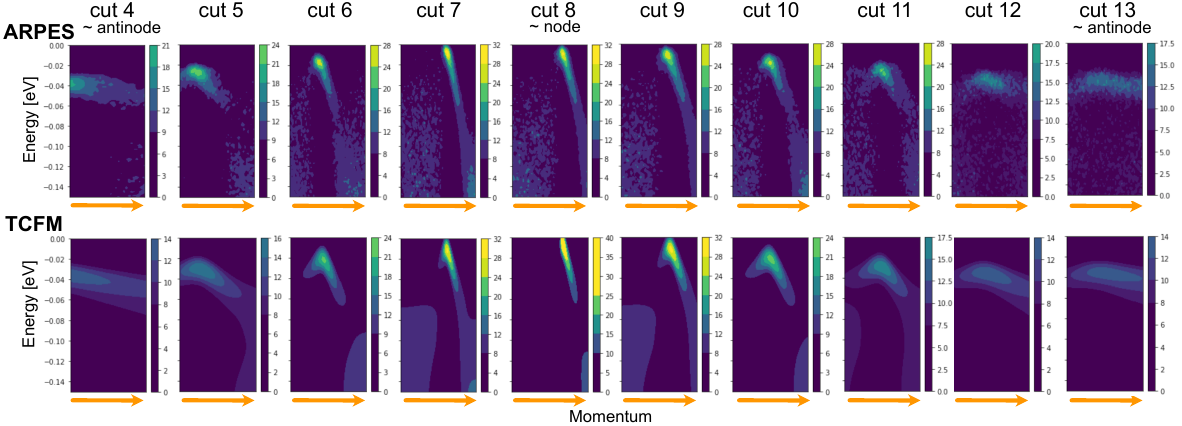}
\caption{Comparison between $\tilde{I}_{\text{ARPES}}(\kk,\w)$ and $\tilde{I}_{\text{TCFM}}(\kk,\w)$. Orange arrows denote the momentum cuts indicated in Fig.~\ref{fig:arpes}(a).} \label{fig:disp}
\end{figure*}

\begin{figure}[tb]
\centering
\includegraphics[width=0.48\textwidth]{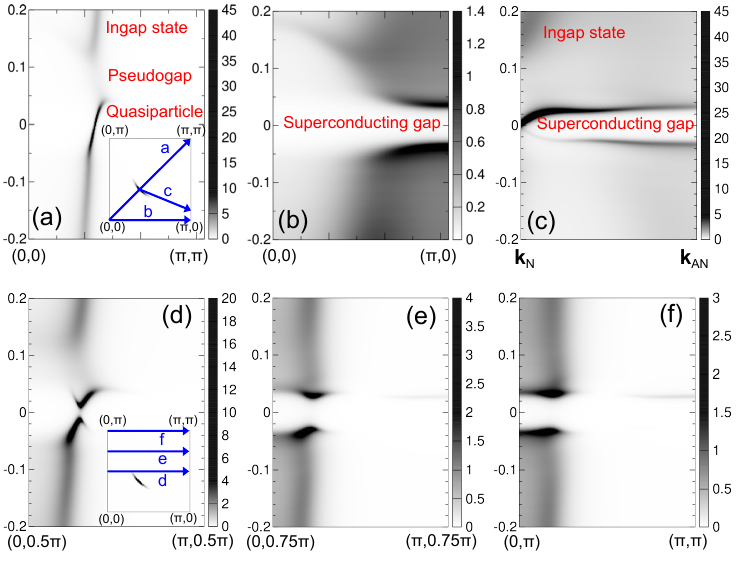}
\caption{$\Akw$ calculated for the TCFM along (a) $(0,0)-(\pi,\pi)$, (b) $(0,0)-(\pi,0)$, (c) $\kN-\kAN$, (d) $(0,0.5\pi)-(\pi,0.5\pi)$, (e) $(0,0.75\pi)-(\pi,0.75\pi)$, and (f) $(0,\pi)-(\pi,\pi)$. Insets to (a) and (d) illustrate the momentum cuts used in (a-f).} \label{fig:akw_disp}
\end{figure}

Figure \ref{fig:disp} compares the energy dispersions along the cuts 4-13 [shown in Fig.~\ref{fig:arpes}(a)] between ARPES \cite{palczewski10} and the TCFM.
We see that the theory reproduces well the dispersion and the normalized intensity of the ARPES result for all the momentum cuts. 
Around the antinode (cuts 4 and 13), we see broad and flat spectra around $-0.04$ eV and a gap opens above them.
With approaching the node (cut 8), the spectra become sharper and the gap diminishes, where a back-bending of the dispersion becomes noticeable (cuts 5-7 and 9-12). 
In the nodal direction (cut 8), the spectra show a strong intensity at low energy, which is a quasiparticle band cutting the Fermi energy.
Note that even the small-intensity structures at high binding energies ($\w\lesssim-0.05$ eV) are well reproduced by our theory.

Based on this agreement for $\w<0$, the TCFM predicts the spectral function for $\w>0$.
Figure \ref{fig:akw_disp} shows $A(\kk,\w)$ calculated for the TCFM up to $\w=\pm 0.2$ eV along various momentum cuts.
In the nodal direction, $(0,0)-(\pi,\pi)$, $A(\kk,\w)$ shows a sharp dispersive band around the Fermi energy, pseudogap for $0.05\lesssim\w\lesssim0.1$ [eV], and the ingap state above it [(a)].
Around $(\pi,0)$, the gap opens for 
$|\w|\lesssim0.04$ eV, and broad spectra appear for $|\w|\gtrsim0.04$ eV [(b,c)]. 
The latter is due to the large background self-energy (see Appendix \ref{sssec:bg}).
In addition, we plot in Figs.~\ref{fig:akw_disp}(d-f) $A(\kk,\w)$ along the cuts often used in ARPES experiments. The results indeed well reproduce the experimental results (see Fig.~2 in Ref.~\cite{li2018coherent} for instance), including the kink between the strong-intensity renormalized band at low binding energies and a steeply-dispersive broad spectra below $\sim -50$meV \cite{kaminski01}.
Qualitatively similar spectra have been obtained by the previous cDMFT calculation \cite{sakai13} for the underdoped regime of the 2D Hubbard model as well while a difference is that the TCFM used in this study fits the ARPES spectra of an optimally-doped sample.

In Appendix \ref{app:disp_ImSbg=0}, we show the spectra without the smearing by the background self-energy (i.e., Im$\Sbg=0$ with keeping Re$\Sbg$), where we find a clear dip above the quasiparticle band also in the antinodal region.
In Appendix \ref{app:disp_Sbg=0}, we further plot the spectra by switching off all of the background self-energy (i.e., $\Sbg=0$), to see the electronic structure inherent to the TCFM.

\subsubsection{Energy-distribution curve}\label{sssec:edc}
\begin{figure}[tb]
\centering
\includegraphics[width=0.48\textwidth]{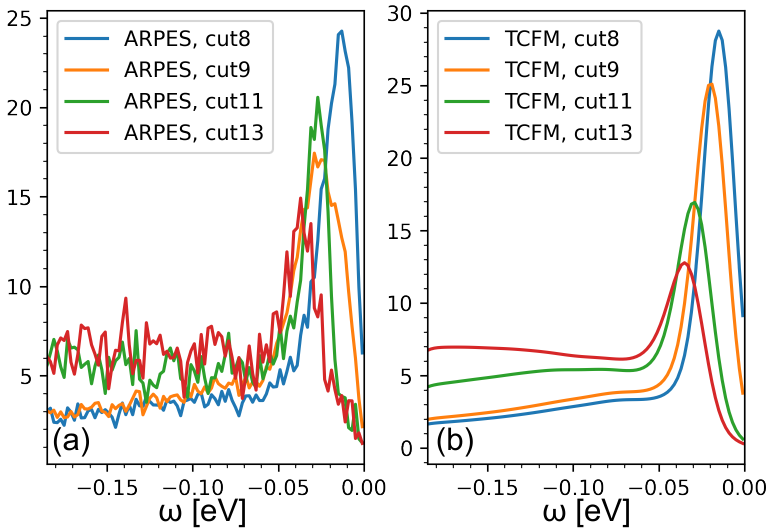}
\caption{Energy distribution curves obtained by (a) ARPES ($\tilde{I}_{\rm ARPES}$) and (b) the TCFM ($\tilde{I}_{\rm TCFM}$) at $\kk=\kk_{\rm max}$ in each cut.}  \label{fig:edc_kf}
\end{figure}

\begin{figure}[tb]
\centering
\includegraphics[width=0.48\textwidth]{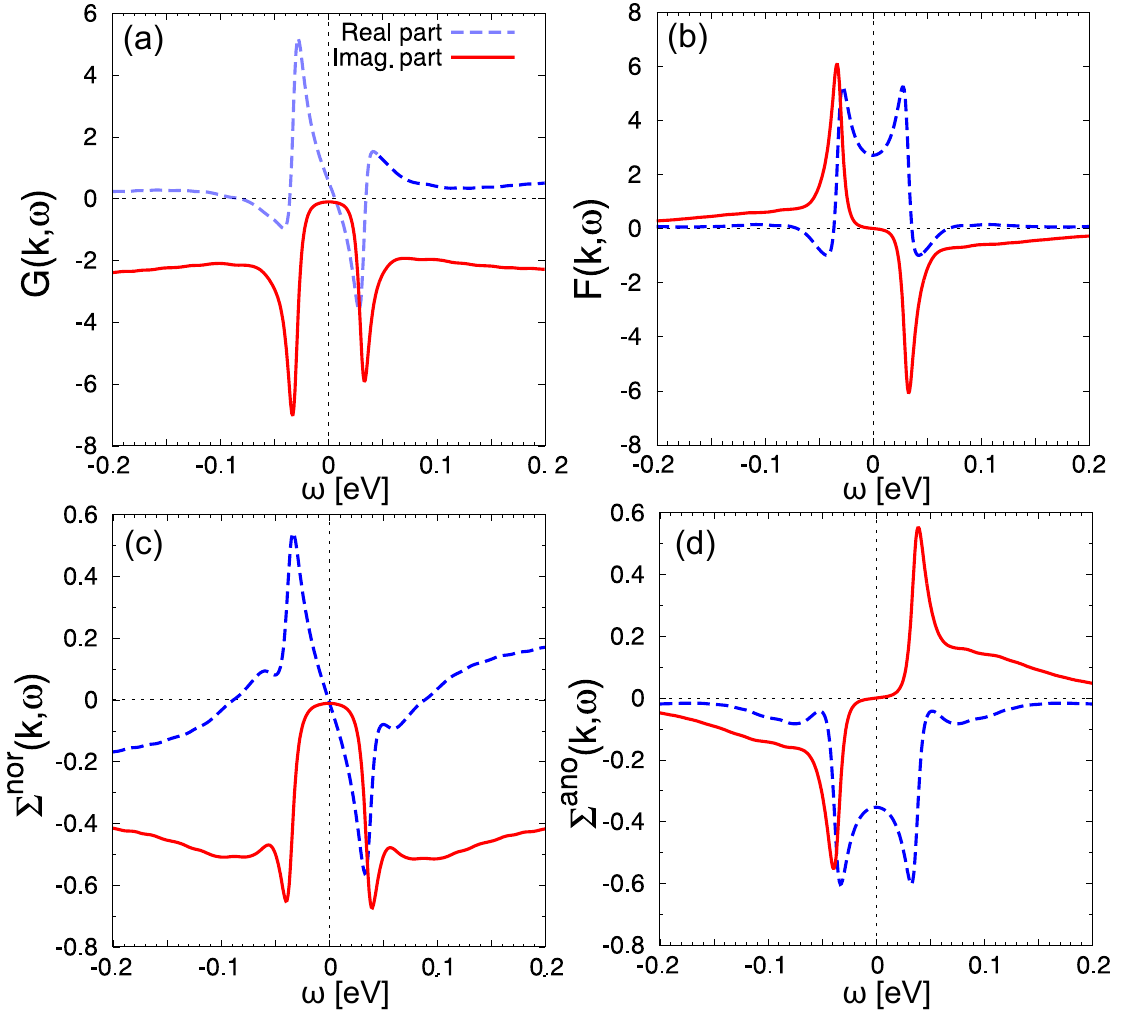}
\caption{Real (blue dashed curve) and imaginary (red solid curve) parts of (a) $G$, (b) $F$, (c) $\Snor$, and (d) $\Sano$ at the antinode, $\kk=\kAN=(\pi,0.14\pi)$, calculated by the fitted TCFM.} \label{fig:edc_an}
\end{figure}

Figure \ref{fig:edc_kf} plots the normalized EDCs at $\kk=\kk_{\rm max}$ on several cuts, obtained by (a) ARPES and (b) the theory.
The overall structure of the experimental spectra is well reproduced by our theory.
Notice that the ARPES data do not always show a monotonic decrease of the low-energy intensity with moving from the node (cut 8) to the antinode (cut 13): For instance, cut 11 shows a higher maximum than cut 9.
This would be attributed to errors arising from the limited statistics of electrons counted in the ARPES measurement as is implied by the large fluctuation of the data, and limits the fitting by the TCFM assuming a simple $\kk$ dependence [i.e., Eqs.~(\ref{eq:ecdk}), (\ref{eq:vk}), (\ref{eq:dcdk}), (\ref{eq:sbg}), and (\ref{eq:edk})].
The EDCs for other momenta along these cuts and corresponding self-energies are presented in Figs.~\ref{fig:edc} and \ref{fig:edc_sig}, respectively, in Appendix \ref{app:EDC_comparison}.

Figure \ref{fig:edc_an} shows the $\w$ dependence of the Green's function and the self-energy of the TCFM at $\kk=\kAN$. 
According to Eq.~(\ref{eq:akw}),  the spectral function can be inferred from Im$G$, which shows (negative) peaks at $\w\simeq\pm 35$ meV.
This is consistent with the experimental value ($-38$ meV) [see cut 13 in Fig.~\ref{fig:edc_kf}(a)].
Im$F$ shows sharp peaks at the same energies, and Re$F$ is significantly enhanced in the lower-energy region.
Im$\Snor$ and Im$\Sano$ also show low-energy peaks at electron-hole symmetric positions, $\w\simeq\pm40$ meV.
This is consistent with the self-energy derived by a machine learning analysis \cite{yamaji21} of the same ARPES data. 
As a notable property of the TCFM, the trace of the self-energy peaks disappears in $G$ and $F$ due to a cancellation between $\Snor$ and $W$ at these peak energies \cite{sakai16,sakai16PRB} [see Eq.~(\ref{eq:w}) for the definition of $W$].
The $\w$-dependent structure of the self-energy for other momenta is presented in Figs.~\ref{fig:edc_sig} and \ref{fig:sig_kf} in Appendix \ref{app:EDC_comparison}.

\subsubsection{Momentum-space maps}\label{sssec:kmap}
\begin{figure*}[tb]
\centering
\includegraphics[width=0.96\textwidth]{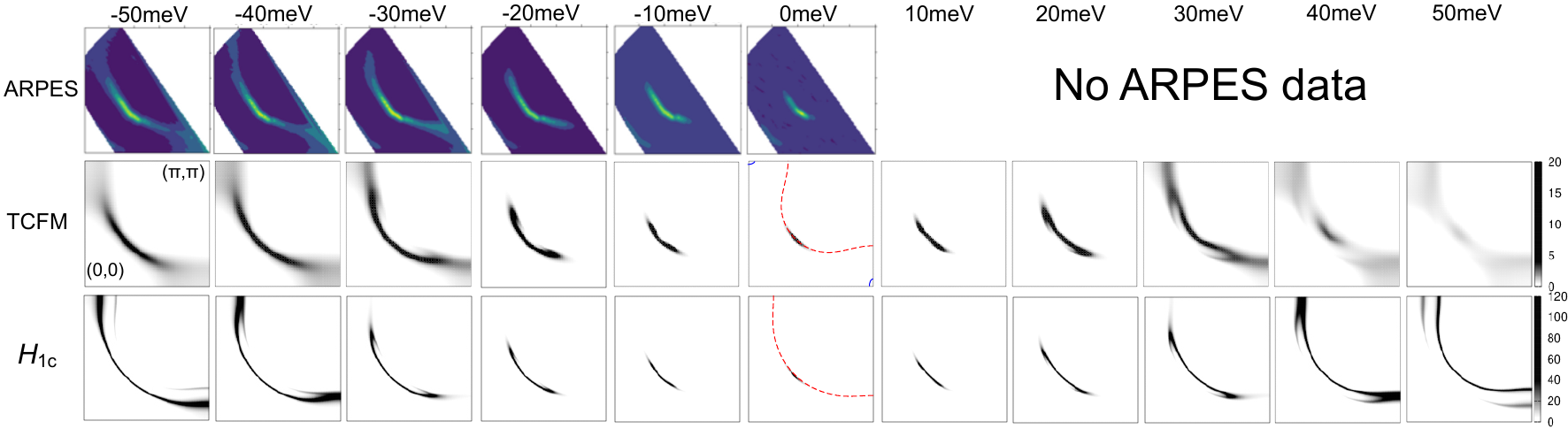}
\caption{Comparison between momentum-space maps of ARPES spectra (top panels) and of $\Akw$  calculated for the TCFM (middle panels) and the single-component Hamiltonian, Eq.~(\ref{eq:H1c}), (bottom panels) at various energies.
Red dashed curves at $0$meV in the theoretical results depict the Fermi surface of the normal state ($D_c=D_d=0$). In the TCFM result, we also plot tiny normal-state Luttinger surfaces (zeros of the Green's function) by blue solid curves around $(\pi,0)$ and $(0,\pi)$.}
 \label{fig:akw_kmap}
\end{figure*}

The upper and middle panels in Fig.~\ref{fig:akw_kmap} compare the momentum-space maps of spectral intensity between ARPES and the TCFM at various energies. 
The ARPES data are available only for $\w\leq 0$.
At $\w=0$, we can see a small arc around the nodal direction. This is due to a finite broadening of the spectra.
As $|\w|$ increases, the arc expands because of the $d$-wave structure of the superconducting gap.
For $\w\leq -30$ meV, the arc reaches the Brillouin-zone boundary, where the spectra are rather broadened.
These are all consistent between the experiment and the theory.
A notable observation in the theory data is that the edges of the arc (seen for $|\w|\leq30$ meV) wrap around inwards (outwards) for $\w<0$ ($\w>0$).
This electron-hole asymmetric behavior results from a combination of the electron-hole asymmetric position of the Fermi surface and the symmetric superconducting gap.

The bottom panels show $\Akw$ calculated with an ordinary mean-field (single-component) Hamiltonian,
\begin{align}
 H_{1c}= \sum_{\kk,\s} \eck \cksd \cks 
         &-\sum_{\kk}\Dck\left(  c_{\kk\ua} c_{-\kk\da} +\text{h.c.}\right),   \label{eq:H1c}
\end{align}
without $d$ degrees of freedom.
The parameters, $\eck$ and $\Dck$, in Eq.~(\ref{eq:H1c}) follow Ref.~\cite{norman95}, which fitted the ARPES results for an optimally-doped Bi2212 with the Hamiltonian, Eq.~(\ref{eq:H1c}).
For $|\w|\lesssim 20$meV, the momentum-space structure is qualitatively similar to that of the TCFM although the spectra are much sharper in $H_{1c}$. 
However, for $|\w|\gtrsim 30$meV, the $H_{1c}$ spectra show strong intensity even in the antinodal region, in contrast to the broadened weak spectra in the TCFM and ARPES in the same region. 
This weak and broad signal in the TCFM, as well as in the ARPES, is ascribed to a large imaginary part of the self-energy (shown in Fig.~\ref{fig:S_kmap}) in this region. This self-energy effect is nothing but the presence of the $d$ fermion: While it mainly resides above the Fermi level in the normal state, the particle-hole counterpart appears below the Fermi level, too, in the superconducting state.

\subsection{Density of states}\label{ssec:stm}
\begin{figure}[tb]
\centering
\includegraphics[width=0.48\textwidth]{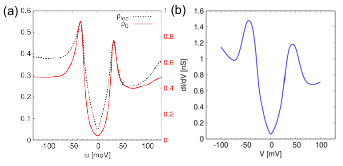}
\caption{Comparison between (a) the calculated DOS and (b) the STM experimental result of $dI/dV$. 
In panel (a), $\rho_{\rm loc}$ is the DOS of the 
TCFM, defined by Eq.~(\ref{eq:rloc}), while $\rho_0$ is the DOS projected onto the surface (and hence more suitable for the comparison with the STM experimental result) as defined by Eq.~(\ref{eq:rho0}).}
 \label{fig:dos}
\end{figure}

Figure \ref{fig:dos} compares the calculated DOS with the site-averaged $dI/dV$ measured by the STM/STS.
We find good agreement between them: (i) peaks at the gap edges, (ii) electron-hole asymmetry, and (iii) V shape around zero bias.
Although the TCFM parameters are determined to reproduce the ARPES data that are available only on the occupied side, the theory remarkably reproduces the DOS not only on the occupied side but also on the unoccupied side. 
Note that the difference between $\rho_{\rm loc}$ [Eq.~(\ref{eq:rloc})] and 
$\rho_0$ [Eq.~(\ref{eq:rho0})] comes from the projection onto the $d$-wave Wannier function [Eq.~(\ref{eq:wannier})], which vanishes in the nodal direction: $\rho_0$ is less contributed from the nodal region, which has low-energy excitations as well as the ingap states around 100 meV.

A discrepancy is found, where the STM data show a clear dip around -70 meV and a hump below it 
whereas they are less prominent in the TCFM.
This discrepancy originates from the apparent difference 
between the STM and the ARPES: the dip-hump structure is substantially weaker in ARPES at all the momenta, as seen in Figs.~\ref{fig:arpes}(b) and \ref{fig:edc_kf}(a), in contrast to Fig.~\ref{fig:dos}(b) of the STM.
Because the TCFM parameters are based on the ARPES data, 
Fig. \ref{fig:dos}(a) show only a weak dip-hump structure.
Note that even in a combined experiment of ARPES and STM for the same sample \cite{fujita20}, the angle-integrated photoemission DOS exhibits a weaker dip-hump structure than the STM result.
The origin of this quantitative discrepancy is unclear but several possibilities, like different matrix-element effects and surface sensitivities, can be envisaged.
Even for ARPES spectra only, a variety of experimental data have been reported with some sensitivities including sample, doping concentration, temperature, momentum dependence and photon energy, where the origin of the variation of the peak-dip-hump structure has been debated~\cite{damascelli03}. 

In our TCFM framework, the hump is generated by the contribution from the fractionalized hidden fermion $d$, although it is superimposed somehow by a large background, particularly around the antinode. Although the main properties discussed in this work are robust, we note that the depth of the dip and the height of the hump are subject to change by a small modification of the model parameters for the $d$ fermions, resulting in a structure being more similar to the STM result. These parameter variations could come from the difference in the surface measurement conditions of QPI from ARPES.

\subsection{Comparison with QPI}\label{ssec:qpi}
\begin{figure}[tb]
  \centering
\includegraphics[width=0.48\textwidth]{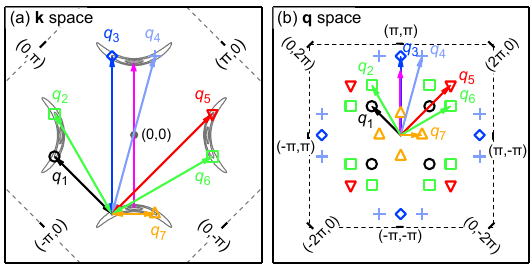}
  \caption{
    Schematic picture of the relation between (a) the $\kk$-space structure of the spectral function and (b) the QPI signals in the $\qq$ space.
    $\qq_i (i=1, ..., 7)$'s denote momenta connecting the tips of the banana-shaped spectral intensity. The vertical magenta arrows passing through $(0,0)$ represent the new branch observed nearby $\qq_3$.
  }\label{fig:octet}
\end{figure}

We now turn to the QPI patterns calculated through the procedure described in detail in Appendix \ref{ssec:method_qpi}.
To specify the QPI signals, we employ a conventional notation, $\qq_i (i=1, ... ,7)$, as shown in Fig.~\ref{fig:octet}.
This notation has been used in the interpretation based on the octet model \cite{hoffman02,mcelroy03}, where the momenta connecting the tips of the banana-shaped spectral intensity in the $\kk$ space [Fig.~\ref{fig:octet}(a)] are considered to make a main contribution and give the strong-intensity spots in the QPI pattern in the $\qq$ space [Fig.~\ref{fig:octet}(b)].

We emphasize that our calculation is not based on the octet model but fully takes into account the contribution of the Green's function from the whole $\kk$ space. We examine the adequacy of the octet model through the comparison between our calculated results and the experiment.

\subsubsection{Overall features of QPI dispersions} \label{sssec:qpidisp}

\begin{figure*}[tb]
  \centering
  \includegraphics[width=0.95\textwidth]{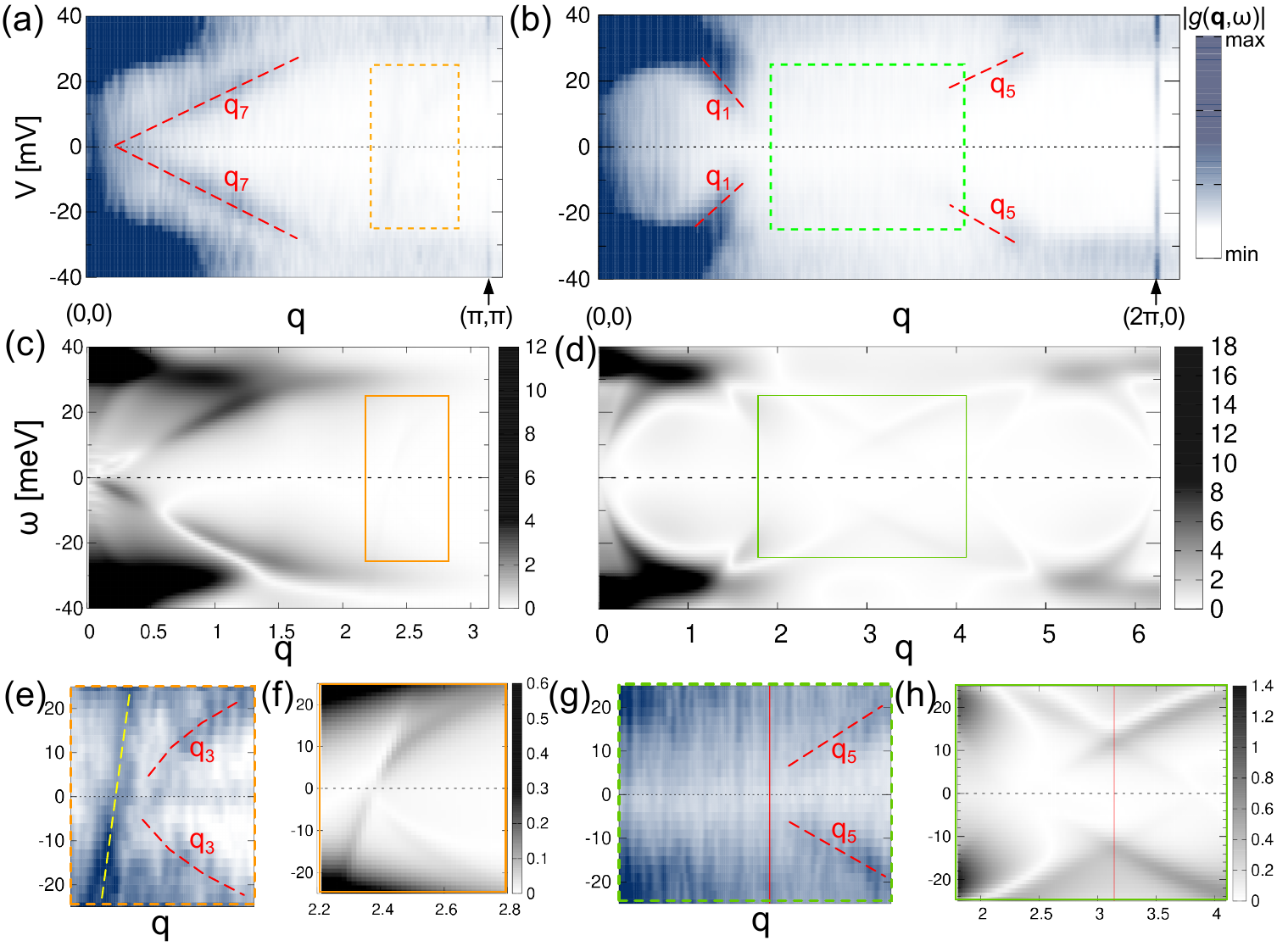}
  \caption{
    Comparison of $|g(\qq,\w)|$ between (a,b) the experiments and (c,d) the TCFM along the $(0,0)-(\pi,\pi)$ and $(0,0)-(2\pi,0)$ lines.
    $V_{\rm set}=-0.1$V. Zoom-in structures are shown in (e), (f), (g), and (h) for areas enclosed by rectangles in (a), (c), (b), and (d), respectively, with different gray scales for better visibility of the dispersions.
    In (e) and (f), $\qq_3$ is visible as electron-hole symmetric dispersions while an additional electron-hole asymmetric dispersion (denoted by yellow dashed line in (e)) is seen on the smaller-$\qq$ side of $\qq_3$.
    In (g) and (h), electron-hole symmetric $\qq_1$ and $\qq_5$ are visible in small and large $\qq$ regions, respectively.
    Red vertical lines denote $(\pi,0)$.
  }\label{fig:gdisp}
\end{figure*}

\begin{figure*}[tb]
  \centering
  \includegraphics[width=0.95\textwidth]{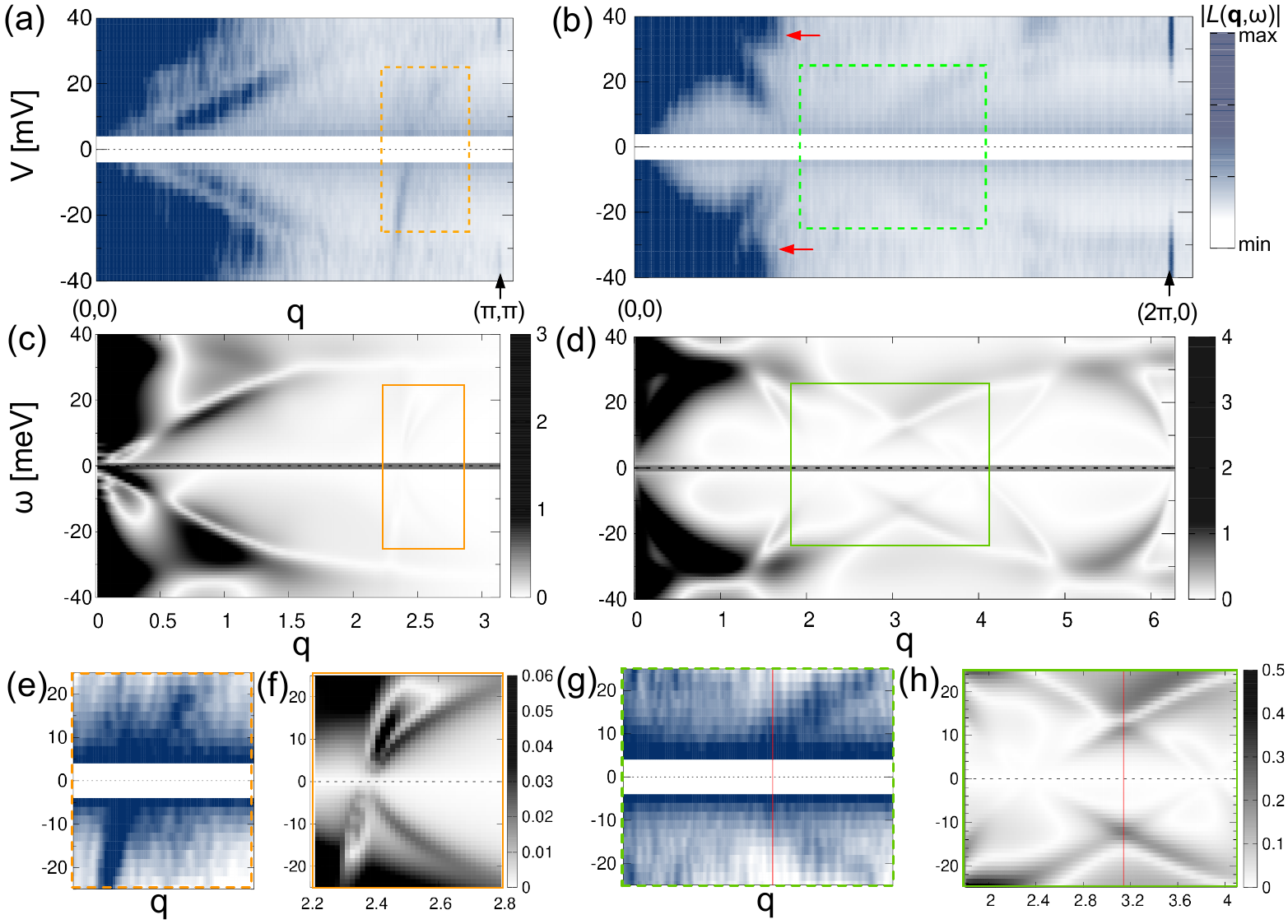}
  \caption{
    Comparison of $|L(\qq,\w)|$ between (a,b) the experiment and (c,d) the TCFM along the $(0,0)-(\pi,\pi)$ and $(0,0)-(2\pi,0)$ lines.
    Zoom-in structures are shown in (e), (f), (g) and (h) for areas enclosed by rectangles in (a), (c), (b) and (d), respectively, for the same purpose as in Fig.~\ref{fig:gdisp}.
    Red arrows in (b) indicate the mirror signals of $\qq_5$ discussed in Appendix \ref{ssec:decom}.
    Red vertical lines in (g) and (h) denote $(\pi,0)$.
    In (a), (b), (e), and (g), it is difficult to obtain accurate experimental data for small $|V|$ because both $\dv{I}{V}$ and $\frac{I}{V}$ in $L=\frac{V}{I}\dv{I}{V}$ are not large enough compared to noise.
  }\label{fig:ldisp}
\end{figure*}

\begin{figure*}[tb]
  \centering
  \includegraphics[width=0.95\textwidth]{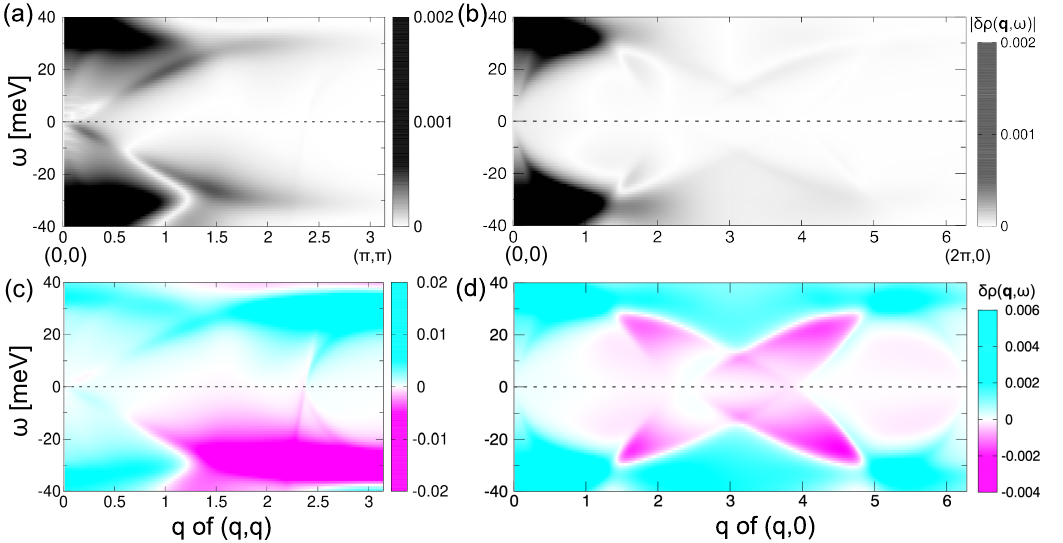}
  \caption{
    (a, b) $|\d \rho (\qq,\w)|$ calculated with the TCFM for $r_0=r_{0SC}=0.4a$ along the $(0,0)-(\pi,\pi)$ and $(0,0)-(2\pi,0)$ lines.
    (c, d) Corresponding $\d \rho$ (with signs) calculated with the TCFM for $\qq$-independent $\Vi$.
  } \label{fig:drhodisp}
\end{figure*}

Figures \ref{fig:gdisp}(a) and \ref{fig:gdisp}(b) show experimental $|\gqw|$ along high-symmetry directions.
The signals $\qq_7$ and $\qq_3$ are observed along the $(0,0)-(\pi,\pi)$ direction [Figs.~\ref{fig:gdisp}(a) and ~\ref{fig:gdisp}(e)], while $\qq_1$ and $\qq_5$ appear along the $(0,0)-(2\pi,0)$ direction [Figs.~\ref{fig:gdisp}(b) and ~\ref{fig:gdisp}(g)].
The signals $\qq_3$, $\qq_5$, and $\qq_7$ merge into a flat background for $|V|\gtrsim$ \SI{30}{\milli\volt} and large $|\qq|$.

In addition to these previously known signals, we identified an electron-hole asymmetric branch across the Fermi energy at wavenumbers slightly smaller than $|\qq_3|$, as shown in Fig.~\ref{fig:gdisp}(e) (yellow dashed line).
This branch originates from a scattering between the opposite nodal points, as indicated by the magenta arrow in Fig.~\ref{fig:octet}(a).
The disappearance of the $d$-wave superconducting gap in the nodal direction is consistent with the electron-hole asymmetric dispersion of this signal.
According to the stationary phase approximation of QPI \cite{liu12}, it is reasonable to have this branch because the scattering between the nodal points satisfies the stationary condition.
However, this branch has been overlooked in experiments for more than two decades, probably because of its low intensity.
Although this electron-hole asymmetric branch goes beyond the octet model, our theoretical model accurately reproduces it, as we will show below.
Details of this branch will be discussed elsewhere \cite{kohsaka_unpub}.

For $|V|\lesssim 20$mV, a strong intensity around $\qq=0$ is limited to small $|\qq|$ while it extends to much larger $|\qq|$ for $|V|\gtrsim 20$mV.
All these features are well reproduced in the calculated results in Figs.~\ref{fig:gdisp}(c), \ref{fig:gdisp}(d), \ref{fig:gdisp}(f), and \ref{fig:gdisp}(h).
The strong intensity for $\qq\simeq \Vec{0}$ in the experimental results will be attributed to the effect of multiple impurities or mesoscopic fluctuation of the doping concentration \cite{dellanna05}.

Figure \ref{fig:ldisp} shows the corresponding maps of $|\Lqw|$.
All the above-mentioned features can also be seen here, which means that these features are not due to the set-point effect and are inherent to the cuprate.

Figures \ref{fig:drhodisp}(a) and \ref{fig:drhodisp}(b) show $|\d\rho(\qq,\w)|$, which is more directly connected to $\hat{G}(\kk,\w)$ [see Eq.~(\ref{eq:drhoqw})] than $|g(\qq,\w)|$ and $|L(\qq,\w)|$ though not directly measurable in experiments.
The resemblance of Figs.~\ref{fig:drhodisp}(a,b) to Figs.~\ref{fig:gdisp} and \ref{fig:ldisp} ensures that the origin of the signals observed in the $g$ and $L$ maps can be discussed in terms of $\hat{G}(\kk,\w)$ in the same way as that of the $\d\rho$ map.   

As the amplitude at $\qq\neq\Vec{0}$ in these plots is roughly proportional to $V_{\{n,s\}}(\qq)$ and its Fourier transform $V_{\{n,s\}}(\rr)$ extends up to $r_{\rm imp}$ for our realistic choice of Eqs.~(\ref{eq:vn}) and (\ref{eq:vs}), the plots are faint for large $|\qq|>\hbar/r_{\rm imp}$.
Then, it will be instructive to see $\d\rho$ calculated with a $\qq$-independent $\Vi$, in which the structures at large $\qq$'s are more clearly seen. 
This is shown in Figs.~\ref{fig:drhodisp}(c) and \ref{fig:drhodisp}(d), where the sign of $\d\rho$ is distinguished by red and blue colors.
Along $(0,0)-(\pi,\pi)$, the $\qq_3$ dispersion and neighboring electron-hole asymmetric dispersion can be clearly seen (though the latter appears as a white curve for $\w>0$ for the reason explained below). In this high-$\qq$ region, the overall sign is opposite between $\w>0$ and $\w<0$ while $\d\rho$ is always positive for $\qq\sim0$.
Along $(0,0)-(2\pi,0)$, the $\qq_1$ and $\qq_5$ dispersions are seen while these are overlapped with other dispersions: These are mirrors of $\qq_1$ and $\qq_5$ starting from the Bragg point $(2\pi,0)$. [Note that the mirror symmetry is, though, incomplete
due to the Wannier projection, Eq.~(\ref{eq:gwan}).]
In this direction, the sign of $\d\rho(\qq,-\w)$ is mostly the same as that of $\d\rho(\qq,\w)$.
For more details, see Appendix~\ref{ssec:decom}.

In the subsequent sections, we mainly discuss $L(\qq,\w)$, because it is directly accessible in experiments, in contrast to $\d\rho(\qq,\w)$. In addition, $L(\qq,\w)$, which is free from the set-point effect, reflects the features inherent to the system more than $g(\qq,\w)$.
 
\subsubsection{Signal of fractionalization in QPI}
\label{sssec:gLsign}

\begin{figure*}[tb]
\centering
\includegraphics[width=0.95\textwidth]{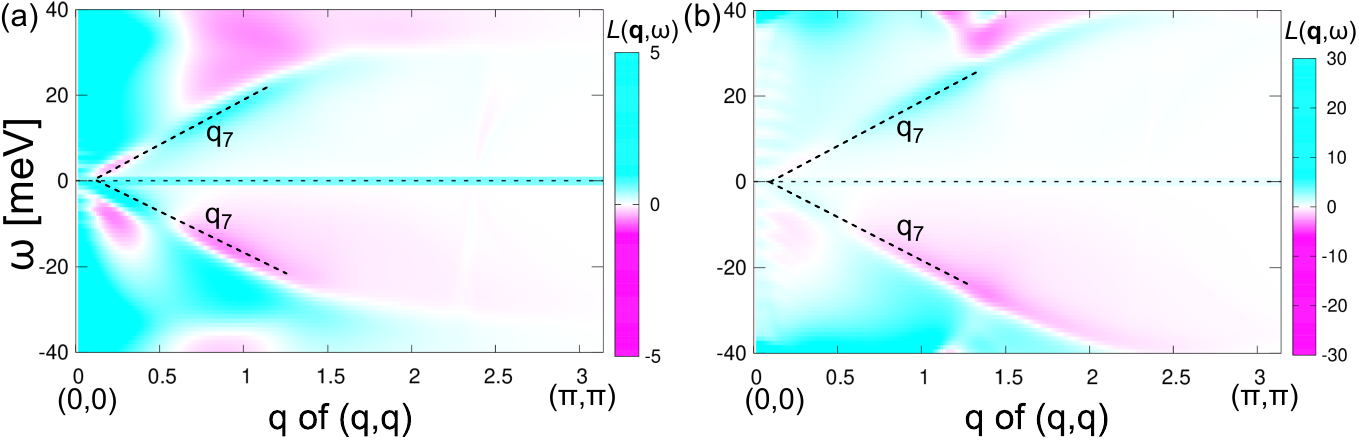}
\caption{Sign structure of $L(\qq,\w)$ along $(0,0)-(\pi,\pi)$ calculated with (a) the TCFM and (b) the Hamiltonian Eq.~(\ref{eq:H1c}) without $d$ degrees of freedom. Dashed lines denote the $\qq_7$ dispersions.}\label{fig:L_sign_00-PP}
\end{figure*}

\begin{figure}[tb]
\centering
\includegraphics[width=0.48\textwidth]{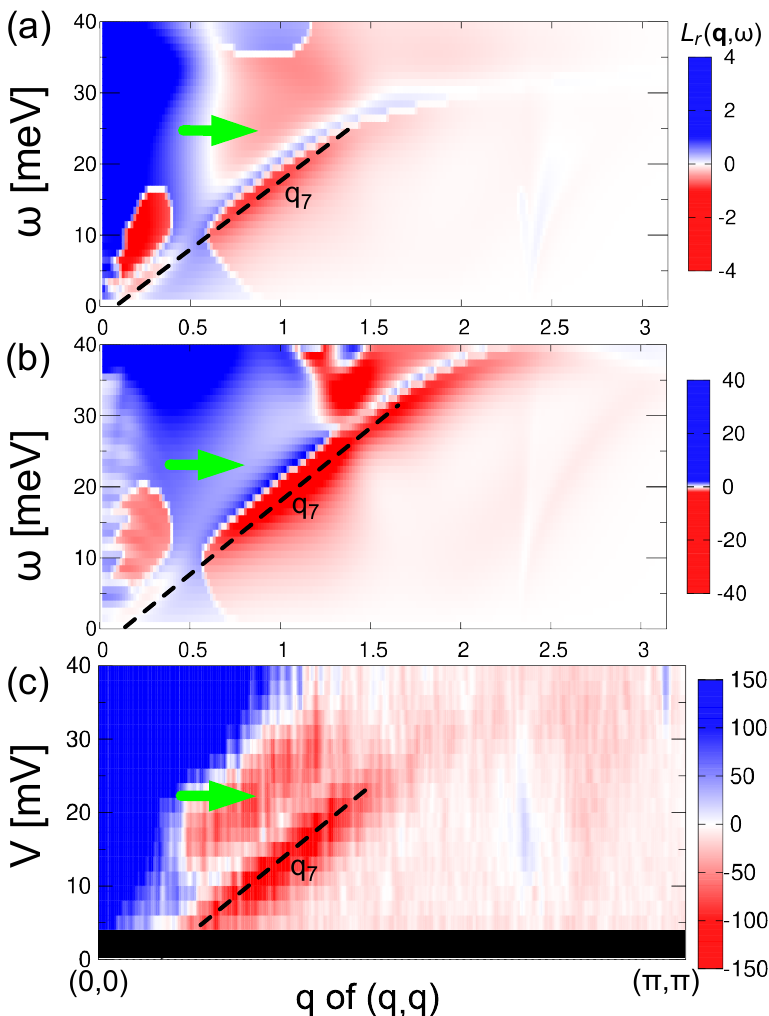}
\caption{Relative-sign structure, $L_r(\qq,\w>0)$, along $(0,0)-(\pi,\pi)$, obtained by (a) the TCFM, (b) the single-component Hamiltonian, Eq.~(\ref{eq:H1c}), and (c) the experiment. The light-green arrows indicate the area discussed in the main text.
Note that the dashed line ($\qq_7$ dispersion) is drawn by following the octet model illustrated in Fig.~\ref{fig:octet}.
Black region in (c) indicates an energy range where accurate data is difficult to obtain because both $\dv{I}{V}$ and $\frac{I}{V}$ in $L=\frac{V}{I}\dv{I}{V}$ are not large enough compared to noise.}
 \label{fig:Lr_00-PP}
\end{figure}

While we have so far discussed only the amplitudes of $g(\qq,\w)$ and $L(\qq,\w)$, a characteristic feature of the QPI signal, distinguishing the TCFM from the single-component models, can be found in the phase (which is reduced to a sign in our case because of the inversion symmetry) of these quantities.
Figure~\ref{fig:L_sign_00-PP}(a) shows the sign structure of $\Lqw$ calculated with the TCFM along $(0,0)-(\pi,\pi)$
[see Appendix \ref{ssec:sign_app} for $(0,0)-(2\pi,0)$].
For comparison, in Fig.~\ref{fig:L_sign_00-PP}(b), we show the same quantity calculated with the ordinary mean-field (single-component) Hamiltonian, 
Eq.~(\ref{eq:H1c}).
For the impurity potential and other parameters not included in Eq.~(\ref{eq:H1c}), we used the same values as those used in the TCFM calculation.

A notable difference between 
Figs.~\ref{fig:L_sign_00-PP}(a) and \ref{fig:L_sign_00-PP}(b) lies in the smaller-$q$ side ($0.5 \lesssim q \lesssim 1.2$) of the $\qq_7$ dispersion: While Fig.~\ref{fig:L_sign_00-PP}(b) shows positive values for both $\w>0$ and $\w<0$ in this region, Fig.~\ref{fig:L_sign_00-PP}(a) shows negative (positive) values for $\w>0$ ($\w<0$). 
This difference is attributed to the pseudogap (or, in other words, a contribution from the $d$ component of the electron) emergent on the positive-$\omega$ ($\sim 0.1$ eV) side in the TCFM, as we explain in the following.

We start from 
\begin{align} 
  &L(\qq,eV) \nonumber\\
  &\simeq\frac{eV}{w_0(eV)}\left[\d\r(\qq,eV)-\frac{\r_0(eV)eV}{w_0(eV)}\frac{\int_0^{eV}\d\r(\qq,\w) d\w}{eV}  \right]
  \label{eq:Lqw}
\end{align}
with $w_0(\w)\equiv \int_0^{\w}\r_0(\w') d\w'$, derived in Appendix \ref{sssec:relation}. 
Since the factor $\frac{\r_0(eV)eV}{w_0(eV)}$ on the right-hand side of Eq.~(\ref{eq:Lqw}) will be of the order of 1, we consider the relative size between $\d\r(\qq,eV)$ and $\frac{\int_0^{eV}\d\r(\qq,\w) d\w}{eV}$, focusing on $\qq\simeq (1,1)$, where the difference between Figs.~\ref{fig:L_sign_00-PP}(a) and \ref{fig:L_sign_00-PP}(b) is most prominent.
For $\w>0$, $\d\r$ increases with energy up to the energy $e_7$ of $\qq_7$, so that
\begin{align}
  \d\r(\qq,eV) > \frac{\int_0^{eV}\d\r(\qq,\w) d\w}{eV} \label{eq:drho_ineq}
\end{align}
i.e., $L>0$, holds for $0<eV<e_7$ as the right-hand side is the average of $\d\r(\qq,\w)$ over $[0,eV]$. Above $e_7$, however, the pseudogap suppresses the spectral intensity, so that $\d\r$ decreases. (We remind the readers that the $s$-wave-like full pseudogap, consistent with various numerical simulations on the Hubbard \cite{sakai13} and $t$-$J$ models \cite{tohyama04} as well as spectroscopic
experiments on cuprates \cite{sakai13}, opens even in the nodal region at $\w>0$ in the TCFM.) This suppression tends to invert the above inequality, resulting in $L<0$ for $\w\gtrsim e_7$. 
For $\w<0$, $\Lqw$ simply reflects the sign structure of $\d\r$ because of the absence of the pseudogap.
As we shall see in Sec.~\ref{sssec:high}, a similar sign change for $\w>0$ can also be seen in the $(0,0)-(2\pi,0)$ direction at a slightly higher energy.

On the other hand, since the one-component model, Eq.~(\ref{eq:H1c}), does not show the pseudogap, there is less tendency to invert the inequality (\ref{eq:drho_ineq}) above $e_7$. This results in a weak but positive sign above $e_7$ in Fig.~\ref{fig:L_sign_00-PP}(b) in contrast to Fig.~\ref{fig:L_sign_00-PP}(a). 

Although it is not straightforward to extract the sign of $L(\qq,\w)$ experimentally (as explained in the second paragraph of Appendix \ref{ssec:method_qpi}), it is still possible to extract a {\it relative sign} from the data taken at $\w$ and $-\w$ at the same $\qq$.
Namely, we define
\begin{align} 
   L_r(\qq,\w) \equiv L(\qq,\w) \text{sign}[L(\qq,-\w)], \label{eq:Lr}
\end{align}
similarly to $g_r(\qq,\w) \equiv g(\qq,\w) \text{sign}[g(\qq,-\w)]$ defined in Refs.~\cite{chi_2017_1,chi_2017_2,gu19}.
$L_r(\qq,\w)$ is positive (negative) if $L(\qq,-\w)$ has the same sign as (opposite sign from) $L(\qq,\w)$.
We show the results of $g_r$ in Supplemental Material \cite{suppl} Fig.~S4.

Figure~\ref{fig:Lr_00-PP}(a) plots the TCFM prediction of  $L_r(\qq,\omega)$, which exhibits
the same negative sign on both sides of $q_7$,
reflecting the sign structure of $L$ in Fig.~\ref{fig:L_sign_00-PP}(a). On the other hand, the single-component Hamiltonian~(\ref{eq:H1c}) predicts a weakly positive sign above $e_7$ as in Fig.~\ref{fig:Lr_00-PP} (b). This difference results in the different sign of $L_r$ in the corresponding region, denoted by a light-green arrow in Figs.~\ref{fig:Lr_00-PP}(a) and \ref{fig:Lr_00-PP}(b). 

We are now able to compare these contrasting theoretical $L_r$'s with experimental data, which is shown in Fig.~\ref{fig:Lr_00-PP}(c). The experimental data support the TCFM prediction and show qualitative disagreement with the single-component-model prediction. In Sec.~\ref{sssec:high}, we will predict that the effect of the fractionalization continues and becomes extended at higher energies.

We note that the difference in the overall amplitudes of Figs.~\ref{fig:Lr_00-PP}(a) and \ref{fig:Lr_00-PP}(c) (see the color scales) is largely attributed to the existence of the multiple impurities in the experiment (in contrast to the single impurity in the theory), as well as the difficulty in evaluating the strength of each impurity potential. However, as is clear from Eqs.~(\ref{eq:drhoqw3}) and (\ref{eq:Lqw2}), the overall intensity of $L$ is roughly proportional to the strength of the impurity potential and hence the relative structures will not be affected by these factors.

\subsubsection{Relation to the ARPES dispersion}\label{sssec:qpi-arpes}

As was mentioned in the introduction, it has been recognized that the $\qq_7$ dispersion in the QPI experimental data shows a deviation from the one expected from ARPES observations.
Namely, while $\qq_7$ extrapolated to $\w=0$ cuts a finite $q$ [see Figs.~\ref{fig:gdisp}(a) and \ref{fig:ldisp}(a)],  
the ARPES shows a clear $d$-wave superconducting gap,  closing at the nodal point at $\w=0$, which should correspond to the $\qq_7$ dispersion crossing the origin $\qq=\omega=0$. This puzzle can be resolved by looking into the calculated results, Figs.~\ref{fig:gdisp}(c) and \ref{fig:ldisp}(c).

In these figures, we find white lines just outside of the $\qq_7$ dispersion both in the experiment and the theory. 
In theory, these are understood as a polelike structure of $G$ and $F$.
Namely, when Im$G$ and Im$F$ show sharp peaks, Re$G$ and Re$F$ change the sign around the peak position, crossing a zero, as seen in Figs.~\ref{fig:edc_an}(a) and \ref{fig:edc_an}(b).
This zero of the real part will explain the white lines because in the $(0,0)-(\pi,\pi)$ direction, $V_s(\qq)=0$ in Eq.~(\ref{eq:drhoqw3}) and therefore $\d\r(\qq,\w)$ is contributed only from
\begin{align}
  \d\r_{GG}(\qq,\w)&= \frac{-2}{\pi N_{\kk}} \sum_{\rr'\kk} w_{\kk}(\rr')w_{\kk+\qq}^\ast(\rr') e^{-i\qq\cdot\rr'} \nonumber\\
                &\times  \text{Re}G(\kk,\w)\text{Im}G(\kk+\qq,\w) \label{eq:GG}
\end{align}
and 
\begin{align}
  \d\r_{FF}(\qq,\w)&= \frac{-2}{\pi N_{\kk}} \sum_{\rr'\kk} w_{\kk}(\rr')w_{\kk+\qq}^\ast(\rr') e^{-i\qq\cdot\rr'} \nonumber\\
                &\times  \text{Re}F(\kk,\w)\text{Im}F(\kk+\qq,\w). \label{eq:FF}
\end{align}
Both of these contributions vanish when the real parts of the Green's functions vanish.
Since the zero of Re$G$ corresponds to the spectral-peak position, the white lines in our calculation correspond to the $\qq_7$ dispersion conventionally estimated by using the autocorrelation of the spectral function.

Around the zero, $|{\rm Re}G|$ and $|{\rm Re}F|$ steeply increase within the width of the sharp peak of the imaginary part and produce a strong intensity in $\d\r$ (and hence in $g$ and $L$, too) on both sides of the white lines.
The strong intensity at $\qq_7$ in Figs.~\ref{fig:gdisp}(c) and ~\ref{fig:ldisp}(c) is ascribed to one of these side peaks.
Therefore, its position shifts from the one expected from the ARPES, i.e., autocorrelation of the spectral function.
In fact, the white lines better cross the origin for both the experiments and the theory in Figs.~\ref{fig:gdisp} and ~\ref{fig:ldisp} in accordance with what is expected from ARPES. 

\subsubsection{High-energy structures} \label{sssec:high}

\begin{figure}[tb]
\centering
\includegraphics[width=0.48\textwidth]{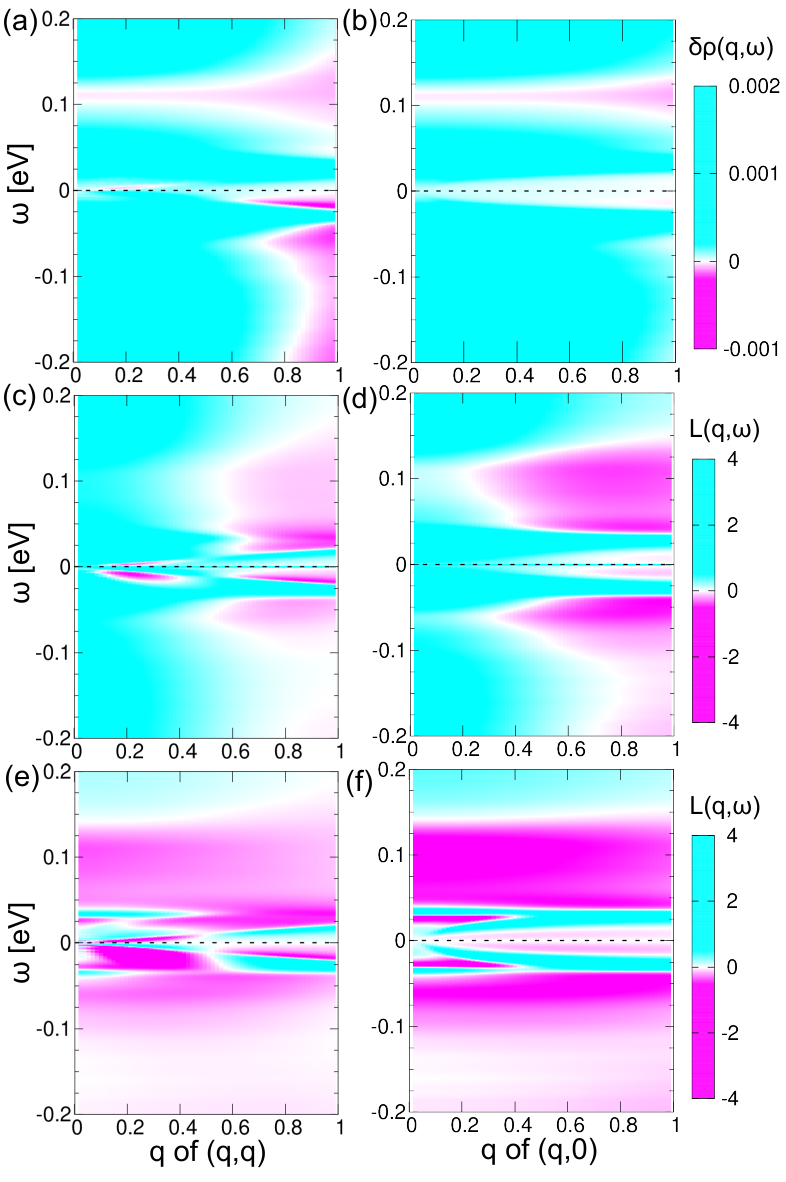}
\caption{High-energy structure of 
 (a,b) $\d\r(\qq,\w)$, (c,d) $L(\qq,\w)$ along $(0,0)-(1,1)$ [(a) and (c)] and $(0,0)-(1,0)$ [(b) and (d)], calculated with the TCFM, and (e,f) the same as (c,d) but without the  Gaussian contribution, Eq.~(\ref{eq:gaussian}), around $\qq\simeq \Vec{0}$.
 Plots for full ranges, $(0,0)-(\pi,\pi)$ and $(0,0)-(2\pi,0)$, are shown in Fig.~S5.} \label{fig:e0.2}
\end{figure}

\begin{figure}[tb]
\centering
\includegraphics[width=0.48\textwidth]{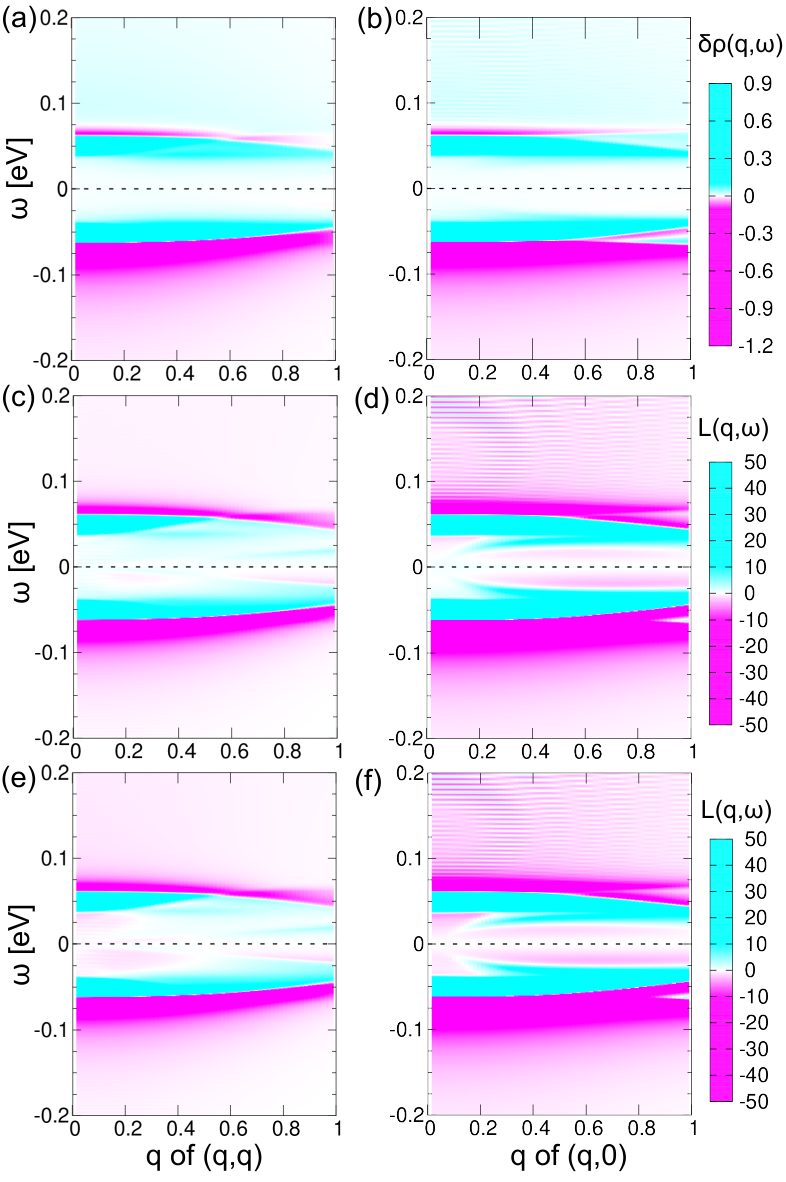}
\caption{High-energy structure of 
 (a,b) $\d\r(\qq,\w)$, (c,d) $L(\qq,\w)$ along $(0,0)-(1,1)$ and $(0,0)-(1,0)$, calculated with the Hamiltonian, Eq.~(\ref{eq:H1c}), and (e,f) the same as (c,d) but without the  Gaussian contribution, Eq.~(\ref{eq:gaussian}), around $\qq\simeq \Vec{0}$. Plots for $(0,0)-(\pi,\pi)$ and $(0,0)-(2\pi,0)$ are shown in Fig.~S5.}
  \label{fig:1c_e0.2}
\end{figure}

Figure \ref{fig:e0.2} shows the high-energy structures of $\d\r$ and $L$ calculated for a small-$|\qq|$ region ($|\qq|<1$) along the $(0,0)-(\pi,\pi)$ and $(0,0)-(2\pi,0)$ directions. 
As we show in Supplemental Material \cite{suppl} Fig.~S5, 
the strong intensity can be seen only in this small-$|\qq|$ region or around $(2\pi,0)$ for $|\w|>0.05$ eV.
This is consistent with the so-called extinction known in the QPI experiments \cite{kohsaka08} and here in the combined analysis, it turns out to have the same origin as the broad background in ARPES. 

In $\d\r$ [Figs.~\ref{fig:e0.2}(a,b)], we see sign changes around $\w=0.1$ eV. This is attributed to the pseudogap while the intensity above this energy is due to the ingap states [see Fig.~\ref{fig:akw_disp}].
Since $\d\r$ is not directly measurable in experiments, we look for a related feature in $g$ and $L$.
In particular, because $g$ is subject to the set-point effect, we focus on $L$ in the following while the results of $g$ are presented in the Supplemental Material \cite{suppl} Fig.~S7.

In Figures \ref{fig:e0.2}(c) and \ref{fig:e0.2}(d), we find a negative-sign region in $L$ for $0.05\lesssim \w\lesssim 0.1$ [eV] at small but finite $q$'s followed back by the positive-sign region for $0.1$eV $\lesssim \w$, which is more pronounced for the $(0,0)-(2\pi,0)$ direction.
If we erase the effect of the Gaussian contribution, Eq.~(\ref{eq:gaussian}), around $q\simeq 0$, we can see the sign change more clearly as an $\w$-dependent (rather than $\qq$-dependent) feature, as shown in Figs.~\ref{fig:e0.2}(e) and \ref{fig:e0.2}(f). 
These sign changes are more pronounced and are an extended continuation of the sign change found in Fig.~\ref{fig:L_sign_00-PP}(a) to higher energy region: The negative-sign region for
$\omega \lesssim 0.1$ eV identifies the pseudogap and the positive sign at higher energy $\omega \gtrsim 0.1$ eV
reflects the ingap states as was explained above from the structure of Eq.~(\ref{eq:Lqw}).
Such a sign change in $L$ cannot be seen in the results for the single-component model, Eq.~(\ref{eq:H1c}), as shown in Fig.~\ref{fig:1c_e0.2}.

Because only the relative sign between two different energies can be obtained experimentally, we define 
\begin{align} 
   L_r^{(0.2{\rm eV})}(\qq,\w) \equiv L(\qq,\w) \text{sign}[L(\qq,0.2{\rm eV})]. \label{eq:Lr_0.2}
\end{align}
As $L(\qq,0.2{\rm eV})$ is always positive for $|\qq|<1$ in Figs.~\ref{fig:e0.2}(c) and \ref{fig:e0.2}(d), $L_r^{(0.2{\rm eV})}$ has the same sign structure as $L$ for $|\qq|<1$. The experimental detection of the sign change through the analysis of $L_r^{(0.2{\rm eV})}$  is an intriguing future issue.
In addition, it would be possible to detect the corresponding feature just by looking at the amplitude of $L(\qq\simeq\Vec{0},\w)$ since the sign change necessarily involves a zero intensity in between.

\section{Summary and Outlook} \label{sec:conclude}
We have performed an integrated analysis of combined ARPES, STM and QPI experimental data for an optimally doped Bi2212 with the theory based on the TCFM, which describes the electron fractionalization. 
The TCFM fitted with ARPES data reproduces the
QPI patterns even for details, which makes a substantial step forward toward a unified understanding of the high-$T_c$ cuprates. 
In particular, distinctive QPI patterns, obtained for the TCFM but absent in the conventional electronic structure without the fractionalization, agree with the experimental data, supporting the unconventional fractionalization of electrons taking place in the cuprates. At the same time, the seemingly controversial data between ARPES and QPI for the quasiparticle dispersion are understood by considering the structure of the TCFM solutions.

The numerical results are obtained by faithfully
calculating the experimentally-measurable quantities, such as the differential conductance, which has often been replaced with the local density of states in theoretical analysis in the literature, despite their mutual inconsistencies often taking place.
We have also revealed that the relative-sign structure in QPI data holds a key to unraveling the pseudogap physics.

We also point out that the background intensity in ARPES is related to the extinction in QPI.
Its microscopic origin, as well as that
of the marginal Fermi liquid, is left for future studies. These elements have been taken into account as phenomenological self-energies in this study and are desired to be derived microscopically from the interactions of $c$ and $d$ as well as those with other relevant degrees of freedom, such as excitons and magnetic/charge/phonon fluctuations, in an extended framework of the TCFM.

The results of fitting to the ARPES and QPI data are consistent with the fractionalization scenario, where the ingap state emerging as the antibonding state of hybridized $c$ and $d$ shows up as one of the isolated poles of the $c$ Green’s function for each momentum and represents a nearly localized bound state inside the Mott gap. 
This work has thus established that this nearly localized bound state proposed in Ref.~\cite{sakai18} is found as a component of the fractionalized fermion in the real cuprate.

These results suggest that the novel type of electron fractionalization, emergent from the Mott physics, is at the heart of the pseudogap and high-temperature superconducting mechanism in cuprates. This is distinct from the spin-fluctuation scenario, where a bosonic excitation plays a major role. Our fitting further imposes strong constraints on the character of the fractionalized electron, or more generally, the momentum and frequency dependence of the self-energy, in particular in the region above the Fermi energy, which the ARPES cannot see.
We have phenomenologically assumed that the impurity couples only to the $c$ fermion in the TCFM and have shown good agreements with the experimental QPI results. Since the microscopic identity of the $d$ fermion has not been established yet, this agreement imposes a helpful constraint in future pursuits of the nature of the $d$ fermion.

We have performed a combined analysis of ARPES, STM, and QPI. To further enhance quantitative reliability of the electron fractionalization theory, it would be important to combine other experimental data, for instance from neutron, optics, magnetic resonance, muon, electron energy-loss spectroscopy, and resonant inelastic X-ray scattering, with the help of machine learning analysis, which is left for an important future study.

The significance of the electron fractionalization lies in its link to the mechanism of the superconductivity itself, for which some proposals were made~\cite{schmid23} from ab initio calculations, though it is not fully understood. Further pursuit using integrated spectroscopy analyses is another intriguing future subject.

The success of the integrated spectroscopy analyses in this study opens a new avenue of experimental data analysis for other challenging issues in strongly correlated electron systems in general.

\begin{acknowledgments}
This work was supported by JSPS KAKENHI Grant No.\ JP23H04528 (SS), No.\ JP24H00198 (TH), No.\ JP23K17351 (TK), No.\ JP23H03818 (YY), No.\ JP23H04524 (YY), and No.\ JP25K07233 (YY). 
This work was also financially supported by 
MEXT KAKENHI, Grant-in-Aid for Transformative Research Area
Grant No.\ JP22H05111 (MI), No.\ JP22H05114 (MI), No.\ JP25H01249 (TH), No.\ JP25H01246 (TK), and No.\ JP25H01250 (TK). This work was supported by MEXT as well in the project ``Program for Promoting Researches on the Supercomputer Fugaku" [``Basic Science for Emergence and Functionality in Quantum Matter--Innovative Strongly-Correlated Electron Science by Integration of ``Fugaku" and Frontier Experiments--" (No.\ JPMXP1020200104) as well as ``Simulation for basic science: approaching the new quantum era" (No.\ JPMXP1020230411) and ``AI Numerical Spectroscopy for Analyzing Emergent Structures of Quantum Entanglement in Correlated Quantum Materials" (No.\ JPMXP1020230410)],
and by the facilities provided by the RIKEN Center for Computational Science (Project IDs: hp230207, hp240213, hp250224) (MI).
We also acknowledge partial support from the RIKEN TRIP initiative (Many-body Electron Systems).
This work (subset of ARPES data) was partially supported by the U.S. Department of Energy, Office of Basic Energy Sciences, Division of Materials Science and Engineering. Ames National Laboratory is operated for the U.S. Department of Energy by Iowa State University under Contract No. DE-AC02-07CH11358.
Part of the computation was done using the facilities of the Supercomputer Center, the Institute for Solid State Physics, the University of Tokyo.
\end{acknowledgments}

\appendix

\section{Details of the Method}\label{sec:method_detail}
In Appendix \ref{ssec:sig}, we construct a self-energy form capable of explaining the ARPES data presented in Sec.~\ref{sec:arpes}.
In Appendix \ref{ssec:fit}, we explain how we fit the ARPES data with the constructed self-energy form and determine the TCFM parameters. 
In Appendix \ref{ssec:method_qpi}, we describe the method to calculate the QPI patterns based on the TCFM, where we have assumed that the impurity couples only to the $c$ fermion in the TCFM, which leads to the bubble approximation of Eq.~(\ref{eq:drhorw}).

\subsection{Form of self-energy}\label{ssec:sig}
While the TCFM Hamiltonian has been successful in describing the low-energy structures of the spectral function and the self-energy \cite{sakai16,sakai16PRB,yamaji21,imada21,singh22,sakai23} in the underdoped region of the 2D Hubbard model at least qualitatively,
more detailed quantitative comparisons with the experimental spectra of the cuprates become possible by incorporating additional structures to
the self-energy $\Sigma_{\rm TCFM}$.
Here, along the line addressed at the end of Sec.~\ref{sec:model}, we provide additional self-energy structures close to zero energy and at high energies.
The former is related to a so-called marginal Fermi-liquid (MFL) behavior \cite{varma89}, which gives a linear-in-energy scattering rate around the Fermi energy.
The latter is related to the flat feature of the ARPES spectra for $\w<-0.1$ eV, mentioned in Sec.~\ref{sec:arpes}.
Indeed, the less-singular part of the self-energy
of the cuprate superconductors
can be decomposed into a
MFL-type component and the remaining flat structure, as explained later~\cite{yamaji21}.
In the following, we elaborate on these contributions one by one. 

\subsubsection{TCFM part}\label{sssec:tcfm}
Equations (\ref{eq:snor}) and (\ref{eq:sano}) define the TCFM part of the self-energy, describing the correlation effect of our main interest. 
They have poles at the same energies, $\w=\pm\wpk$ with $\wpk\equiv\sqrt{\edk^2+\Ddk^2}$.
The contributions of these poles are canceled out in the Green's functions [see Eqs.~(\ref{eq:g}) and (\ref{eq:w})] at the pole energies but still significantly influence the quasiparticle dynamics \cite{sakai16,sakai16PRB}.

\subsubsection{Marginal Fermi-liquid part}\label{sssec:mfl}
As examined in the literature,
low-energy ARPES spectra, at least along the nodal direction, require a term proportional to $|\omega|$ in the imaginary part of normal self-energy~\cite{smit2024momentum}. This non-analytic structure is indeed found not only along the nodal direction but also in the antinodal region~\cite{yamaji21}.
Even though a small deviation from the MFL was reported along the nodal direction~\cite{smit2024momentum}, the $\omega$-linear term is enough to capture the essential behaviors of the low-energy spectra in the entire Brillouin zone.
The non-analytical term has been examined to describe the momentum-independent anomalous behavior of the normal-state spectra known as the MFL behavior~\cite{varma89}.

As discussed in Ref.~\onlinecite{varma89}, the MFL component of the self-energy can arise from electron scatterings by intensive low-energy bosonic fluctuations not contained in Eq.~(\ref{eq:tcfm}).
In Ref.~\cite{yamaji21}, the formation of the self-energy peaks causes such electron scatterings, giving the MFL component.
The excitonic excitations that are not taken into account in the TCFM Hamiltonian, Eq.~(\ref{eq:tcfm}), coupled to the $d$ fermions, are possibly responsible for the MFL component of the self-energy.
However, the microscopic origin of the MFL is not well established.

Since it is observed universally in experiments on cuprates, without specifying its origin, we model the MFL component, $\Sigma_{\rm MFL}(\w)$, with a soft cutoff as
\begin{align}
{\rm Im} \Sigma_{\rm MFL}(\w) =&
-(1/\pi) \sqrt{\w^2 + (\pi/\tau_0)^2 }\nonumber\\
&\times \{1+\exp [(\w - \Lambda)/\sigma_{\rm MFL}]\}^{-1}\nonumber\\
&\times \{1+\exp [-(\w + \Lambda)/\sigma_{\rm MFL}]\}^{-1}.
\end{align}
Here, a small but finite inverse lifetime, $\tau_0^{-1}=1$ meV is introduced.
The soft cutoff is defined by sigmoid functions with the step width $\sigma_{\rm MFL}=30$ meV and the cutoff energy $\Lambda$ =0.4 eV.
The real part of the $\Smfl$ is obtained by the Kramers-Kronig transformation of ${\rm Im}\Smfl$.

\subsubsection{Background part}\label{sssec:bg}

\begin{figure}[tb]
\centering
\includegraphics[width=0.48\textwidth]{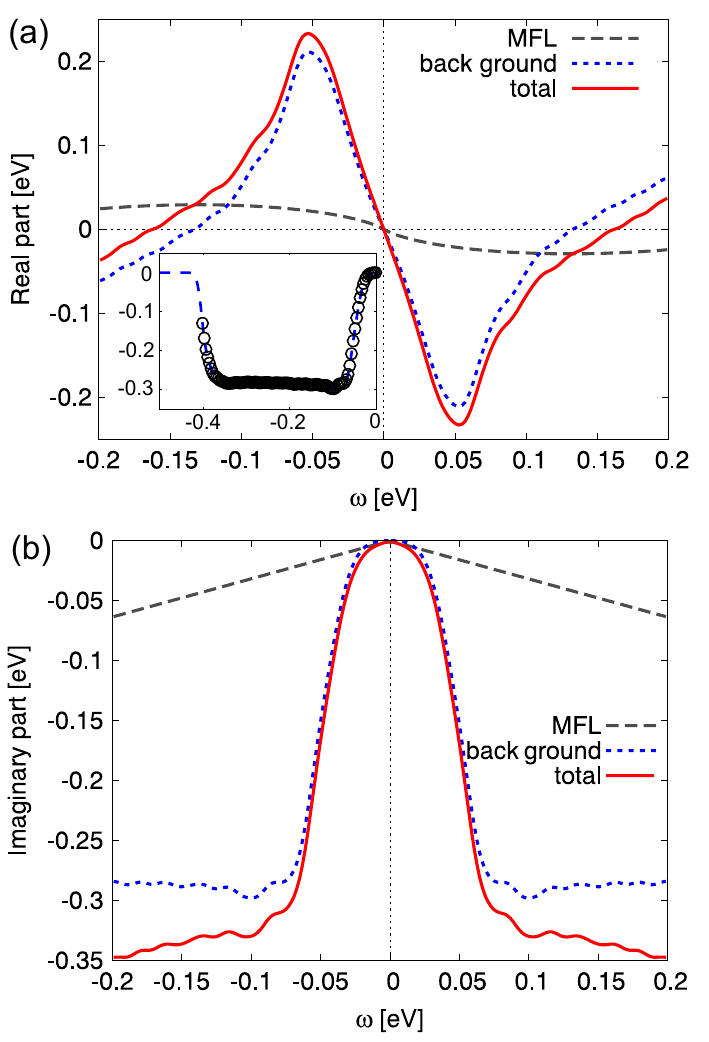}
\caption{(a) Real and (b) imaginary parts of the background self-energy $\Sbg$ (blue dashed curve), marginal Fermi-liquid self-energy $\Smfl$ (gray dashed curve), and their sum (red solid curve), derived from the ARPES data at the antinodal point, $\kk=\kAN=(\pi,0.14\pi)$.
Inset to (a) shows ${\rm Im}\Sbg$ interpolated by the Gaussian process regression (blue dashed curve) and that obtained from the ARPES data (black open circles).} 
\label{fig:sbg}
\end{figure}

As one can see in Fig.~\ref{fig:arpes}(b), the ARPES data show a nearly flat intensity below $\sim-0.1$ eV for all the momenta.
This flat spectrum has been attributed to a broad structure of Im$\Snor$ in the corresponding energy range~\cite{bok10,li2018coherent,yamaji21}.
In particular, according to the machine-learning analysis of the ARPES spectrum at $\kk=\kAN$~\cite{yamaji21}, the self-energy exhibits a component that is nearly constant over the high-energy region for $\w \lesssim -0.1$ eV.
Hereafter, this flat structure in 
Im$\Snor$ is referred to as the background, while its microscopic origin remains debated.

The possible origin of the background has been discussed in the literature.
For instance, when bosonic excitations are coupled to fermions, a typical self-energy structure appears at $|\omega|$ larger than the bosonic energy scales through a mechanism different from the origin of the MFL.
A well-studied example is the self-energy generated by electron-phonon couplings~\cite{schrieffer63,scalapino66}.
For $|\omega|$ larger than the typical boson energy scale, flat structures appear in Im$\Snor$, irrespective
of the origin of the boson~\cite{chubukov20}.
Even in the electron gas, where plasmons play an important role, the broad structure in the self-energy was also found~\cite{PhysRevLett.117.206402}.
In the spectral function of the Hubbard model, the incoherent continuum is found  
between the quasiparticle dispersion and
the incoherent lower Hubbard band~\cite{macridin07,sakai10,kohno14,charlebois20}, which is possibly relevant
to the flat structure in ARPES.
Though electron-phonon couplings in the cuprates have been examined in detail~\cite{PhysRevLett.126.146401}, contributions of spin fluctuations are also discussed based on the analysis of the ARPES spectra~\cite{bok10}.
 The flat structure may alternatively be induced by the excitonic fluctuations, in addition to the MFL component.
In the TCFM picture, excitonic fluctuations may arise because it was proposed that the $d$ fermions represent the fermionic degrees of freedom contained in the weakly bound excitons~\cite{yamaji11PRL,yamaji11PRB,sakai18,imada19}, which might be responsible for the incoherent continuum in the Hubbard model mentioned above.
However, the couplings between the $d$ fermion and neither the exciton, phonon, nor the spin fluctuation are explicitly included in Eq.~(\ref{eq:tcfm}).
Therefore, we add the flat component to ${\rm Im}\Sigma^{\rm nor}_{\rm TCFM}$ in the TCFM description of the ARPES spectra and QPI patterns to reproduce the ARPES flat structure without specifying its origin.

Irrespective of its origin, the background component is nearly constant over the high-energy region for $\w \lesssim -0.1$ eV in Bi2212~\cite{yamaji21}.
Because the background is not contained in the TCFM Hamiltonian (which describes only a few low-energy spectral peaks), we incorporate this component as follows.

We derive the $\w$ dependence of this additional self-energy, $\Sbg$, from the ARPES data at $\kk=\kAN$, where we have high-resolution data. 
The self-energy derived from the ARPES data~\cite{yamaji21} contains peak structures and MFL-like contributions as well.
Therefore, we need to subtract these contributions when we construct the background component to avoid double counting.
We start from the imaginary part of the total self-energy, ${\rm Im}\Sigma^{\rm tot}$, for the optimally doped Bi2212.
While the peak structures are canceled in ${\rm Im}\Sigma^{\rm tot}$ as in the TCFM Hamiltonian, observed in Ref.~\onlinecite{yamaji21}, ${\rm Im}\Sigma^{\rm tot}$ contains a Lorentzian component
generating the superconducting gap.
By eliminating the Lorentzian component and $\w$ independent impurity scattering 
contribution, we obtain the imaginary part of the background component, ${\rm Im}\Sbg$, as shown in Fig.~\ref{fig:sbg}(b).
The real part of $\Sbg$ is obtained through the Kramers-Kronig relation.
To perform the Kramers-Kronig transformation, the imaginary part is interpolated using Gaussian process regression based on 512 normal distributions with a standard deviation of 0.01 eV, arranged at equal intervals from -0.4 eV to 0.4 eV, which naturally introduces a soft cutoff as shown in the inset of Fig.~\ref{fig:sbg}(a).
Here, for simplicity, we assume electron-hole symmetry for $\Sbg$.
This background contribution is relevant to the so-called extinction of QPI signals, as shown in Fig.~\ref{fig:e0.2}.

The obtained self-energy is shown by blue dashed curves in Fig.~\ref{fig:sbg}.
As the momentum dependence of this self-energy is not known in experiments, we assume the following simple form to account for the difference between nodal and antinodal regions, introducing a single fitting parameter $c_{\bg}$:
\begin{align}
 \Sbg(\kk,\w)=\Sbg(\kAN,\w)\left[1+c_{\bg} h(\kk) \right] \label{eq:sbg}
\end{align}
with
\begin{align}
  h(\kk)=(\cos k_x-\cos k_y)^2 - (\cos k_{\AN x}-\cos k_{\AN y})^2.
\end{align}
$h(\kk)=0$ for $\kk=\kAN$ while $h(\kk)=-(\cos k_{\AN x}-\cos k_{\AN y})^2\simeq -3.63$ in the nodal direction.

\subsection{Fitting the ARPES data} \label{ssec:fit}
\subsubsection{Expression of the spectral function}\label{sssec:spectra}

Equation (\ref{eq:tcfm}) can be rewritten into the following form,
\begin{align}
H&=\sum_{\kk}
\begin{pmatrix}
c_{\kk,\ua}^\dagger & c_{-\kk,\da} & d_{\kk,\ua}^\dagger & d_{-\kk,\da}
\end{pmatrix}
\textbf{H}_{\kk}
\begin{pmatrix}
c_{\kk,\ua} \\
c_{-\kk,\da}^\dagger \\
d_{\kk,\ua} \\
d_{-\kk,\da}^\dagger
\end{pmatrix}
\label{eq:tcfm2}
\end{align}
with
\begin{align}
\textbf{H}_{\kk}\equiv
\begin{pmatrix}
\eck & \Dck & \Vk & 0 \\
\Dck & -\eck & 0   & \Vk \\
\Vk & 0 & \edk & \Ddk \\
0  & \Vk & \Ddk & -\edk
\end{pmatrix}.
\label{eq:hk}
\end{align}

With $\Sbg$ and $\Smfl$, the $(1,1)$ and $(3,3)$ (i.e., up-spin) components of Eq.~(\ref{eq:hk}) are modified as
\begin{align}
  \eck &\to \eck +\Sbg(\kk,\w)+\Smfl(\w),\nonumber\\
  \edk &\to \edk+f_{\bg,d}\Sbg(\kk,\w) \equiv \tilde{\e}_{d\kk}(\w).\label{eq:edk}
\end{align}
Here, we have introduced the MFL self-energy only for the $c$ fermion while we have introduced a fitting parameter $f_{\bg,d}$ for the background self-energy of the $d$ fermion.
We modify the $(2,2)$ and $(4,4)$ (i.e., down-spin) components of Eq.~(\ref{eq:hk}) accordingly.

Integrating out the $d$ degrees of freedom, we obtain 
\begin{align}
  \Snor(\kk,\w)&=
  \frac{\Vk^2[\w+i\eta+\tilde{\e}_{d\kk}^{\ast}(-\w)]}
  {[\w+i\eta - \tilde{\e}_{d\kk}(\w)][\w+i\eta + \tilde{\e}_{d\kk}^{\ast}(-\w)]-\Ddk^2}
\nonumber\\
  &+ \Sbg(\kk,\w)+\Smfl(\w) \label{eq:snor2}
\end{align}  
and
\begin{align}
  \Sano(\kk,\w)&=  \frac{-\Vk^2 \Ddk}
{[\w+i\eta - \tilde{\e}_{d\kk}(\w)][\w+i\eta + \tilde{\e}_{d\kk}^{\ast}(-\w)]-\Ddk^2}
\label{eq:sano2}
\end{align}
as the self-energy for the $c$ fermion. Here, $\eta$ is an energy- and momentum-independent smearing factor.

The Green's function matrix $\hat{G}$ for the $c$ fermion is then given by
\begin{align}
 \hat{G}(\kk,\w)&=
 \left(
    \begin{array}{cc}
     \Gkw & \Fkw \\
     \Fkw & G^\ast(\kk,-\w)
    \end{array}
   \right)   \nonumber\\ 
 &=\left(
    \begin{array}{cc}
       \w+i\eta-\eck-\Snor(\kk,\w)  & \quad \\ 
        -\Dck -\Sano(\kk,\w)  &   \quad
    \end{array}
    \right.\nonumber\\
    &\left.
    \begin{array}{cc}
       \quad  &  -\Dck -\Sano(\kk,\w) \\ 
       \quad &  \w+i\eta+\eck+{\Snor}^{\ast}(\kk,-\w)
    \end{array}
   \right)^{-1}.
\label{eq:nambu}
\end{align}

This defines the spectral function  as
\begin{align}
  \Akw = -\frac{1}{\pi}{\rm Im}\Gkw \label{eq:akw}
\end{align}
with 
\begin{align}
  \Gkw=[\w+i\eta-\eck-\Stot(\kk,\w)]^{-1}.\label{eq:g}
\end{align}
Here, $\Stot(\kk,\w)\equiv\Snor(\kk,\w)+\Wkw$ and
\begin{align}
  \Wkw\equiv\frac{\{\Dck+\Sano(\kk,\w)\}^2}{\w+i\eta+\eck+\Snor(\kk,-\w)^*} \label{eq:w}
\end{align}
represents the contribution from the anomalous part. 
The density of states (DOS) is defined by 
\begin{align}
\rloc(\w)\equiv
\frac{1}{N_\kk}\sum_\kk \Akw
=-\frac{1}{\pi}{\rm Im}[\Gloc(\w)]_{11},\label{eq:rloc}
\end{align}
where 
$\Gloc(\w) \equiv \frac{1}{N_\kk}\sum_\kk \hat{G}(\kk,\w)$ 
is the local Green's function.

\subsubsection{Cost function for fitting}\label{sssec:cost}
To fit the ARPES data, we define
\begin{align}
  I_{\rm TCFM}(\kk,\w)=\int_{-\infty}^0 A(\kk,\w')f(T,\w')R(\w-\w')d\w',
\end{align}
where $f(T,\w)$ is the Fermi distribution function at temperature $T$ ($=12$K) and
\begin{align}
 R(\w)=\frac{1}{\sqrt{2\pi}\sr}\exp(-\frac{\w^2}{2\sr^2})\label{eq:resol}
\end{align}
represents the energy resolution of the experimental facility.
Since the ARPES intensity depends on the matrix-element effect, which may give an additional momentum dependence in the intensity, we further normalize the data as
\begin{align}
  \tilde{I}_{\rm TCFM}(\kk,\w)=I_{\rm TCFM}(\kk,\w)/\int_{E_{\rm cut}}^0 I_{\rm TCFM}(\kk,\w')d\w'
\end{align}
at each momentum $\kk$ and compare it with 
\begin{align}
  \tilde{I}_{\rm ARPES}(\kk,\w)=I_{\rm ARPES}(\kk,\w)/\int_{E_{\rm cut}}^0 I_{\rm ARPES}(\kk,\w')d\w',
\end{align}
where $I_{\rm ARPES}(\kk,\w)$ is the raw ARPES data.
We set $E_{\rm cut}=-0.15$ eV in the following.

We determine the fitting parameters, $t_{c,\{0,1,2,3\}}$, $t_{d,\{0,1,2\}}$, $V_{\{0,1,2\}}$, $D_{c,0}$, $D_{d,0}$, $f_\bg$, $c_\bg$, $\sr$, and $\eta$ defined in Eqs.~(\ref{eq:ecdk}), (\ref{eq:vk}),  (\ref{eq:dcdk}), (\ref{eq:sbg}), (\ref{eq:edk}), and (\ref{eq:resol}) to minimize the following cost function,
\begin{align}
  C=\frac{1}{\tilde{N}_\kk}\sum_{\kk} \int_{E_{\rm cut}}^0 | \tilde{I}_{\rm TCFM}(\kk,\w)-\tilde{I}_{\rm ARPES}(\kk,\w)|^2 d\w.\label{eq:c}
\end{align}
Here, $\kk$ runs over cuts 4-13 shown by orange arrows in Fig.~\ref{fig:arpes}(a) and $\tilde{N}_\kk=580$ is the number of the $\kk$ points.
Note that we set $t_{d3}=0$ because the optimized value for it is always found to be quite small even when we start from nonzero values for it.

The noise in the ARPES data determines the lower bound of the cost function $C$.
The amplitude of the noise also enables us to evaluate the quality of the fitting.
We assume that the ARPES data can be decomposed into a smooth inherent spectrum $f(\kk,\w)$ and noise $\varepsilon (\kk,\w)$ as $\tilde{I}_{\rm ARPES}(\kk,\w)=f(\kk,\w)+\varepsilon (\kk,\w)$.
Then, the cost function $C$ is decomposed as
\begin{align}
C\simeq& \frac{1}{\tilde{N}_\kk}\sum_{\kk} \int_{E_{\rm cut}}^0 | \tilde{I}_{\rm TCFM}(\kk,\w)-f(\kk,\w)|^2 d\w \nonumber\\
&+ \frac{1}{\tilde{N}_\kk}\sum_{\kk} \int_{E_{\rm cut}}^0 |\varepsilon (\kk,\w)|^2 d\w,
\label{eq:appendix_C_decomp}
\end{align}
where we assume that the noise $\varepsilon (\kk,\w)$ is random and independent at each $(\kk,\w)$.
The second term on the right-hand side of Eq.~(\ref{eq:appendix_C_decomp}) provides an estimate of the lower bound of $C$.

Here, we try to estimate the noise by fitting the experimental ARPES data with a Gaussian process model.
We employ Gaussian distributions with the standard deviation $\sigma=10$ meV at each $\omega$ point and
smoothly interpolate the noisy ARPES data.
Here, the ARPES data are taken at 301 $\omega$ points ranging from -0.405 eV to 0.195 eV.
The standard deviation $\sigma$ of the Gaussian distributions is chosen to be the experimental resolution.
To avoid overfitting to the noise, we fit the ARPES data at five consecutive $\kk$ points by a single curve.
(There are 120 $\kk$ points along each momentum cut shown in Fig.~\ref{fig:arpes}.
Out of the 120 points, each locus indicated by an orange arrow contains 31-81 $\kk$ points.)
We thus obtain a Gaussian process model, $\tilde{I}_{\rm GP}(\kk,\omega)$.
When the Gaussian process model is a decent approximation of $f(\kk,\w)$, the following integral,
\begin{align}
C_{\rm n}=\frac{1}{\tilde{N}_\kk}\sum_{\kk} \int_{E_{\rm cut}}^0 | \tilde{I}_{\rm GP}(\kk,\w)-\tilde{I}_{\rm ARPES}(\kk,\w)|^2 d\w,
\label{eq:cn}
\end{align}
is a decent approximation of the integrated noise, i.e., 
\begin{align}
C_{\rm n}\sim \frac{1}{\tilde{N}_\kk}\sum_{\kk} \int_{E_{\rm cut}}^0 |\varepsilon (\kk,\w)|^2 d\w.
\end{align}

The deviation of the TCFM spectrum from the inherent spectrum, $f(\kk,\w)$, is given by the first term on the right-hand side of Eq.~(\ref{eq:appendix_C_decomp}).
Then, the standard deviation of the TCFM spectrum
is measured by $\sqrt{(C-C_{\rm n})/C_{\rm n}}$.
Fitting the ARPES data with the TCFM, we obtain the optimized cost function of $C=0.452$.
As Eq.~(\ref{eq:cn}) gives $C_{\rm n}=0.214$,
the standard deviation is $\sqrt{(C-C_{\rm n})/C_{\rm n}}\simeq 1.05$.
Although the number of the TCFM parameters is as small as 16, their optimization turns out to reproduce the inherent spectrum within almost one standard deviation in the full $\qq-\omega$ space.

\subsection{Quasiparticle interference}\label{ssec:method_qpi}
Based on the electronic structure, i.e., $\hat{G}(\kk,\w)$, obtained by fitting the ARPES data, we calculate the QPI patterns induced by a nonmagnetic impurity.
To calculate the spatial modulation of the LDOS, we follow the method of a single-impurity scattering \cite{wang03} (Appendix \ref{sssec:ldos}) and a projection onto a surface Wannier function \cite{dellanna05,choubey14,kreisel15,torre16,choubey17} (Appendix \ref{sssec:wannier}). 
Because the experimentally measurable quantities are not the LDOS itself, we further calculate the quantities [$g$, $L$ and $Z$ defined in Eqs.~(\ref{eq:grw}), (\ref{eq:lrw}), and (\ref{eq:zrw}) below, respectively] which can be directly measured in experiments (Appendix \ref{sssec:gandL}).

In the calculation, we place a single impurity at the origin.
Since this conserves the inversion symmetry, the Fourier-transformed quantities like $\hat{V}_{\rm imp}(\qq)$, $\d\rho(\qq,\w)$, $g(\qq,\w)$, and $L(\qq,\w)$ (defined below) are real.
In experiments, however, these quantities can have a phase because the overall phase depends on the impurity position, which is difficult to locate experimentally, and because multiple impurities can contribute. 
This prevents us from a direct comparison of the phase of $g(\qq,\w)$ and $L(\qq,\w)$ between experiments and the theory.
We therefore use $L_r$ [defined by Eq.~(\ref{eq:Lr})] and $g_r$ when we compare the calculated sign structure with the experimental results.

\subsubsection{Projection onto a surface Wannier function}\label{sssec:wannier}
To make a better comparison with STM experiments, which have a spatial resolution of 0.1nm scale, we project the TCFM lattice Green's function onto a  Wannier function respecting the $d_{x^2-y^2}$ symmetry \cite{choubey14,kreisel15,torre16,choubey17} and thereby take into account the structure inside a unit cell. 
Since STM measures a current from the surface of a cleaved sample, which consists of a Bi-O plane in the case of Bi2212, it essentially measures a tail of the Wannier function centering on the CuO$_2$ plane closest to the surface. 
To represent this tail, we employ a $d_{x^2-y^2}$-symmetry Gaussian function,
\begin{align}
    w_{\RR}(\rr)&\equiv A\frac{(x-X)^2-(y-Y)^2}{a^2}\exp\left[ -\frac{B(\rr-\RR)^2}{2a^2} \right],
\label{eq:wannier}
\end{align}
following Ref.~\onlinecite{torre16}. Here, $a=1$ is the lattice constant, $\RR=(X,Y)$ represents the coordinate of the Wannier center projected on the surface, and $\rr=(x,y)$ represents the continuous-space coordinate on the surface.
We choose $B=4$, with which the unperturbed DOS projected on the surface, $\rho_0$ defined below, well reproduces the STM experimental result.
Although we choose $A=\sqrt{B^3/(\pi a^2)}$, which normalizes Eq.~(\ref{eq:wannier}), the conclusions obtained below do not essentially depend on this choice since $\rho(\rr,\w)$ defined below is simply proportional to $A^2$ and we are interested only in relative intensity.

With the Fourier transformation, $w_{\kk}(\rr)\equiv\sum_{\RR} w_{\RR}(\rr) e^{-i\kk\cdot\RR}$, we define  
\begin{align} \hat{\mathcal{G}}_0(\rr,\w)&\equiv\frac{1}{N_{\kk}}\sum_{\kk} |w_{\kk}(\rr)|^2 \hat{G}(\kk,\w) \label{eq:g0wan}
\end{align}
and
\begin{align}
  \hat{\mathcal{G}}(\rr,\w)&\equiv\frac{1}{N_{\kk}}\sum_{\kk} w_{\kk}(\rr) \hat{G}(\kk,\w),\label{eq:gwan}
\end{align}
in a real continuous space, where $N_{\kk}$ is the number of $\kk$ points.

\subsubsection{Calculation of LDOS modulation}\label{sssec:ldos}
After the above preparations, we now calculate the LDOS at the surface, modulated by impurity scatterings \cite{wang03,andersen03,capriotti03,pereg-barnea03,zhang03}. It is written as
\begin{align}  \rho(\rr,\w)=\rho_0(\rr,\w)+\d\rho(\rr,\w) \label{eq:rho}
\end{align}
with 
\begin{align}  
\rho_0(\rr,\w)\equiv-\frac{1}{\pi} \Im [\hat{\mathcal{G}}_0(\rr,\w)]_{11}, \label{eq:rho0}
\end{align}
representing the nonperturbed part, and $\d\rho(\rr,\w)$, representing the spatially-modulated part induced by an impurity.
In the lowest order of the nonmagnetic impurity potential, $\Vi$, located at the origin, the spatially modulated part is given by \cite{wang03,boeker20}
\begin{align}
  \d\rho(\rr,\w)=-\frac{1}{\pi} \Im \left[\hat{\mathcal{G}}(\rr,\w)\Vi\hat{\mathcal{G}}(-\rr,\w)\right]_{11}. \label{eq:drhorw}
\end{align}
Note that the electron correlation effect is represented by the self-energy contained in the $c$-Green's function $\hat{\mathcal{G}}$, where the self-energy originates from the coupling to $d$ through integrating out the $d$ degrees of freedom.

For comparison with experiments, we calculate the Fourier transformation of Eq.~(\ref{eq:rho}), i.e., 
\begin{align}
  \r(\qq,\w)&\equiv \sum_{\rr} \r(\rr,\w) e^{-i\qq\cdot\rr}\nonumber\\
                &=\r_0(\qq,\w)+\d\rqw.
\end{align}
Here, $\rr$ is summed over the whole real-space points, where we consider $N_c=4\times4$ points in each unit cell.
We have confirmed that the following results do not change qualitatively for a larger $N_c$. 
Since $\r_0(\rr,\w)$ is periodic with respect to the lattice constant $a$, $\r_0(\qq,\w)$ is nonzero only at $(2\pi m/a,2\pi n/a)$ with integers $m$ and $n$. 
Hence, we focus on $q_x, q_y\in (-2\pi/a,2\pi/a)$, where $\rqw$  
is identical with $\d\rqw$ except for $\qq=(0,0)$.
Note that, while $\rqw$ and $\d\rqw$ can be complex in general, these are real in our calculated results 
since the impurity placed at the origin preserves the inversion symmetry of the system.
Thus, for $\qq\neq (0,0)$, $\rho(\qq,\w)$ is equal to
\begin{align}
  \d\r(\qq,\w)&=\frac{i}{2\pi N_{\kk}N_c} \sum_{\rr'\kk} w_{\kk}(\rr')w_{\kk+\qq}^\ast(\rr') e^{-i\qq\cdot\rr'} \nonumber\\
                &\times[\hat{G}(\kk,\w)\Vi\hat{G}(\kk+\qq,\w) \nonumber\\
                &-\{\hat{G}(\kk+\qq,\w)\Vi\hat{G}(\kk,\w)\}^\ast]_{11}, \label{eq:drhoqw}
\end{align}
where $\rr'$ is summed over the $N_c$ positions in the unit cell. We have used $w_{\kk}(-\rr)=w_{\kk}^\ast(\rr)$ and $w_{\kk}(\RR+\rr')=w_{\kk}(\rr')e^{-i\kk\cdot\RR}$. 
 
To further take into account a realistic form of the impurity, we employ a finite spread of the impurity potential in real space by replacing $\Vi$ with $\int \Vi(\rr_0)d\rr_0$ \cite{zhang03,capriotti03,zhu04,nunner06,vishik09}.
This is reasonable because the source of the impurity potential in cuprates is considered to be located in the dopant layer away from the CuO$_2$ plane.
In fact, as we see later in Sec.~\ref{ssec:qpi}, this finite spread causes a noticeable decay of the QPI intensity as $|\qq|$ increases, consistently with the experimental QPI data.
Then, instead of Eq.~(\ref{eq:drhorw}), we start from 
\begin{align}
  &\d\rho(\rr,\w)\nonumber\\
  &\equiv-\frac{1}{\pi} \int d\rr_0 \Im \left[\hat{\mathcal{G}}(\rr-\rr_0,\w)\Vi(\rr_0)\hat{\mathcal{G}}(-\rr+\rr_0,\w)\right]_{11},\label{eq:drhorw2}
\end{align}
and obtain 
\begin{align}
  \d\r(\qq,\w)&=\frac{i}{2\pi N_{\kk}N_c} \sum_{\rr'\kk} w_{\kk}(\rr')w_{\kk+\qq}^\ast(\rr') e^{-i\qq\cdot\rr'} \nonumber\\
                &\times[\hat{G}(\kk,\w)\Vi(\qq)\hat{G}(\kk+\qq,\w) \nonumber\\
                &-\{\hat{G}(\kk+\qq,\w)\Vi(\qq)\hat{G}(\kk,\w)\}^\ast]_{11}, \label{eq:drhoqw2}
\end{align}
where $\Vi(\qq)=\int \Vi(\rr_0)e^{-i\qq\cdot\rr_0}d\rr_0$.

Experimental data also indicate that both normal and anomalous (i.e., superconducting) components of $\Vi$ are necessary:
The importance of the anomalous component has been 
highlighted in previous theoretical works as well
\cite{cheng05,nunner05,nunner06}.
We therefore decompose $\Vi(\rr)$ as $\Vi(\rr)=V_n(\rr)\hat{\s}_z+V_s(\rr)\hat{\s}_x$ with the Pauli matrices $\hat{\s}_{x}$ and $\hat{\s}_{z}$.
Following Ref.~\onlinecite{sulangi17}, to mimic the realistic impurity potential, we take the Yukawa form,
\begin{align} 
  V_n(\rr)&=V_{n0}\exp(-\frac{\sqrt{\rr^2+z_0^2}-z_0}{r_{\rm imp}})\frac{z_0}{\sqrt{r_{\rm imp}^2+z_0^2}},\label{eq:vn}\\
  V_s(\rr)&=V_{s0}\left(\cos \frac{x}{a} -\cos \frac{y}{a}\right)\nonumber\\
  &\times\exp(-\frac{\sqrt{\rr^2+z_0^2}-z_0}{r_{\rm imp}})
  \frac{z_0}{\sqrt{r_{\rm imp}^2+z_0^2}}. \label{eq:vs}
\end{align}
with $r_{\rm imp}=0.5a$ and $z_0=2a$.
Note that the $d$-wave form factor in $V_s$ is necessary to make a finite contribution in Eq.~(\ref{eq:drhoqw2}) since $V_s$ connects the normal and anomalous Green's functions.

As the above impurity potentials, Eqs.~(\ref{eq:vn}) and (\ref{eq:vs}), hold the inversion symmetry, their Fourier transformation, $V_n(\qq)$ and $V_s(\qq)$, are real. In this case, Eq.~(\ref{eq:drhoqw2}) can be recast into
\begin{align}
  \d\r(\qq,\w)&=V_n(\qq) \{ \d\r_{GG}(\qq,\w) -\d\r_{FF}(\qq,\w) \} \nonumber\\
                    &+ V_s(\qq) \{ \d\r_{GF}(\qq,\w) +\d\r_{FG}(\qq,\w) \}  \label{eq:drhoqw3}
\end{align}
with
\begin{align}
  \d\r_{XY}(\qq,\w)&= \frac{-1}{\pi N_{\kk}} \sum_{\rr'\kk} w_{\kk}(\rr')w_{\kk+\qq}^\ast(\rr') e^{-i\qq\cdot\rr'} \nonumber\\
                &\times \text{Im}[ X(\kk,\w)Y(\kk+\qq,\w)]  \label{eq:XY}
\end{align}
for $X,Y=G,F$.

The amplitude of the impurity potential is not known experimentally, and we use small values within the realistic range, namely, $V_{n0}=0.1$ eV and $V_{s0}=0.01$ eV in the following unless otherwise mentioned. 
Our main interest is in the relative intensity and sign of the QPI pattern in the energy-momentum space, which are insensitive to the overall impurity strength. 
The small impurity strength ensures the validity of the lowest-order approximation in Eq.~(\ref{eq:drhorw})
while the $T$-matrix theory has been extended to the momentum-dependent impurity potentials \cite{kampf97,vishik09}.
This choice of the small values of $V_{\{n0,s0\}}$ is indeed justified by the good agreement with the experimental results as shown in Sec.~\ref{sec:result}.
The remaining discrepancy from the experimental results, in particular the relative intensity of the QPI signals at $q_{1-7}$, will mainly be ascribed either to higher-order scattering processes or to the error in fitting the ARPES data.

\subsubsection{Calculation of experimentally measurable data}\label{sssec:gandL}
The experimental data obtained by STS are not equal to the LDOS, $\rho(\rr,\w)$, but require an appropriate analysis.
The STS measures the current $I$ flowing between the tip and the sample when the bias voltage $V$ is applied between them.
Assuming that the LDOS of the tip does not significantly depend on energy in the range of our interest, we can write the current at a lateral position $\rr$ on the sample surface as
\begin{align} 
  I(\rr,z,V) =C e^{-\k(\rr)z} \int_0^{eV} \rho(\rr,\w)d\w, \label{eq:I}
\end{align}
where $\k(\rr)>0$ describes the decay of the current against the tip-sample distance $z$ and $C$ is a constant. 
In practice, when measuring the tunneling current and the differential conductance, $z$ is adjusted to maintain a constant current $\Iset$ at a fixed bias voltage $\Vset$ at each $\rr$ through
\begin{align}
  \Iset =C e^{-\k(\rr)z_\mathrm{cc}(\rr; \Vset, \Iset)} \int_0^{e\Vset} \rho(\rr,\w)d\w, \label{eq:Iset}
\end{align}
  where $z_\mathrm{cc}$ denotes a constant-current topographic image obtained at the set-point voltage $\Vset$ and current $\Iset$.
  Accordingly, the measured current is
  \begin{align}
    I(\rr,z_\mathrm{cc},V) = Ce^{-\k(\rr) z_\mathrm{cc}(\rr; \Vset, \Iset)} \int_0^{eV} \rho(\rr,\w)d\w. \label{eq:Imeasure}
  \end{align}
Combining Eqs.~(\ref{eq:Iset}) and (\ref{eq:Imeasure}), we can eliminate the factor $Ce^{-\k(\rr)z_\mathrm{cc}}$, to obtain
\begin{align} 
 I(\rr,eV;\Vset,\Iset) = \dfrac{\Iset \int_0^{eV} \rho(\rr,\w)d\w}{\int_0^{e\Vset} \rho(\rr,\w)d\w}. \label{eq:I2}
\end{align}
The differential conductance $\frac{dI}{dV}$ is then given by
\begin{align}
   g(\rr,eV;\Vset,\Iset) \equiv \frac{dI}{dV} =e\Iset \frac{\rho(\rr,eV)}{\int_0^{e\Vset}\rho(\rr,\w)d\w}. \label{eq:grw}
\end{align}
The presence of the denominator in Eq.~(\ref{eq:grw}), known as the set-point effect, prevents a direct comparison between $g(\rr,eV;\Vset,\Iset)$ and $\rho(\rr,\w)$.
The following functions are used in the analysis of experimental data to mitigate the set-point effect:
\begin{align}
  L(\rr,eV) \equiv \frac{g(\rr,eV;\Vset,\Iset)}{I(\rr,eV;\Vset,\Iset)/V} = \frac{eV\rho(\rr,eV)}{\int_0^{eV} \rho(\rr,\w)d\w}  \label{eq:lrw}
\end{align}
and 
\begin{align}
   Z(\rr,eV) \equiv \frac{g(\rr,eV;\Vset,\Iset)}{g(\rr,-eV;\Vset,\Iset)}=\frac{\rho(\rr,eV)}{\rho(\rr,-eV)}.   \label{eq:zrw}
\end{align}
For simplicity, we write $g(\rr,eV;\Vset,\Iset)$ as $g(\rr,eV)$ in the rest of the manuscript, 
abbreviating $\Vset$ and $\Iset$ in the arguments.

To compare with the experimental results, we calculate the Fourier transformation of these quantities as
\begin{align} 
  g(\qq,\w) \equiv \sum_{\rr} g(\rr,\w) e^{-i\qq \cdot \rr},\label{eq:gqw3}\\
  L(\qq,\w) \equiv \sum_{\rr} L(\rr,\w) e^{-i\qq \cdot \rr}\label{eq:lqw3},
\end{align}
and
\begin{align} 
  Z(\qq,\w) \equiv \sum_{\rr} Z(\rr,\w) e^{-i\qq \cdot \rr}.\label{eq:zqw3}
\end{align}

According to the above Fourier transformations, the spatially unmodulated part will not contribute to the QPI signals at finite $\qq$'s in the thermodynamic limit. 
However, because of the finite size of the sample or more strictly the domain where the assumption of a single impurity is valid, the experimental data of $g(\qq,\w)$ and $L(\qq,\w)$ show strong intensities for non-zero but small $\qq$'s, as seen in Figs.~\ref{fig:gdisp}(a,b) and ~\ref{fig:ldisp}(a,b).
Although these $\qq\simeq \Vec{0}$ structures are  not of our interest, in order to make a comparison with the experimental data, we have added the following Gaussian functions to Eqs.~(\ref{eq:gqw3}) and (\ref{eq:lqw3}), respectively:
\begin{align} 
&A_g\bar{\rho}_0(\w)\exp(-(q_x^2+q_y^2)/q_0^2),\\
&A_L\frac{\bar{\rho}_0(\w)}{\int_0^{eV}\bar{\rho}_0(\w)d\w}\exp(-(q_x^2+q_y^2)/q_0^2).
  \label{eq:gaussian}
\end{align}
Here, $\bar{\rho}_0(\w)$ is the average of $\rho_0(\rr,\w)$ over the unit cell.
We used $A_g=2000$, $A_L=0.7$, and $q_0=0.5/a$ to fit the experimental data.
Note that since Eq.~(\ref{eq:gaussian}) decays rapidly with $|\qq|$, it does not affect the behavior for $|\qq|\gtrsim q_0$ of our main interest. The resemblance of the obtained $\qq\simeq\Vec{0}$ structure [Figs.~\ref{fig:gdisp}(c,d) and ~\ref{fig:ldisp}(c,d)] to the experimental results suggests that the $\qq\simeq\Vec{0}$ signals in the experimental data indeed come from the above-mentioned size effect.

\subsubsection{Simple relations between $g,L$ and $\d\r$}\label{sssec:relation}
When $|\d\r|\ll|\r_0|$ holds in real space (as is indeed satisfied for the small $V_n$ and $V_s$ used in this study), 
\begin{align} 
  \frac{g(\rr,eV)}{e\Iset} &= \frac{ \r_0(\rr,eV)+\d\r(\rr,eV)}{\int_0^{e\Vset}\{ \r_0(\rr,\w)+\d\r(\rr,\w)\} d\w} \nonumber\\
  &\simeq \frac{ \r_0(\rr,eV)+\d\r(\rr,eV)}{\int_0^{e\Vset}\r_0(\rr,\w) d\w} \nonumber\\
  &-\r_0(\rr,eV) \frac{ \int_0^{e\Vset}\d\r(\rr,\w) d\w }{\left\{\int_0^{e\Vset}\r_0(\rr,\w) d\w\right\}^2}
\end{align}
Assuming $\r_0(\rr,eV)=\r_0(eV)$,
the Fourier transformation of the above equation is given for $\qq\neq \Vec{0}$ by 
\begin{align} 
  \frac{g(\qq,eV)}{e\Iset} \simeq \frac{ \d\r(\qq,eV)}{w_0(e\Vset)} -\r_0(eV) \frac{ \int_0^{e\Vset}\d\r(\qq,\w) d\w }{w_0(e\Vset)^2} \label{eq:gqw2}
\end{align}
with $w_0(\w)\equiv \int_0^{\w}\r_0(\w') d\w'$. Similarly, we can derive
\begin{align} 
  \frac{L(\qq,eV)}{eV} \simeq\frac{ \d\r(\qq,eV)}{w_0(eV)} -\r_0(eV) \frac{ \int_0^{eV}\d\r(\qq,\w) d\w }{w_0(eV)^2}.\label{eq:Lqw2}
\end{align}
These equations are useful to understand the difference in the $\qq-\w$ space maps between $g$, $L$, and $\d\r$ and indeed used in Secs.~\ref{sssec:gLsign} and \ref{sssec:high}.

\section{Single-particle spectra of the TCFM} \label{app:tcfm}

\subsection{Additional information of EDCs} \label{app:EDC_comparison}

\begin{figure}[tb]
\centering
\includegraphics[width=0.48\textwidth]{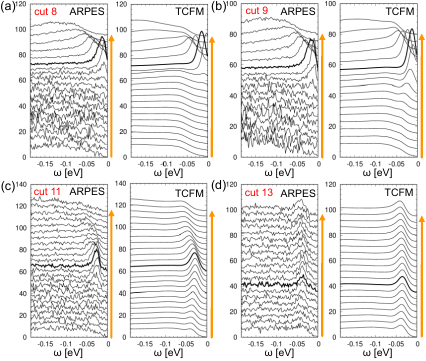}
\caption{(a-d) Comparison of the EDCs along the cuts 8, 9, 11, and 13 between ARPES ($\tilde{I}_{\rm ARPES}$) and the TCFM ($\tilde{I}_{\rm TCFM}$).
Each curve is shifted vertically by 5 for clarity.
Orange arrows correspond to those in Fig.~\ref{fig:arpes} and bold curves denote the one closest to $\kk_{\rm max}$.} \label{fig:edc}
\end{figure}

\begin{figure}[tb]
\centering
\includegraphics[width=0.48\textwidth]{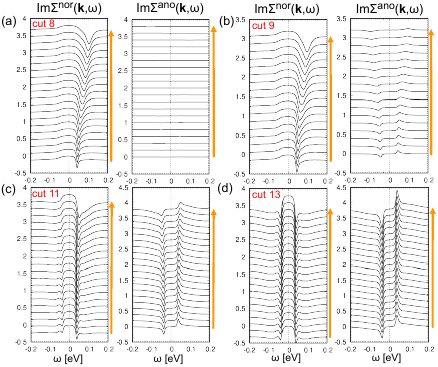}
\caption{The $\w$ dependence of Im$\Snor$ and Im$\Sano$ along the cuts 8, 9, 11, and 13 [(a), (b), (c), and (d), respectively].
Each curve is shifted vertically by 0.2 for clarity.
The orange arrows correspond to those in Fig.~\ref{fig:arpes}.} \label{fig:edc_sig}
\end{figure}

\begin{figure}[tb]
\centering
\includegraphics[width=0.48\textwidth]{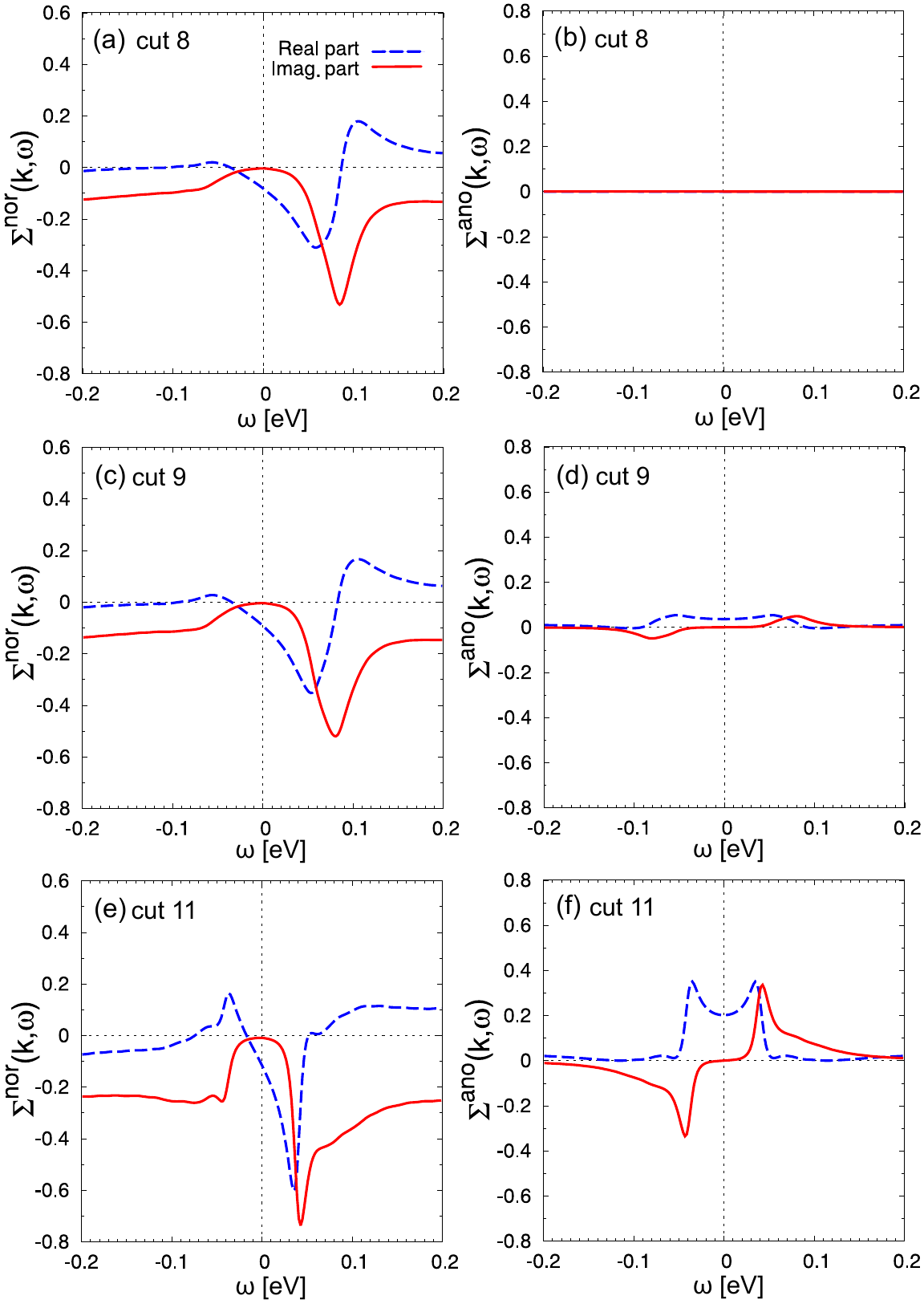}
\caption{The $\w$ dependence of the TCFM self-energy at $\kk=\kk_{\rm max}$ in the cuts 8 (a,b), 9 (c,d), and 11 (e,f), corresponding to the spectra in Fig.~\ref{fig:edc_kf}(b). Note that the corresponding self-energy for the cut 13 is presented in Fig.~\ref{fig:edc_an}(c,d), to which the scale of the ordinate is adjusted.}\label{fig:sig_kf}
\end{figure}

Figure \ref{fig:edc} displays the TCFM results of the EDCs of the normalized spectra ($\tilde{I}_{\rm TCFM}$) along the cuts 8, 9, 11, and 13, and compares them with the ARPES results ($\tilde{I}_{\rm ARPES}$). We see a good agreement, even quantitatively, between the two methods though the ARPES spectra exhibit a larger noise. 
Notably, the sharpness of the spectral peak strongly depends on the momentum along each cut, in particular near the node (cuts 8 and 9). 
This means that the imaginary part of the self-energy strongly depends on the momentum in the direction crossing the Fermi surface, beyond the constant assumption used in Ref.~\cite{vekhter03}. 

This is indeed seen in Fig.~\ref{fig:edc_sig}, which plots the imaginary part of the self-energies at the corresponding momenta.
The amplitude of the self-energy peaks significantly depends on the momentum along the cuts, as well as on the cuts themselves. In the nodal direction (cut 8), Im$\Snor$ shows a peak for $\w>0$ and its peak position depends on the momentum. 
On the other hand, in the antinodal region (cut 13), both Im$\Snor$ and Im$\Sano$ show peaks at electron-hole symmetric positions, which do not significantly depend on the momentum along the cut.

Figure \ref{fig:sig_kf} shows both the real and imaginary parts of the TCFM self-energy at $\kk=\kk_{\rm max}$ in the cuts 8, 9, and 11. Note that the corresponding self-energy for the cut 13 is presented in Fig.~\ref{fig:edc_an}(c,d). As one goes from the antinode (cut 13) to the node (cut 8), $\Sano$ diminishes while $\Snor$ keeps a substantial intensity even at the node, producing a pseudogap above the Fermi energy and renormalizing the electron dispersion around the Fermi energy. Notably, when Im$\Sano$ shows a low-energy peak (as in the cuts 11 and 13), Im$\Snor$ also shows a peak at the same energy. This is characteristic of the TCFM \cite{sakai16} and is consistent with the CDMFT calculation for the Hubbard model, as well as the self-energy structure of the cuprates extracted with a machine-learning analysis \cite{yamaji21}. 

\subsection{Dispersions for Im$\Sbg=0$} \label{app:disp_ImSbg=0}

\begin{figure}[tb]
\centering
\includegraphics[width=0.48\textwidth]{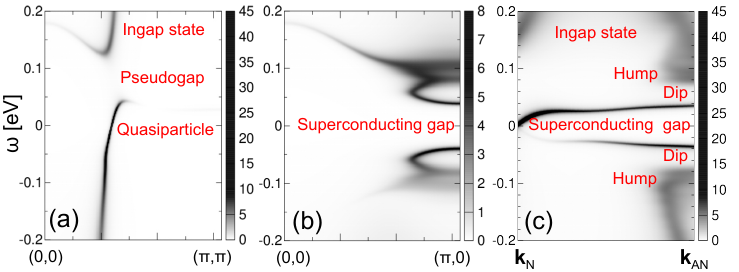}
\caption{$\Akw$ calculated for the TCFM along (a) $(0,0)-(\pi,\pi)$, (b) $(0,0)-(\pi,0)$, and (c) 
$\kN-\kAN$, with setting Im$\Sbg=0$.} 
\label{fig:akw_disp_ImSbg=0}
\end{figure}

Figure \ref{fig:akw_disp_ImSbg=0} shows $\Akw$ [Eq.~(\ref{eq:akw})]
calculated for the TCFM with Im$\Sbg$=0 (see Appendix \ref{sssec:bg} for the definition of $\Sbg$).
This result represents the spectral structure inherent to the TCFM, without the smearing by the background self-energy. 
Note that Re$\Sbg$ remains in this calculation, giving a large renormalization around the Fermi energy and a deformation of the dispersion at higher energies.

The most prominent difference from Fig.~\ref{fig:akw_disp}(a-c) is seen in the antinodal region, where $\Sbg$ is particularly strong due to the momentum dependence of Eq.~(\ref{eq:sbg}) and $c_{\rm bg}=0.218$.
We can see a clear dip (gap) between the Bogoliubov quasiparticle band and the ingap state, which corresponds to a hump.
A similar peak-dip-hump structure appears for $\w<0$ due to the particle-hole symmetric nature of the superconductivity.

\subsection{Dispersions for $\Sbg=0$} \label{app:disp_Sbg=0}

\begin{figure}[tb]
\centering
\includegraphics[width=0.48\textwidth]{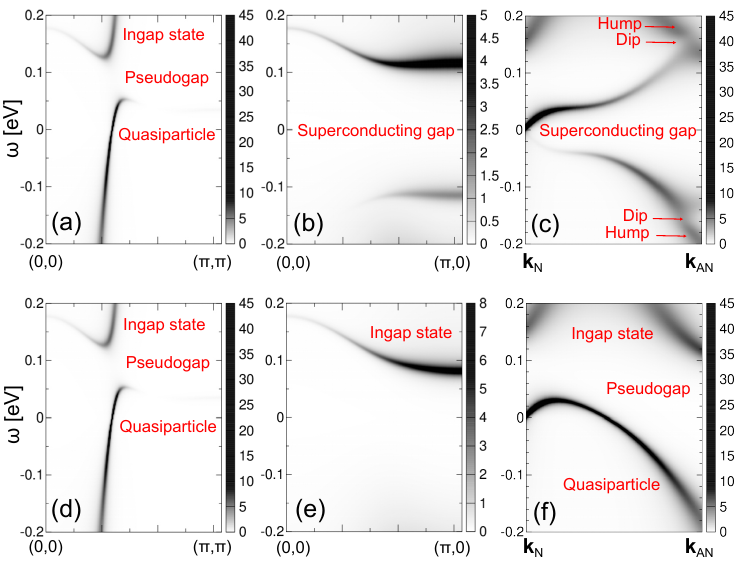}
\caption{(a-c) The same as Fig.~\ref{fig:akw_disp_ImSbg=0} but with $\Sbg=0$.
(d-f) Normal-state dispersions corresponding to (a-c), calculated with $\Sbg=0$ and $D_{c,0}=D_{d,0}=0$.} 
\label{fig:akw_disp_Sbg=0}
\end{figure}

To see the intrinsic electronic structure of the TCFM, without the smearing and renormalization by the background self-energy, we plot in 
Fig.~\ref{fig:akw_disp_Sbg=0}  $\Akw$ [Eq.~(\ref{eq:akw})]
calculated with $\Sbg$=0. Panels (a-c) show the results for the superconducting state while panels (d-f) show those for the normal state calculated with $D_{c,0}=D_{d,0}=0$.

In the nodal direction [(a,d)], the results are similar to Fig.~\ref{fig:akw_disp_ImSbg=0}(a) because of the small renormalization by Re$\Sbg$, where the pseudogap opens above the Fermi energy.
Along $(0,0)-(\pi,0)$ [(b,e)], quasiparticle band is located at $\w<-0.2$ eV (not shown) in the normal state and the ingap state is seen for $0<\w<0.2$ eV. In the superconducting state [(b)], a particle-hole counterpart of the ingap state appears around $\w=-0.1$ eV. Note that the absolute value of the ingap energy is smaller than that of the quasiparticle in this region unlike the other regions. When the background self-energy is added, this is strongly modified and becomes broadened as in Fig.~\ref{fig:akw_disp}(b).
From the node to the antinode [(c,f)], the quasiparticle band has a small dispersion (of $\sim 0.2$ eV) and the ingap state disperses nearly in parallel with the quasiparticle band [(f)]. When the superconducting gap opens [(c)], the Bogoliubov bands appear. Around $\kAN$, the upper Bogoliubov band almost touches the ingap state, resulting in a weak dip between them. For $\w<0$, the particle-hole counterpart of this structure also shows a weak dip. Note that the dip-hump structure will become more prominent in the underdoped region as the previous cDMFT studies showed \cite{sakai16,sakai16PRB}.

\subsection{$\kk$-space maps of Green's function and self-energy} \label{app:k-space_map}
\begin{figure*}[tb]
\centering
\includegraphics[width=0.96\textwidth]{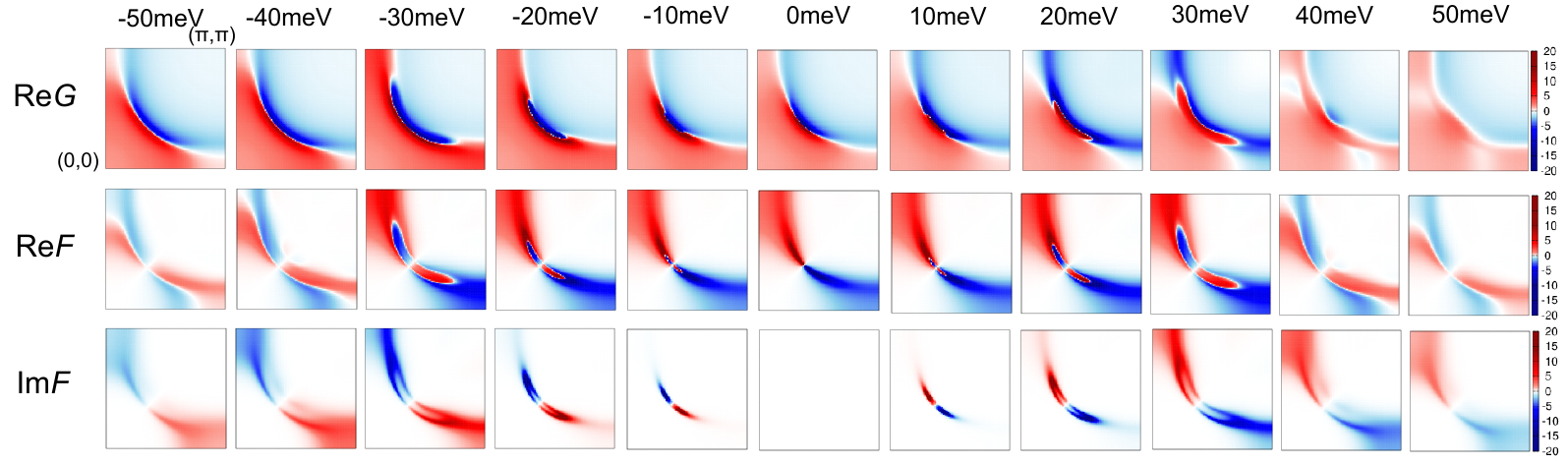}
\caption{Isoenergy momentum-space maps of Re$G$, Re$F$, and Im$F$. For Im$G$, see $\Akw=-\frac{1}{\pi}$Im$G$ in Fig.~\ref{fig:akw_kmap} lower panels.} \label{fig:G_kmap}
\end{figure*}

\begin{figure*}[tb]
\centering
\includegraphics[width=0.96\textwidth]{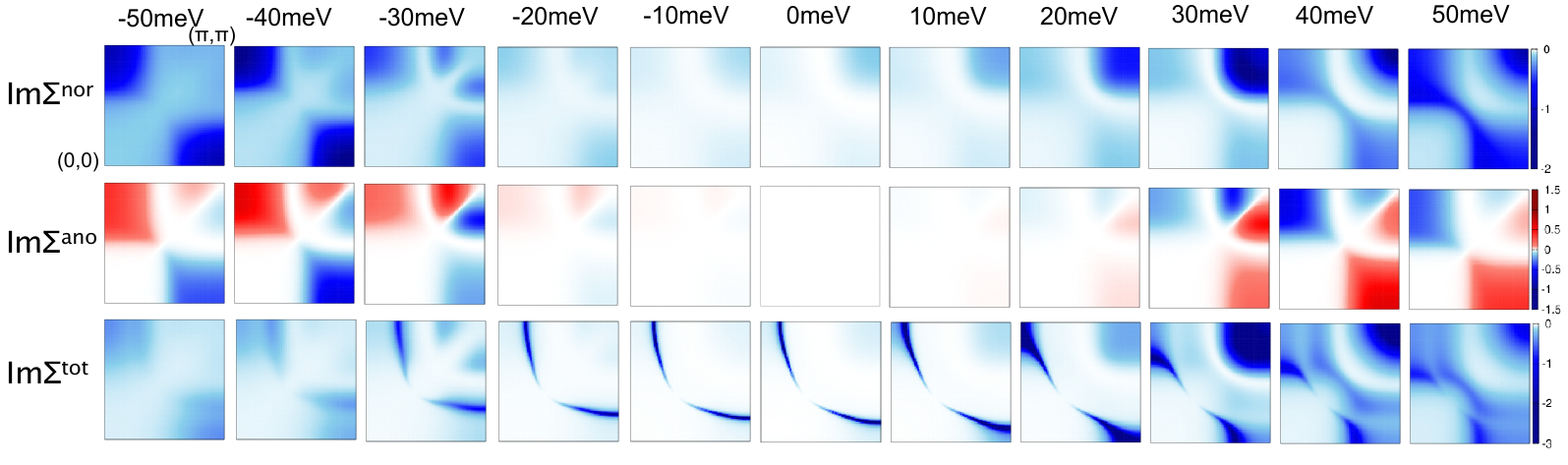}
\caption{Isoenergy momentum-space maps of Im$\Snor$, Im$\Stot$, and Im$\Sano$.} \label{fig:S_kmap}
\end{figure*}

Figure \ref{fig:akw_kmap} shows that the edges of the arc (for $|\w|\leq30$ meV) curve inwards (outwards) for $\w<0$ ($\w>0$).
Correspondingly, Re$G=0$ (white curves) in Fig.~\ref{fig:G_kmap} encompasses $(0,0)$ [$(\pi,\pi)$] for $\w=20$ and $30$ meV [$-20$ and $-30$ meV]. 
We show in Fig.~\ref{fig:S_kmap} the corresponding maps of the imaginary part of the self-energy.
Im$\Snor$ shown in the upper panels exhibits a strong negative intensity around $(\pi,\pi)$ 
particularly for $\w>0$, enhancing the electron-hole asymmetric behavior of the Green's function. Im$\Snor$ and Im$\Stot$ also show a relatively strong intensity around $(\pi,0)-(0,\pi)$ line, particularly on the positive-energy side.
This reduces and broadens the spectral intensity in the antinodal region as seen in Fig.~\ref{fig:akw_kmap} middle panels, in consistency with the ARPES results (top panels). This suppression and broadening of the antinodal spectra are ascribed to the presence of the $d$ fermion, and hence cannot be seen in the results of the single-component Hamiltonian, $H_{1c}$, shown in Fig.~\ref{fig:akw_kmap} bottom panels.

\section{Additional results of QPI}
\subsection{Decomposition of the contributions to $\d\rho$ }\label{ssec:decom}

\begin{figure}[tb]
\centering
\includegraphics[width=0.48\textwidth]{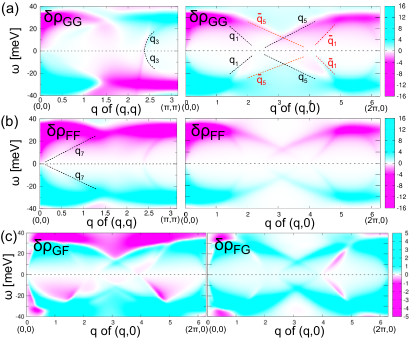}
\caption{Each component, Eq.~(\ref{eq:XY}), contributing to $\d \r (\qq,\w)$ along the $(0,0)-(\pi,\pi)$ and $(0,0)-(2\pi,0)$ lines.
Note that $\d \r_{GF} (\qq,\w)=\d \r_{FG} (\qq,\w)\equiv 0$ along $(0,0)-(\pi,\pi)$.} \label{fig:decom}
\end{figure}

To understand the sign structure of $\d\rho$ shown in Figs.~\ref{fig:drhodisp}(c) and \ref{fig:drhodisp}(d),  we plot in Fig.~\ref{fig:decom} each component, Eq.~(\ref{eq:XY}), contributing to $\d\rho$.
First, we notice that $\d\rho_{FF}$ is antisymmetric with respect to $\w=0$. This is because in Eq.~(\ref{eq:FF}) Re$F$ is symmetric about $\w=0$ while Im$F$ is antisymmetric.
As for $\d\rho_{GG}$, the overall sign structure for $20<|\w|<40$ [meV] is antisymmetric. This can be understood by an approximately electron-hole symmetric electronic structure at the optimal doping, where Im$G$ is negative-definite while Re$G$ changes sign around $\w=0$ [as is seen in Fig.~\ref{fig:edc_an}(a)]. 
For $\d\rho_{GF}$ and $\d\rho_{FG}$ [Fig.~\ref{fig:decom}(c)], we plot them only along $(0,0)-(2\pi,0)$ since they vanish for $(0,0)-(\pi,\pi)$:
As $G$ ($F$) is (anti)symmetric around the $(0,0)-(\pi,\pi)$ line, the $\kk$ summation in Eq.~(\ref{eq:XY}) vanishes.
While the amplitudes of $\d\rho_{GF}$ and $\d\rho_{FG}$ are smaller than those of $\d\rho_{GG}$ and $\d\rho_{FF}$, it is interesting that the sign structures of 
$\d\rho_{GF}$ and $\d\rho_{FG}$ are overall {\it symmetric} about $\w=0$. This is because the combination of Re$G$ and Im$F$, as well as of Re$F$ and Im$G$, keeps its sign across $\w=0$ at most $\kk$ points.
Because $\d\rho_{GF}$ and $\d\rho_{FG}$ contribute to $\d\rho$ only through $V_s$, the sign structure of $\d\rho$ could inform us of the relative strength of $V_s$ to $V_n$:
In experiment, it is possible to see the relative sign between $\d\rho(\qq,\w)$ and $\d\rho(\qq,-\w)$, which is the same as that of $g_r(\qq,\w)$ discussed in Sec.~\ref{sssec:gLsign}.

As discussed above, for $(0,0)-(\pi,\pi)$, Eq.~(\ref{eq:drhoqw3}) reduces to
\begin{align}
  \d\rho(\qq,\w)=V_n(\qq)\left[ \d\rho_{GG}(\qq,\w) - \d\rho_{FF}(\qq,\w) \right].\label{eq:GG-FF}
\end{align}
In Figs.~\ref{fig:decom}(a) and \ref{fig:decom}(b), we therefore see that both contributions from $G$ and $F$ cooperatively strengthen the $\qq_7$ dispersion through $V_n$,
particularly at $\omega>0$. On the other hand, for $q$ smaller than $|\qq_7|$, the same-sign contributions from $G$ and $F$ give a destructive interference.

The contribution from $\d\rho_{GG}$ to $\qq_3$ is positive on both sides of $\w$ while that from $-\d\rho_{FF}$ is positive (negative) for $\w>0$ ($\w<0$).
This difference leads to a stronger amplitude of $\qq_3$ for $\w>0$ than $\w<0$, consistently with the experimental result for $g$ and $L$ [Figs.~\ref{fig:gdisp}(e) and \ref{fig:ldisp}(e)].

The electron-hole asymmetric dispersion next to the $\qq_3$ signal comes from $\d\rho_{GG}$, which is negative on both sides of $\w$ and stronger in amplitude for $\w<0$ than $\w>0$. Although $-\d\rho_{FF}$ gives only a broad intensity in the corresponding region, it is positive (negative) for $\w>0$ ($\w<0$). 
This further strengthens the $\w<0$ part compared to the $\w>0$ part, explaining the white curve seen for $\d\rho(\qq,\w>0)$ in Fig.~\ref{fig:drhodisp}(c).
In the experimental results [Figs.~\ref{fig:gdisp}(a) and \ref{fig:ldisp}(a)], too, the electron-hole asymmetric dispersion has a stronger intensity for $\w<0$ than $\w>0$.

For $(0,0)-(2\pi,0)$,  $\d\rho_{GG}$ and $\d\rho_{FF}$ have the same sign in most $(\qq,\w)$ points, so that they mostly cancel with each other in Eq.~(\ref{eq:GG-FF}).
Then, $\d\rho_{GF}$ and $\d\rho_{FG}$ [Fig.~\ref{fig:decom}(c)], which are weaker than $\d\rho_{GG}$ and $\d\rho_{FF}$ but have a more symmetric sign structure between $\w>0$ and $\w<0$ for $|\w|\lesssim30$ meV, contribute to $\d\r$ through $V_s$, resulting in a nearly electron-hole symmetric sign structure of $\d\r$ shown in Fig.~\ref{fig:drhodisp}(d).

The plots for $(0,0)-(2\pi,0)$
show the $\qq_1$ and $\qq_5$ dispersions and their mirror ones. 
The bifurcated signals denoted by red arrows in Fig.~\ref{fig:ldisp}(a) would be explained by the mirror of $\qq_5$ while the mirror of $\qq_1$ is not visible in the experimental data because it is located at high momenta where the impurity scattering is weak.

\subsection{Sign structure along $(0,0)-(2\pi,0)$}\label{ssec:sign_app}

\begin{figure}[tb]
\centering
\includegraphics[width=0.48\textwidth]{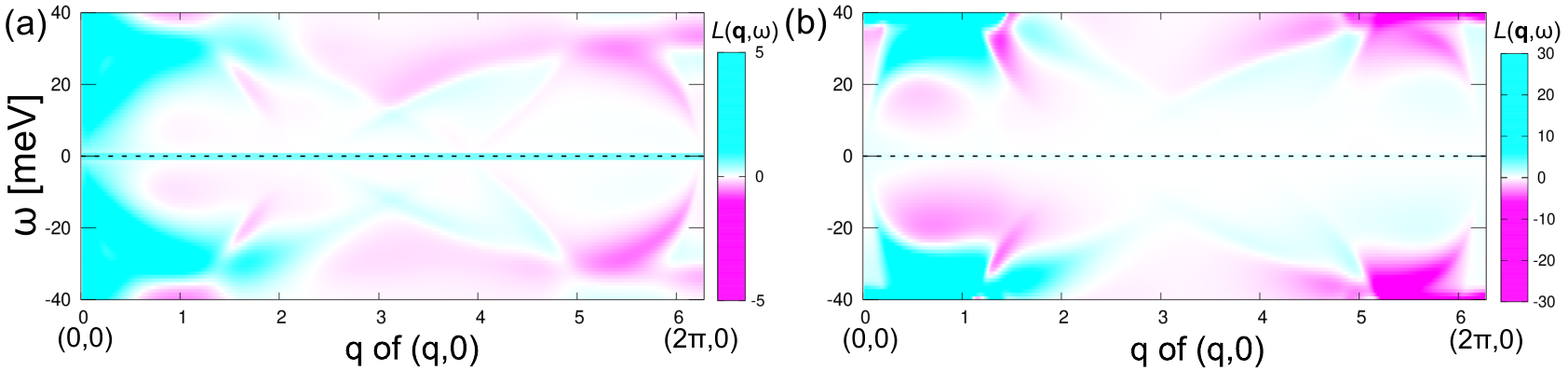}
\caption{Sign structure of $L(\qq,\w)$ along $(0,0)-(2\pi,0)$ calculated with (a) the TCFM and (b) the single-component Hamiltonian, Eq.~(\ref{eq:H1c}).} \label{fig:L_sign_00-2P0}
\end{figure}

Figure \ref{fig:L_sign_00-2P0} shows $L(\qq,\w)$ along $(0,0)-(2\pi,0)$ calculated with (a) the TCFM and (b) the single-component Hamiltonian, Eq.~(\ref{eq:H1c}).
In both cases, the sign is almost symmetric about $\w=0$. 
Thereby, the relative sign, $L_r(\qq,\w)$, is positive at most energy-momentum points in both cases in accordance with experimental results [see Supplemental Material \cite{suppl} Fig.~S2].

\subsection{Phase-reference maps}

\begin{figure}[tb]
\centering
\includegraphics[width=0.48\textwidth]{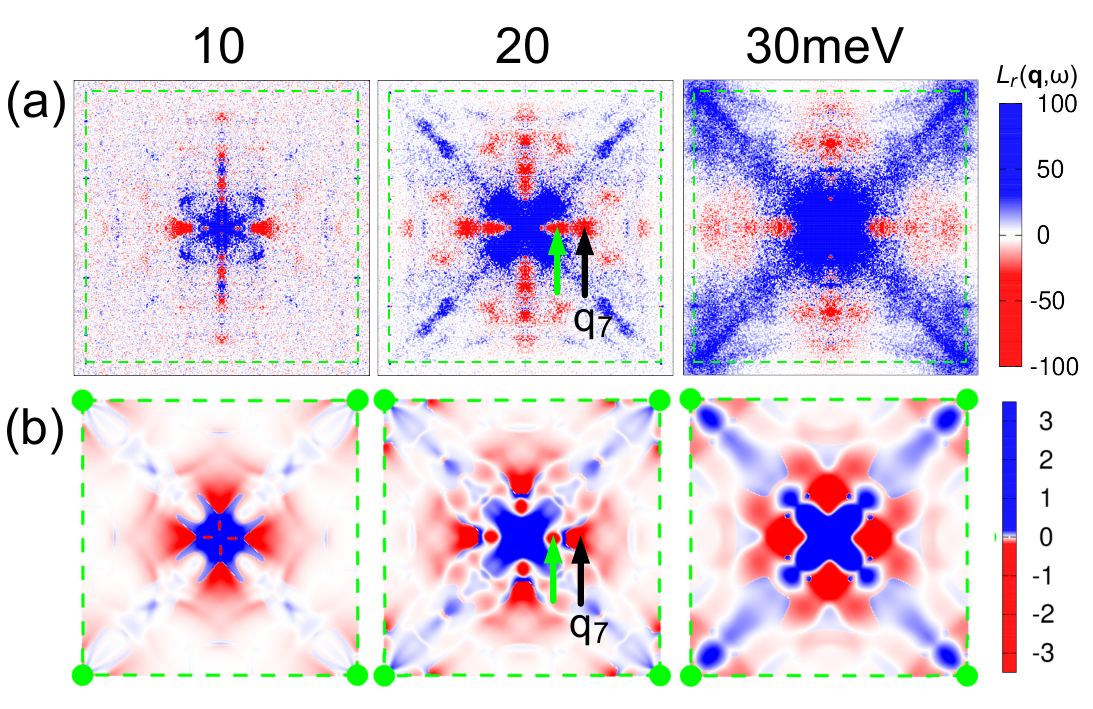}
\caption{Comparison of $\qq$-space maps of $L_r$ 
between (a) the experiment and (b) the TCFM.
Light-green circles in (b) denote the Bragg points, $(\pm2\pi,0)$ and $(0,\pm2\pi)$.
Light-green arrows denote the signal discussed in Sec.~\ref{sssec:gLsign}.
Black arrows denote the $\qq_7$ spots.} \label{fig:Lrmap}
\end{figure}

Figure \ref{fig:Lrmap} compares the calculated $\qq$-space maps of $L_r(\qq,\w)$ with the experimental results at $\w=10$, $20$, and $30$ meV. 
Corresponding results for $g_r(\qq,\w)$ are shown in Supplemental Material \cite{suppl} Fig.~S12.
Overall, all the results show a positive sign around the diagonal directions [$(0,0)-(\pm 2\pi,0)$ and $(0,0)-(0,\pm 2\pi)$] and a negative sign around the $(0,0)-(\pm\pi,\pm\pi)$ directions.
This means that the $\qq_{1,4,5}$ signals are positive while $\qq_{2,6,7}$ signals are negative, in accordance with the previous work \cite{gu19} for $g_r(\qq,\w)$, where the sign was understood as the relative phase of the $d$-wave superconducting gaps connected by $\qq$.

The amplitude in a large-$|\qq|$ region increases as $\w$ increases from 10 to 30 meV. The strong intensity in the diagonal directions corresponds to the broad flat intensity for $\w=30-40$ meV in Fig.~\ref{fig:ldisp}.

In the $L_r$ map at $\w=20$ and 30 meV, red spots (denoted by light-green arrow in Fig.~\ref{fig:Lrmap}) appear inside $\qq_7$ spots. These correspond to the negative-sign signal mentioned for Fig.~\ref{fig:Lr_00-PP}. Such signals are not seen in $g_r(\qq,\w)$ shown in Supplemental Material \cite{suppl} Fig.~S12.

Note that speckle patterns observed in experiments can be reproduced by introducing multiple impurities \cite{sulangi17} or mesoscopic inhomogeneity of the doping concentration \cite{dellanna05}. While these effects will significantly improve the appearance of the calculated results, providing a better agreement with the experimental results, the center positions of the speckled signals are the same as those of the single-impurity problem.

\bibliography{ref}

\begin{thebibliography}{128}%
\makeatletter
\providecommand \@ifxundefined [1]{%
 \@ifx{#1\undefined}
}%
\providecommand \@ifnum [1]{%
 \ifnum #1\expandafter \@firstoftwo
 \else \expandafter \@secondoftwo
 \fi
}%
\providecommand \@ifx [1]{%
 \ifx #1\expandafter \@firstoftwo
 \else \expandafter \@secondoftwo
 \fi
}%
\providecommand \natexlab [1]{#1}%
\providecommand \enquote  [1]{``#1''}%
\providecommand \bibnamefont  [1]{#1}%
\providecommand \bibfnamefont [1]{#1}%
\providecommand \citenamefont [1]{#1}%
\providecommand \href@noop [0]{\@secondoftwo}%
\providecommand \href [0]{\begingroup \@sanitize@url \@href}%
\providecommand \@href[1]{\@@startlink{#1}\@@href}%
\providecommand \@@href[1]{\endgroup#1\@@endlink}%
\providecommand \@sanitize@url [0]{\catcode `\\12\catcode `\$12\catcode
  `\&12\catcode `\#12\catcode `\^12\catcode `\_12\catcode `\%12\relax}%
\providecommand \@@startlink[1]{}%
\providecommand \@@endlink[0]{}%
\providecommand \url  [0]{\begingroup\@sanitize@url \@url }%
\providecommand \@url [1]{\endgroup\@href {#1}{\urlprefix }}%
\providecommand \urlprefix  [0]{URL }%
\providecommand \Eprint [0]{\href }%
\providecommand \doibase [0]{https://doi.org/}%
\providecommand \selectlanguage [0]{\@gobble}%
\providecommand \bibinfo  [0]{\@secondoftwo}%
\providecommand \bibfield  [0]{\@secondoftwo}%
\providecommand \translation [1]{[#1]}%
\providecommand \BibitemOpen [0]{}%
\providecommand \bibitemStop [0]{}%
\providecommand \bibitemNoStop [0]{.\EOS\space}%
\providecommand \EOS [0]{\spacefactor3000\relax}%
\providecommand \BibitemShut  [1]{\csname bibitem#1\endcsname}%
\let\auto@bib@innerbib\@empty
\bibitem [{\citenamefont {Bednorz}\ and\ \citenamefont
  {M{\"u}ller}(1986)}]{bednorz86}%
  \BibitemOpen
  \bibfield  {author} {\bibinfo {author} {\bibfnamefont {J.~G.}\ \bibnamefont
  {Bednorz}}\ and\ \bibinfo {author} {\bibfnamefont {K.~A.}\ \bibnamefont
  {M{\"u}ller}},\ }\bibfield  {title} {\bibinfo {title} {Possible high ${T}_c$
  superconductivity in the {Ba}-{La}-{Cu}-{O} system},\ }\href
  {https://doi.org/10.1007/BF01303701} {\bibfield  {journal} {\bibinfo
  {journal} {Zeitschrift f{\"u}r Physik B Condensed Matter}\ }\textbf {\bibinfo
  {volume} {64}},\ \bibinfo {pages} {189} (\bibinfo {year} {1986})}\BibitemShut
  {NoStop}%
\bibitem [{\citenamefont {Keimer}\ \emph {et~al.}(2015)\citenamefont {Keimer},
  \citenamefont {Kivelson}, \citenamefont {Norman}, \citenamefont {Uchida},\
  and\ \citenamefont {Zaanen}}]{keimer15}%
  \BibitemOpen
  \bibfield  {author} {\bibinfo {author} {\bibfnamefont {B.}~\bibnamefont
  {Keimer}}, \bibinfo {author} {\bibfnamefont {S.~A.}\ \bibnamefont
  {Kivelson}}, \bibinfo {author} {\bibfnamefont {M.~R.}\ \bibnamefont
  {Norman}}, \bibinfo {author} {\bibfnamefont {S.}~\bibnamefont {Uchida}},\
  and\ \bibinfo {author} {\bibfnamefont {J.}~\bibnamefont {Zaanen}},\
  }\bibfield  {title} {\bibinfo {title} {From quantum matter to
  high-temperature superconductivity in copper oxides},\ }\href
  {https://doi.org/10.1038/nature14165} {\bibfield  {journal} {\bibinfo
  {journal} {Nature}\ }\textbf {\bibinfo {volume} {518}},\ \bibinfo {pages}
  {179} (\bibinfo {year} {2015})}\BibitemShut {NoStop}%
\bibitem [{\citenamefont {Damascelli}\ \emph {et~al.}(2003)\citenamefont
  {Damascelli}, \citenamefont {Hussain},\ and\ \citenamefont
  {Shen}}]{damascelli03}%
  \BibitemOpen
  \bibfield  {author} {\bibinfo {author} {\bibfnamefont {A.}~\bibnamefont
  {Damascelli}}, \bibinfo {author} {\bibfnamefont {Z.}~\bibnamefont
  {Hussain}},\ and\ \bibinfo {author} {\bibfnamefont {Z.-X.}\ \bibnamefont
  {Shen}},\ }\bibfield  {title} {\bibinfo {title} {Angle-resolved photoemission
  studies of the cuprate superconductors},\ }\href
  {https://doi.org/10.1103/RevModPhys.75.473} {\bibfield  {journal} {\bibinfo
  {journal} {Rev. Mod. Phys.}\ }\textbf {\bibinfo {volume} {75}},\ \bibinfo
  {pages} {473} (\bibinfo {year} {2003})}\BibitemShut {NoStop}%
\bibitem [{\citenamefont {Sobota}\ \emph {et~al.}(2021)\citenamefont {Sobota},
  \citenamefont {He},\ and\ \citenamefont {Shen}}]{sobota21}%
  \BibitemOpen
  \bibfield  {author} {\bibinfo {author} {\bibfnamefont {J.~A.}\ \bibnamefont
  {Sobota}}, \bibinfo {author} {\bibfnamefont {Y.}~\bibnamefont {He}},\ and\
  \bibinfo {author} {\bibfnamefont {Z.-X.}\ \bibnamefont {Shen}},\ }\bibfield
  {title} {\bibinfo {title} {Angle-resolved photoemission studies of quantum
  materials},\ }\href {https://doi.org/10.1103/RevModPhys.93.025006} {\bibfield
   {journal} {\bibinfo  {journal} {Rev. Mod. Phys.}\ }\textbf {\bibinfo
  {volume} {93}},\ \bibinfo {pages} {025006} (\bibinfo {year}
  {2021})}\BibitemShut {NoStop}%
\bibitem [{\citenamefont {Fischer}\ \emph {et~al.}(2007)\citenamefont
  {Fischer}, \citenamefont {Kugler}, \citenamefont {Maggio-Aprile},
  \citenamefont {Berthod},\ and\ \citenamefont {Renner}}]{fischer07}%
  \BibitemOpen
  \bibfield  {author} {\bibinfo {author} {\bibfnamefont {{\O}.}~\bibnamefont
  {Fischer}}, \bibinfo {author} {\bibfnamefont {M.}~\bibnamefont {Kugler}},
  \bibinfo {author} {\bibfnamefont {I.}~\bibnamefont {Maggio-Aprile}}, \bibinfo
  {author} {\bibfnamefont {C.}~\bibnamefont {Berthod}},\ and\ \bibinfo {author}
  {\bibfnamefont {C.}~\bibnamefont {Renner}},\ }\bibfield  {title} {\bibinfo
  {title} {Scanning tunneling spectroscopy of high-temperature
  superconductors},\ }\href {https://doi.org/10.1103/RevModPhys.79.353}
  {\bibfield  {journal} {\bibinfo  {journal} {Rev. Mod. Phys.}\ }\textbf
  {\bibinfo {volume} {79}},\ \bibinfo {pages} {353} (\bibinfo {year}
  {2007})}\BibitemShut {NoStop}%
\bibitem [{\citenamefont {Hasegawa}\ and\ \citenamefont
  {Avouris}(1993)}]{hasegawa93}%
  \BibitemOpen
  \bibfield  {author} {\bibinfo {author} {\bibfnamefont {Y.}~\bibnamefont
  {Hasegawa}}\ and\ \bibinfo {author} {\bibfnamefont {P.}~\bibnamefont
  {Avouris}},\ }\bibfield  {title} {\bibinfo {title} {Direct observation of
  standing wave formation at surface steps using scanning tunneling
  spectroscopy},\ }\href {https://doi.org/10.1103/PhysRevLett.71.1071}
  {\bibfield  {journal} {\bibinfo  {journal} {Phys. Rev. Lett.}\ }\textbf
  {\bibinfo {volume} {71}},\ \bibinfo {pages} {1071} (\bibinfo {year}
  {1993})}\BibitemShut {NoStop}%
\bibitem [{\citenamefont {Crommie}\ \emph {et~al.}(1993)\citenamefont
  {Crommie}, \citenamefont {Lutz},\ and\ \citenamefont {Eigler}}]{crommie93}%
  \BibitemOpen
  \bibfield  {author} {\bibinfo {author} {\bibfnamefont {M.~F.}\ \bibnamefont
  {Crommie}}, \bibinfo {author} {\bibfnamefont {C.~P.}\ \bibnamefont {Lutz}},\
  and\ \bibinfo {author} {\bibfnamefont {D.~M.}\ \bibnamefont {Eigler}},\
  }\bibfield  {title} {\bibinfo {title} {Imaging standing waves in a
  two-dimensional electron gas},\ }\href {https://doi.org/10.1038/363524a0}
  {\bibfield  {journal} {\bibinfo  {journal} {Nature}\ }\textbf {\bibinfo
  {volume} {363}},\ \bibinfo {pages} {524} (\bibinfo {year}
  {1993})}\BibitemShut {NoStop}%
\bibitem [{\citenamefont {Hoffman}\ \emph {et~al.}(2002)\citenamefont
  {Hoffman}, \citenamefont {McElroy}, \citenamefont {Lee}, \citenamefont
  {Lang}, \citenamefont {Eisaki}, \citenamefont {Uchida},\ and\ \citenamefont
  {Davis}}]{hoffman02}%
  \BibitemOpen
  \bibfield  {author} {\bibinfo {author} {\bibfnamefont {J.~E.}\ \bibnamefont
  {Hoffman}}, \bibinfo {author} {\bibfnamefont {K.}~\bibnamefont {McElroy}},
  \bibinfo {author} {\bibfnamefont {D.-H.}\ \bibnamefont {Lee}}, \bibinfo
  {author} {\bibfnamefont {K.~M.}\ \bibnamefont {Lang}}, \bibinfo {author}
  {\bibfnamefont {H.}~\bibnamefont {Eisaki}}, \bibinfo {author} {\bibfnamefont
  {S.}~\bibnamefont {Uchida}},\ and\ \bibinfo {author} {\bibfnamefont {J.~C.}\
  \bibnamefont {Davis}},\ }\bibfield  {title} {\bibinfo {title} {Imaging
  quasiparticle interference in {Bi}$_2${Sr}$_2${CaCu}$_2${O}$_{8+\delta}$},\
  }\href {https://doi.org/10.1126/science.1072640} {\bibfield  {journal}
  {\bibinfo  {journal} {Science}\ }\textbf {\bibinfo {volume} {297}},\ \bibinfo
  {pages} {1148} (\bibinfo {year} {2002})}\BibitemShut {NoStop}%
\bibitem [{\citenamefont {McElroy}\ \emph {et~al.}(2003)\citenamefont
  {McElroy}, \citenamefont {Simmonds}, \citenamefont {Hoffman}, \citenamefont
  {Lee}, \citenamefont {Orenstein}, \citenamefont {Eisaki}, \citenamefont
  {Uchida},\ and\ \citenamefont {Davis}}]{mcelroy03}%
  \BibitemOpen
  \bibfield  {author} {\bibinfo {author} {\bibfnamefont {K.}~\bibnamefont
  {McElroy}}, \bibinfo {author} {\bibfnamefont {R.~W.}\ \bibnamefont
  {Simmonds}}, \bibinfo {author} {\bibfnamefont {J.~E.}\ \bibnamefont
  {Hoffman}}, \bibinfo {author} {\bibfnamefont {D.-H.}\ \bibnamefont {Lee}},
  \bibinfo {author} {\bibfnamefont {J.}~\bibnamefont {Orenstein}}, \bibinfo
  {author} {\bibfnamefont {H.}~\bibnamefont {Eisaki}}, \bibinfo {author}
  {\bibfnamefont {S.}~\bibnamefont {Uchida}},\ and\ \bibinfo {author}
  {\bibfnamefont {J.~C.}\ \bibnamefont {Davis}},\ }\bibfield  {title} {\bibinfo
  {title} {Relating atomic-scale electronic phenomena to wave-like
  quasiparticle states in superconducting
  {Bi}$_2${Sr}$_2${CaCu}$_2${O}$_{8+\delta}$},\ }\href
  {https://doi.org/10.1038/nature01496} {\bibfield  {journal} {\bibinfo
  {journal} {Nature}\ }\textbf {\bibinfo {volume} {422}},\ \bibinfo {pages}
  {592} (\bibinfo {year} {2003})}\BibitemShut {NoStop}%
\bibitem [{\citenamefont {Hanaguri}\ \emph {et~al.}(2004)\citenamefont
  {Hanaguri}, \citenamefont {Lupien}, \citenamefont {Kohsaka}, \citenamefont
  {Lee}, \citenamefont {Azuma}, \citenamefont {Takano}, \citenamefont
  {Takagi},\ and\ \citenamefont {Davis}}]{hanaguri04}%
  \BibitemOpen
  \bibfield  {author} {\bibinfo {author} {\bibfnamefont {T.}~\bibnamefont
  {Hanaguri}}, \bibinfo {author} {\bibfnamefont {C.}~\bibnamefont {Lupien}},
  \bibinfo {author} {\bibfnamefont {Y.}~\bibnamefont {Kohsaka}}, \bibinfo
  {author} {\bibfnamefont {D.-H.}\ \bibnamefont {Lee}}, \bibinfo {author}
  {\bibfnamefont {M.}~\bibnamefont {Azuma}}, \bibinfo {author} {\bibfnamefont
  {M.}~\bibnamefont {Takano}}, \bibinfo {author} {\bibfnamefont
  {H.}~\bibnamefont {Takagi}},\ and\ \bibinfo {author} {\bibfnamefont {J.~C.}\
  \bibnamefont {Davis}},\ }\bibfield  {title} {\bibinfo {title} {A
  `checkerboard' electronic crystal state in lightly hole-doped
  {Ca}$_{2-x}${Na}$_x${CuO}$_2${Cl}$_2$},\ }\href
  {https://doi.org/10.1038/nature02861} {\bibfield  {journal} {\bibinfo
  {journal} {Nature}\ }\textbf {\bibinfo {volume} {430}},\ \bibinfo {pages}
  {1001} (\bibinfo {year} {2004})}\BibitemShut {NoStop}%
\bibitem [{\citenamefont {Hanaguri}\ \emph {et~al.}(2007)\citenamefont
  {Hanaguri}, \citenamefont {Kohsaka}, \citenamefont {Davis}, \citenamefont
  {Lupien}, \citenamefont {Yamada}, \citenamefont {Azuma}, \citenamefont
  {Takano}, \citenamefont {Ohishi}, \citenamefont {Ono},\ and\ \citenamefont
  {Takagi}}]{hanaguri07}%
  \BibitemOpen
  \bibfield  {author} {\bibinfo {author} {\bibfnamefont {T.}~\bibnamefont
  {Hanaguri}}, \bibinfo {author} {\bibfnamefont {Y.}~\bibnamefont {Kohsaka}},
  \bibinfo {author} {\bibfnamefont {J.~C.}\ \bibnamefont {Davis}}, \bibinfo
  {author} {\bibfnamefont {C.}~\bibnamefont {Lupien}}, \bibinfo {author}
  {\bibfnamefont {I.}~\bibnamefont {Yamada}}, \bibinfo {author} {\bibfnamefont
  {M.}~\bibnamefont {Azuma}}, \bibinfo {author} {\bibfnamefont
  {M.}~\bibnamefont {Takano}}, \bibinfo {author} {\bibfnamefont
  {K.}~\bibnamefont {Ohishi}}, \bibinfo {author} {\bibfnamefont
  {M.}~\bibnamefont {Ono}},\ and\ \bibinfo {author} {\bibfnamefont
  {H.}~\bibnamefont {Takagi}},\ }\bibfield  {title} {\bibinfo {title}
  {Quasiparticle interference and superconducting gap in
  {Ca}$_{2-x}${Na}$_x${CuO}$_2${Cl}$_2$},\ }\href
  {https://doi.org/10.1038/nphys753} {\bibfield  {journal} {\bibinfo  {journal}
  {Nature Physics}\ }\textbf {\bibinfo {volume} {3}},\ \bibinfo {pages} {865}
  (\bibinfo {year} {2007})}\BibitemShut {NoStop}%
\bibitem [{\citenamefont {Kohsaka}\ \emph {et~al.}(2008)\citenamefont
  {Kohsaka}, \citenamefont {Taylor}, \citenamefont {Wahl}, \citenamefont
  {Schmidt}, \citenamefont {Lee}, \citenamefont {Fujita}, \citenamefont
  {Alldredge}, \citenamefont {McElroy}, \citenamefont {Lee}, \citenamefont
  {Eisaki}, \citenamefont {Uchida}, \citenamefont {Lee},\ and\ \citenamefont
  {Davis}}]{kohsaka08}%
  \BibitemOpen
  \bibfield  {author} {\bibinfo {author} {\bibfnamefont {Y.}~\bibnamefont
  {Kohsaka}}, \bibinfo {author} {\bibfnamefont {C.}~\bibnamefont {Taylor}},
  \bibinfo {author} {\bibfnamefont {P.}~\bibnamefont {Wahl}}, \bibinfo {author}
  {\bibfnamefont {A.}~\bibnamefont {Schmidt}}, \bibinfo {author} {\bibfnamefont
  {J.}~\bibnamefont {Lee}}, \bibinfo {author} {\bibfnamefont {K.}~\bibnamefont
  {Fujita}}, \bibinfo {author} {\bibfnamefont {J.~W.}\ \bibnamefont
  {Alldredge}}, \bibinfo {author} {\bibfnamefont {K.}~\bibnamefont {McElroy}},
  \bibinfo {author} {\bibfnamefont {J.}~\bibnamefont {Lee}}, \bibinfo {author}
  {\bibfnamefont {H.}~\bibnamefont {Eisaki}}, \bibinfo {author} {\bibfnamefont
  {S.}~\bibnamefont {Uchida}}, \bibinfo {author} {\bibfnamefont {D.-H.}\
  \bibnamefont {Lee}},\ and\ \bibinfo {author} {\bibfnamefont {J.~C.}\
  \bibnamefont {Davis}},\ }\bibfield  {title} {\bibinfo {title} {How cooper
  pairs vanish approaching the mott insulator in
  {Bi}$_2${Sr}$_2${CaCu}$_2${O}$_{8+\delta}$},\ }\href
  {https://doi.org/10.1038/nature07243} {\bibfield  {journal} {\bibinfo
  {journal} {Nature}\ }\textbf {\bibinfo {volume} {454}},\ \bibinfo {pages}
  {1072} (\bibinfo {year} {2008})}\BibitemShut {NoStop}%
\bibitem [{\citenamefont {Lee}\ \emph {et~al.}(2009)\citenamefont {Lee},
  \citenamefont {Fujita}, \citenamefont {Schmidt}, \citenamefont {Kim},
  \citenamefont {Eisaki}, \citenamefont {Uchida},\ and\ \citenamefont
  {Davis}}]{lee09}%
  \BibitemOpen
  \bibfield  {author} {\bibinfo {author} {\bibfnamefont {J.}~\bibnamefont
  {Lee}}, \bibinfo {author} {\bibfnamefont {K.}~\bibnamefont {Fujita}},
  \bibinfo {author} {\bibfnamefont {A.~R.}\ \bibnamefont {Schmidt}}, \bibinfo
  {author} {\bibfnamefont {C.~K.}\ \bibnamefont {Kim}}, \bibinfo {author}
  {\bibfnamefont {H.}~\bibnamefont {Eisaki}}, \bibinfo {author} {\bibfnamefont
  {S.}~\bibnamefont {Uchida}},\ and\ \bibinfo {author} {\bibfnamefont {J.~C.}\
  \bibnamefont {Davis}},\ }\bibfield  {title} {\bibinfo {title} {Spectroscopic
  fingerprint of phase-incoherent superconductivity in the underdoped
  {Bi}$_2${Sr}$_2${CaCu}$_2${O}$_{8+\delta}$},\ }\href
  {https://doi.org/10.1126/science.1176369} {\bibfield  {journal} {\bibinfo
  {journal} {Science}\ }\textbf {\bibinfo {volume} {325}},\ \bibinfo {pages}
  {1099} (\bibinfo {year} {2009})}\BibitemShut {NoStop}%
\bibitem [{\citenamefont {Alldredge}\ \emph {et~al.}(2012)\citenamefont
  {Alldredge}, \citenamefont {Fujita}, \citenamefont {Eisaki}, \citenamefont
  {Uchida},\ and\ \citenamefont {McElroy}}]{alldredge12}%
  \BibitemOpen
  \bibfield  {author} {\bibinfo {author} {\bibfnamefont {J.~W.}\ \bibnamefont
  {Alldredge}}, \bibinfo {author} {\bibfnamefont {K.}~\bibnamefont {Fujita}},
  \bibinfo {author} {\bibfnamefont {H.}~\bibnamefont {Eisaki}}, \bibinfo
  {author} {\bibfnamefont {S.}~\bibnamefont {Uchida}},\ and\ \bibinfo {author}
  {\bibfnamefont {K.}~\bibnamefont {McElroy}},\ }\bibfield  {title} {\bibinfo
  {title} {Three-component electronic structure of the cuprates derived from
  spectroscopic-imaging scanning tunneling microscopy},\ }\href
  {https://doi.org/10.1103/PhysRevB.85.174501} {\bibfield  {journal} {\bibinfo
  {journal} {Phys. Rev. B}\ }\textbf {\bibinfo {volume} {85}},\ \bibinfo
  {pages} {174501} (\bibinfo {year} {2012})}\BibitemShut {NoStop}%
\bibitem [{\citenamefont {He}\ \emph {et~al.}(2014)\citenamefont {He},
  \citenamefont {Yin}, \citenamefont {Zech}, \citenamefont {Soumyanarayanan},
  \citenamefont {Yee}, \citenamefont {Williams}, \citenamefont {Boyer},
  \citenamefont {Chatterjee}, \citenamefont {Wise}, \citenamefont {Zeljkovic},
  \citenamefont {Kondo}, \citenamefont {Takeuchi}, \citenamefont {Ikuta},
  \citenamefont {Mistark}, \citenamefont {Markiewicz}, \citenamefont {Bansil},
  \citenamefont {Sachdev}, \citenamefont {Hudson},\ and\ \citenamefont
  {Hoffman}}]{he14}%
  \BibitemOpen
  \bibfield  {author} {\bibinfo {author} {\bibfnamefont {Y.}~\bibnamefont
  {He}}, \bibinfo {author} {\bibfnamefont {Y.}~\bibnamefont {Yin}}, \bibinfo
  {author} {\bibfnamefont {M.}~\bibnamefont {Zech}}, \bibinfo {author}
  {\bibfnamefont {A.}~\bibnamefont {Soumyanarayanan}}, \bibinfo {author}
  {\bibfnamefont {M.~M.}\ \bibnamefont {Yee}}, \bibinfo {author} {\bibfnamefont
  {T.}~\bibnamefont {Williams}}, \bibinfo {author} {\bibfnamefont {M.~C.}\
  \bibnamefont {Boyer}}, \bibinfo {author} {\bibfnamefont {K.}~\bibnamefont
  {Chatterjee}}, \bibinfo {author} {\bibfnamefont {W.~D.}\ \bibnamefont
  {Wise}}, \bibinfo {author} {\bibfnamefont {I.}~\bibnamefont {Zeljkovic}},
  \bibinfo {author} {\bibfnamefont {T.}~\bibnamefont {Kondo}}, \bibinfo
  {author} {\bibfnamefont {T.}~\bibnamefont {Takeuchi}}, \bibinfo {author}
  {\bibfnamefont {H.}~\bibnamefont {Ikuta}}, \bibinfo {author} {\bibfnamefont
  {P.}~\bibnamefont {Mistark}}, \bibinfo {author} {\bibfnamefont {R.~S.}\
  \bibnamefont {Markiewicz}}, \bibinfo {author} {\bibfnamefont
  {A.}~\bibnamefont {Bansil}}, \bibinfo {author} {\bibfnamefont
  {S.}~\bibnamefont {Sachdev}}, \bibinfo {author} {\bibfnamefont {E.~W.}\
  \bibnamefont {Hudson}},\ and\ \bibinfo {author} {\bibfnamefont {J.~E.}\
  \bibnamefont {Hoffman}},\ }\bibfield  {title} {\bibinfo {title} {Fermi
  surface and pseudogap evolution in a cuprate superconductor},\ }\href
  {https://doi.org/10.1126/science.1248221} {\bibfield  {journal} {\bibinfo
  {journal} {Science}\ }\textbf {\bibinfo {volume} {344}},\ \bibinfo {pages}
  {608} (\bibinfo {year} {2014})}\BibitemShut {NoStop}%
\bibitem [{\citenamefont {Fujita}\ \emph {et~al.}(2014)\citenamefont {Fujita},
  \citenamefont {Kim}, \citenamefont {Lee}, \citenamefont {Lee}, \citenamefont
  {Hamidian}, \citenamefont {Firmo}, \citenamefont {Mukhopadhyay},
  \citenamefont {Eisaki}, \citenamefont {Uchida}, \citenamefont {Lawler},
  \citenamefont {Kim},\ and\ \citenamefont {Davis}}]{fujita14}%
  \BibitemOpen
  \bibfield  {author} {\bibinfo {author} {\bibfnamefont {K.}~\bibnamefont
  {Fujita}}, \bibinfo {author} {\bibfnamefont {C.~K.}\ \bibnamefont {Kim}},
  \bibinfo {author} {\bibfnamefont {I.}~\bibnamefont {Lee}}, \bibinfo {author}
  {\bibfnamefont {J.}~\bibnamefont {Lee}}, \bibinfo {author} {\bibfnamefont
  {M.~H.}\ \bibnamefont {Hamidian}}, \bibinfo {author} {\bibfnamefont {I.~A.}\
  \bibnamefont {Firmo}}, \bibinfo {author} {\bibfnamefont {S.}~\bibnamefont
  {Mukhopadhyay}}, \bibinfo {author} {\bibfnamefont {H.}~\bibnamefont
  {Eisaki}}, \bibinfo {author} {\bibfnamefont {S.}~\bibnamefont {Uchida}},
  \bibinfo {author} {\bibfnamefont {M.~J.}\ \bibnamefont {Lawler}}, \bibinfo
  {author} {\bibfnamefont {E.-A.}\ \bibnamefont {Kim}},\ and\ \bibinfo {author}
  {\bibfnamefont {J.~C.}\ \bibnamefont {Davis}},\ }\bibfield  {title} {\bibinfo
  {title} {Simultaneous transitions in cuprate momentum-space topology and
  electronic symmetry breaking},\ }\href
  {https://doi.org/10.1126/science.1248783} {\bibfield  {journal} {\bibinfo
  {journal} {Science}\ }\textbf {\bibinfo {volume} {344}},\ \bibinfo {pages}
  {612} (\bibinfo {year} {2014})}\BibitemShut {NoStop}%
\bibitem [{\citenamefont {Machida}\ \emph {et~al.}(2016)\citenamefont
  {Machida}, \citenamefont {Kohsaka}, \citenamefont {Matsuoka}, \citenamefont
  {Iwaya}, \citenamefont {Hanaguri},\ and\ \citenamefont
  {Tamegai}}]{machida16}%
  \BibitemOpen
  \bibfield  {author} {\bibinfo {author} {\bibfnamefont {T.}~\bibnamefont
  {Machida}}, \bibinfo {author} {\bibfnamefont {Y.}~\bibnamefont {Kohsaka}},
  \bibinfo {author} {\bibfnamefont {K.}~\bibnamefont {Matsuoka}}, \bibinfo
  {author} {\bibfnamefont {K.}~\bibnamefont {Iwaya}}, \bibinfo {author}
  {\bibfnamefont {T.}~\bibnamefont {Hanaguri}},\ and\ \bibinfo {author}
  {\bibfnamefont {T.}~\bibnamefont {Tamegai}},\ }\bibfield  {title} {\bibinfo
  {title} {Bipartite electronic superstructures in the vortex core of
  {Bi}$_2${Sr}$_2${CaCu}$_2${O}$_{8+\delta}$},\ }\href
  {https://doi.org/10.1038/ncomms11747} {\bibfield  {journal} {\bibinfo
  {journal} {Nature Communications}\ }\textbf {\bibinfo {volume} {7}},\
  \bibinfo {pages} {11747} (\bibinfo {year} {2016})}\BibitemShut {NoStop}%
\bibitem [{\citenamefont {Gu}\ \emph {et~al.}(2019)\citenamefont {Gu},
  \citenamefont {Wan}, \citenamefont {Tang}, \citenamefont {Du}, \citenamefont
  {Yang}, \citenamefont {Wang}, \citenamefont {Zhong}, \citenamefont {Wen},
  \citenamefont {Gu},\ and\ \citenamefont {Wen}}]{gu19}%
  \BibitemOpen
  \bibfield  {author} {\bibinfo {author} {\bibfnamefont {Q.}~\bibnamefont
  {Gu}}, \bibinfo {author} {\bibfnamefont {S.}~\bibnamefont {Wan}}, \bibinfo
  {author} {\bibfnamefont {Q.}~\bibnamefont {Tang}}, \bibinfo {author}
  {\bibfnamefont {Z.}~\bibnamefont {Du}}, \bibinfo {author} {\bibfnamefont
  {H.}~\bibnamefont {Yang}}, \bibinfo {author} {\bibfnamefont {Q.-H.}\
  \bibnamefont {Wang}}, \bibinfo {author} {\bibfnamefont {R.}~\bibnamefont
  {Zhong}}, \bibinfo {author} {\bibfnamefont {J.}~\bibnamefont {Wen}}, \bibinfo
  {author} {\bibfnamefont {G.~D.}\ \bibnamefont {Gu}},\ and\ \bibinfo {author}
  {\bibfnamefont {H.-H.}\ \bibnamefont {Wen}},\ }\bibfield  {title} {\bibinfo
  {title} {Directly visualizing the sign change of d-wave superconducting gap
  in {Bi}$_2${Sr}$_2${CaCu}$_2${O}$_{8+\delta}$ by phase-referenced
  quasiparticle interference},\ }\href
  {https://doi.org/10.1038/s41467-019-09340-5} {\bibfield  {journal} {\bibinfo
  {journal} {Nature Communications}\ }\textbf {\bibinfo {volume} {10}},\
  \bibinfo {pages} {1603} (\bibinfo {year} {2019})}\BibitemShut {NoStop}%
\bibitem [{\citenamefont {Yu}\ \emph {et~al.}(2019)\citenamefont {Yu},
  \citenamefont {Ma}, \citenamefont {Cai}, \citenamefont {Zhong}, \citenamefont
  {Ye}, \citenamefont {Shen}, \citenamefont {Gu}, \citenamefont {Chen},\ and\
  \citenamefont {Zhang}}]{yu19}%
  \BibitemOpen
  \bibfield  {author} {\bibinfo {author} {\bibfnamefont {Y.}~\bibnamefont
  {Yu}}, \bibinfo {author} {\bibfnamefont {L.}~\bibnamefont {Ma}}, \bibinfo
  {author} {\bibfnamefont {P.}~\bibnamefont {Cai}}, \bibinfo {author}
  {\bibfnamefont {R.}~\bibnamefont {Zhong}}, \bibinfo {author} {\bibfnamefont
  {C.}~\bibnamefont {Ye}}, \bibinfo {author} {\bibfnamefont {J.}~\bibnamefont
  {Shen}}, \bibinfo {author} {\bibfnamefont {G.~D.}\ \bibnamefont {Gu}},
  \bibinfo {author} {\bibfnamefont {X.~H.}\ \bibnamefont {Chen}},\ and\
  \bibinfo {author} {\bibfnamefont {Y.}~\bibnamefont {Zhang}},\ }\bibfield
  {title} {\bibinfo {title} {High-temperature superconductivity in monolayer
  {Bi}$_2${Sr}$_2${CaCu}$_2${O}$_{8+\delta}$},\ }\href
  {https://doi.org/10.1038/s41586-019-1718-x} {\bibfield  {journal} {\bibinfo
  {journal} {Nature}\ }\textbf {\bibinfo {volume} {575}},\ \bibinfo {pages}
  {156} (\bibinfo {year} {2019})}\BibitemShut {NoStop}%
\bibitem [{\citenamefont {Ding}\ \emph {et~al.}(1996)\citenamefont {Ding},
  \citenamefont {Yokoya}, \citenamefont {Campuzano}, \citenamefont {Takahashi},
  \citenamefont {Randeria}, \citenamefont {Norman}, \citenamefont {Mochiku},
  \citenamefont {Kadowaki},\ and\ \citenamefont {Giapintzakis}}]{ding96}%
  \BibitemOpen
  \bibfield  {author} {\bibinfo {author} {\bibfnamefont {H.}~\bibnamefont
  {Ding}}, \bibinfo {author} {\bibfnamefont {T.}~\bibnamefont {Yokoya}},
  \bibinfo {author} {\bibfnamefont {J.~C.}\ \bibnamefont {Campuzano}}, \bibinfo
  {author} {\bibfnamefont {T.}~\bibnamefont {Takahashi}}, \bibinfo {author}
  {\bibfnamefont {M.}~\bibnamefont {Randeria}}, \bibinfo {author}
  {\bibfnamefont {M.~R.}\ \bibnamefont {Norman}}, \bibinfo {author}
  {\bibfnamefont {T.}~\bibnamefont {Mochiku}}, \bibinfo {author} {\bibfnamefont
  {K.}~\bibnamefont {Kadowaki}},\ and\ \bibinfo {author} {\bibfnamefont
  {J.}~\bibnamefont {Giapintzakis}},\ }\bibfield  {title} {\bibinfo {title}
  {Spectroscopic evidence for a pseudogap in the normal state of underdoped
  high-{Tc} superconductors},\ }\href {https://doi.org/10.1038/382051a0}
  {\bibfield  {journal} {\bibinfo  {journal} {Nature}\ }\textbf {\bibinfo
  {volume} {382}},\ \bibinfo {pages} {51} (\bibinfo {year} {1996})}\BibitemShut
  {NoStop}%
\bibitem [{\citenamefont {Shen}\ \emph {et~al.}(1993)\citenamefont {Shen},
  \citenamefont {Dessau}, \citenamefont {Wells}, \citenamefont {King},
  \citenamefont {Spicer}, \citenamefont {Arko}, \citenamefont {Marshall},
  \citenamefont {Lombardo}, \citenamefont {Kapitulnik}, \citenamefont
  {Dickinson}, \citenamefont {Doniach}, \citenamefont {DiCarlo}, \citenamefont
  {Loeser},\ and\ \citenamefont {Park}}]{shen93}%
  \BibitemOpen
  \bibfield  {author} {\bibinfo {author} {\bibfnamefont {Z.-X.}\ \bibnamefont
  {Shen}}, \bibinfo {author} {\bibfnamefont {D.~S.}\ \bibnamefont {Dessau}},
  \bibinfo {author} {\bibfnamefont {B.~O.}\ \bibnamefont {Wells}}, \bibinfo
  {author} {\bibfnamefont {D.~M.}\ \bibnamefont {King}}, \bibinfo {author}
  {\bibfnamefont {W.~E.}\ \bibnamefont {Spicer}}, \bibinfo {author}
  {\bibfnamefont {A.~J.}\ \bibnamefont {Arko}}, \bibinfo {author}
  {\bibfnamefont {D.}~\bibnamefont {Marshall}}, \bibinfo {author}
  {\bibfnamefont {L.~W.}\ \bibnamefont {Lombardo}}, \bibinfo {author}
  {\bibfnamefont {A.}~\bibnamefont {Kapitulnik}}, \bibinfo {author}
  {\bibfnamefont {P.}~\bibnamefont {Dickinson}}, \bibinfo {author}
  {\bibfnamefont {S.}~\bibnamefont {Doniach}}, \bibinfo {author} {\bibfnamefont
  {J.}~\bibnamefont {DiCarlo}}, \bibinfo {author} {\bibfnamefont
  {T.}~\bibnamefont {Loeser}},\ and\ \bibinfo {author} {\bibfnamefont {C.~H.}\
  \bibnamefont {Park}},\ }\bibfield  {title} {\bibinfo {title} {Anomalously
  large gap anisotropy in the a-b plane of
  {Bi}$_{2}${Sr}$_{2}${CaCu}$_{2}${O}$_{8+\delta}$},\ }\href
  {https://doi.org/10.1103/PhysRevLett.70.1553} {\bibfield  {journal} {\bibinfo
   {journal} {Phys. Rev. Lett.}\ }\textbf {\bibinfo {volume} {70}},\ \bibinfo
  {pages} {1553} (\bibinfo {year} {1993})}\BibitemShut {NoStop}%
\bibitem [{\citenamefont {Norman}\ \emph {et~al.}(1998)\citenamefont {Norman},
  \citenamefont {Ding}, \citenamefont {Randeria}, \citenamefont {Campuzano},
  \citenamefont {Yokoya}, \citenamefont {Takeuchi}, \citenamefont {Takahashi},
  \citenamefont {Mochiku}, \citenamefont {Kadowaki}, \citenamefont
  {Guptasarma},\ and\ \citenamefont {Hinks}}]{norman98}%
  \BibitemOpen
  \bibfield  {author} {\bibinfo {author} {\bibfnamefont {M.~R.}\ \bibnamefont
  {Norman}}, \bibinfo {author} {\bibfnamefont {H.}~\bibnamefont {Ding}},
  \bibinfo {author} {\bibfnamefont {M.}~\bibnamefont {Randeria}}, \bibinfo
  {author} {\bibfnamefont {J.~C.}\ \bibnamefont {Campuzano}}, \bibinfo {author}
  {\bibfnamefont {T.}~\bibnamefont {Yokoya}}, \bibinfo {author} {\bibfnamefont
  {T.}~\bibnamefont {Takeuchi}}, \bibinfo {author} {\bibfnamefont
  {T.}~\bibnamefont {Takahashi}}, \bibinfo {author} {\bibfnamefont
  {T.}~\bibnamefont {Mochiku}}, \bibinfo {author} {\bibfnamefont
  {K.}~\bibnamefont {Kadowaki}}, \bibinfo {author} {\bibfnamefont
  {P.}~\bibnamefont {Guptasarma}},\ and\ \bibinfo {author} {\bibfnamefont
  {D.~G.}\ \bibnamefont {Hinks}},\ }\bibfield  {title} {\bibinfo {title}
  {Destruction of the {Fermi} surface in underdoped high-${T}_c$
  superconductors},\ }\href {http://dx.doi.org/10.1038/32366} {\bibfield
  {journal} {\bibinfo  {journal} {Nature}\ }\textbf {\bibinfo {volume} {392}},\
  \bibinfo {pages} {157} (\bibinfo {year} {1998})}\BibitemShut {NoStop}%
\bibitem [{\citenamefont {Shen}\ \emph {et~al.}(2005)\citenamefont {Shen},
  \citenamefont {Ronning}, \citenamefont {Lu}, \citenamefont {Baumberger},
  \citenamefont {Ingle}, \citenamefont {Lee}, \citenamefont {Meevasana},
  \citenamefont {Kohsaka}, \citenamefont {Azuma}, \citenamefont {Takano},
  \citenamefont {Takagi},\ and\ \citenamefont {Shen}}]{shen05}%
  \BibitemOpen
  \bibfield  {author} {\bibinfo {author} {\bibfnamefont {K.~M.}\ \bibnamefont
  {Shen}}, \bibinfo {author} {\bibfnamefont {F.}~\bibnamefont {Ronning}},
  \bibinfo {author} {\bibfnamefont {D.~H.}\ \bibnamefont {Lu}}, \bibinfo
  {author} {\bibfnamefont {F.}~\bibnamefont {Baumberger}}, \bibinfo {author}
  {\bibfnamefont {N.~J.~C.}\ \bibnamefont {Ingle}}, \bibinfo {author}
  {\bibfnamefont {W.~S.}\ \bibnamefont {Lee}}, \bibinfo {author} {\bibfnamefont
  {W.}~\bibnamefont {Meevasana}}, \bibinfo {author} {\bibfnamefont
  {Y.}~\bibnamefont {Kohsaka}}, \bibinfo {author} {\bibfnamefont
  {M.}~\bibnamefont {Azuma}}, \bibinfo {author} {\bibfnamefont
  {M.}~\bibnamefont {Takano}}, \bibinfo {author} {\bibfnamefont
  {H.}~\bibnamefont {Takagi}},\ and\ \bibinfo {author} {\bibfnamefont {Z.-X.}\
  \bibnamefont {Shen}},\ }\bibfield  {title} {\bibinfo {title} {Nodal
  quasiparticles and antinodal charge ordering in
  {Ca}$_{2-x}${Na}$_x${CuO}$_2${Cl}$_2$},\ }\href
  {https://doi.org/10.1126/science.1103627} {\bibfield  {journal} {\bibinfo
  {journal} {Science}\ }\textbf {\bibinfo {volume} {307}},\ \bibinfo {pages}
  {901} (\bibinfo {year} {2005})}\BibitemShut {NoStop}%
\bibitem [{\citenamefont {Kondo}\ \emph {et~al.}(2009)\citenamefont {Kondo},
  \citenamefont {Khasanov}, \citenamefont {Takeuchi}, \citenamefont
  {Schmalian},\ and\ \citenamefont {Kaminski}}]{kondo09}%
  \BibitemOpen
  \bibfield  {author} {\bibinfo {author} {\bibfnamefont {T.}~\bibnamefont
  {Kondo}}, \bibinfo {author} {\bibfnamefont {R.}~\bibnamefont {Khasanov}},
  \bibinfo {author} {\bibfnamefont {T.}~\bibnamefont {Takeuchi}}, \bibinfo
  {author} {\bibfnamefont {J.}~\bibnamefont {Schmalian}},\ and\ \bibinfo
  {author} {\bibfnamefont {A.}~\bibnamefont {Kaminski}},\ }\bibfield  {title}
  {\bibinfo {title} {Competition between the pseudogap and superconductivity in
  the high-${T}_c$ copper oxides},\ }\href
  {https://doi.org/10.1038/nature07644} {\bibfield  {journal} {\bibinfo
  {journal} {Nature}\ }\textbf {\bibinfo {volume} {457}},\ \bibinfo {pages}
  {296} (\bibinfo {year} {2009})}\BibitemShut {NoStop}%
\bibitem [{\citenamefont {Kondo}\ \emph {et~al.}(2013)\citenamefont {Kondo},
  \citenamefont {Palczewski}, \citenamefont {Hamaya}, \citenamefont {Takeuchi},
  \citenamefont {Wen}, \citenamefont {Xu}, \citenamefont {Gu},\ and\
  \citenamefont {Kaminski}}]{kondo13}%
  \BibitemOpen
  \bibfield  {author} {\bibinfo {author} {\bibfnamefont {T.}~\bibnamefont
  {Kondo}}, \bibinfo {author} {\bibfnamefont {A.~D.}\ \bibnamefont
  {Palczewski}}, \bibinfo {author} {\bibfnamefont {Y.}~\bibnamefont {Hamaya}},
  \bibinfo {author} {\bibfnamefont {T.}~\bibnamefont {Takeuchi}}, \bibinfo
  {author} {\bibfnamefont {J.~S.}\ \bibnamefont {Wen}}, \bibinfo {author}
  {\bibfnamefont {Z.~J.}\ \bibnamefont {Xu}}, \bibinfo {author} {\bibfnamefont
  {G.}~\bibnamefont {Gu}},\ and\ \bibinfo {author} {\bibfnamefont
  {A.}~\bibnamefont {Kaminski}},\ }\bibfield  {title} {\bibinfo {title}
  {Formation of gapless {Fermi} arcs and fingerprints of order in the pseudogap
  state of cuprate superconductors},\ }\href
  {https://doi.org/10.1103/PhysRevLett.111.157003} {\bibfield  {journal}
  {\bibinfo  {journal} {Phys. Rev. Lett.}\ }\textbf {\bibinfo {volume} {111}},\
  \bibinfo {pages} {157003} (\bibinfo {year} {2013})}\BibitemShut {NoStop}%
\bibitem [{\citenamefont {Tanaka}\ \emph {et~al.}(2006)\citenamefont {Tanaka},
  \citenamefont {Lee}, \citenamefont {Lu}, \citenamefont {Fujimori},
  \citenamefont {Fujii}, \citenamefont {null Risdiana}, \citenamefont
  {Terasaki}, \citenamefont {Scalapino}, \citenamefont {Devereaux},
  \citenamefont {Hussain},\ and\ \citenamefont {Shen}}]{tanaka06}%
  \BibitemOpen
  \bibfield  {author} {\bibinfo {author} {\bibfnamefont {K.}~\bibnamefont
  {Tanaka}}, \bibinfo {author} {\bibfnamefont {W.~S.}\ \bibnamefont {Lee}},
  \bibinfo {author} {\bibfnamefont {D.~H.}\ \bibnamefont {Lu}}, \bibinfo
  {author} {\bibfnamefont {A.}~\bibnamefont {Fujimori}}, \bibinfo {author}
  {\bibfnamefont {T.}~\bibnamefont {Fujii}}, \bibinfo {author} {\bibnamefont
  {null Risdiana}}, \bibinfo {author} {\bibfnamefont {I.}~\bibnamefont
  {Terasaki}}, \bibinfo {author} {\bibfnamefont {D.~J.}\ \bibnamefont
  {Scalapino}}, \bibinfo {author} {\bibfnamefont {T.~P.}\ \bibnamefont
  {Devereaux}}, \bibinfo {author} {\bibfnamefont {Z.}~\bibnamefont {Hussain}},\
  and\ \bibinfo {author} {\bibfnamefont {Z.-X.}\ \bibnamefont {Shen}},\
  }\bibfield  {title} {\bibinfo {title} {Distinct {Fermi}-momentum-dependent
  energy gaps in deeply underdoped {Bi}2212},\ }\href
  {https://doi.org/10.1126/science.1133411} {\bibfield  {journal} {\bibinfo
  {journal} {Science}\ }\textbf {\bibinfo {volume} {314}},\ \bibinfo {pages}
  {1910} (\bibinfo {year} {2006})}\BibitemShut {NoStop}%
\bibitem [{\citenamefont {Kondo}\ \emph {et~al.}(2007)\citenamefont {Kondo},
  \citenamefont {Takeuchi}, \citenamefont {Kaminski}, \citenamefont {Tsuda},\
  and\ \citenamefont {Shin}}]{kondo07}%
  \BibitemOpen
  \bibfield  {author} {\bibinfo {author} {\bibfnamefont {T.}~\bibnamefont
  {Kondo}}, \bibinfo {author} {\bibfnamefont {T.}~\bibnamefont {Takeuchi}},
  \bibinfo {author} {\bibfnamefont {A.}~\bibnamefont {Kaminski}}, \bibinfo
  {author} {\bibfnamefont {S.}~\bibnamefont {Tsuda}},\ and\ \bibinfo {author}
  {\bibfnamefont {S.}~\bibnamefont {Shin}},\ }\bibfield  {title} {\bibinfo
  {title} {Evidence for two energy scales in the superconducting state of
  optimally doped ({Bi},{Pb})$_{2}$({Sr},{La})$_{2}${CuO}$_{6+\delta}$},\
  }\href {https://doi.org/10.1103/PhysRevLett.98.267004} {\bibfield  {journal}
  {\bibinfo  {journal} {Phys. Rev. Lett.}\ }\textbf {\bibinfo {volume} {98}},\
  \bibinfo {pages} {267004} (\bibinfo {year} {2007})}\BibitemShut {NoStop}%
\bibitem [{\citenamefont {Pan}\ \emph {et~al.}(2001)\citenamefont {Pan},
  \citenamefont {O'Neal}, \citenamefont {Badzey}, \citenamefont {Chamon},
  \citenamefont {Ding}, \citenamefont {Engelbrecht}, \citenamefont {Wang},
  \citenamefont {Eisaki}, \citenamefont {Uchida}, \citenamefont {Guptak},
  \citenamefont {Ngk}, \citenamefont {Hudson}, \citenamefont {Lang},\ and\
  \citenamefont {Davis}}]{pan01}%
  \BibitemOpen
  \bibfield  {author} {\bibinfo {author} {\bibfnamefont {S.~H.}\ \bibnamefont
  {Pan}}, \bibinfo {author} {\bibfnamefont {J.~P.}\ \bibnamefont {O'Neal}},
  \bibinfo {author} {\bibfnamefont {R.~L.}\ \bibnamefont {Badzey}}, \bibinfo
  {author} {\bibfnamefont {C.}~\bibnamefont {Chamon}}, \bibinfo {author}
  {\bibfnamefont {H.}~\bibnamefont {Ding}}, \bibinfo {author} {\bibfnamefont
  {J.~R.}\ \bibnamefont {Engelbrecht}}, \bibinfo {author} {\bibfnamefont
  {Z.}~\bibnamefont {Wang}}, \bibinfo {author} {\bibfnamefont {H.}~\bibnamefont
  {Eisaki}}, \bibinfo {author} {\bibfnamefont {S.}~\bibnamefont {Uchida}},
  \bibinfo {author} {\bibfnamefont {A.~K.}\ \bibnamefont {Guptak}}, \bibinfo
  {author} {\bibfnamefont {K.-W.}\ \bibnamefont {Ngk}}, \bibinfo {author}
  {\bibfnamefont {E.~W.}\ \bibnamefont {Hudson}}, \bibinfo {author}
  {\bibfnamefont {K.~M.}\ \bibnamefont {Lang}},\ and\ \bibinfo {author}
  {\bibfnamefont {J.~C.}\ \bibnamefont {Davis}},\ }\bibfield  {title} {\bibinfo
  {title} {Microscopic electronic inhomogeneity in the high-${T}_c$
  superconductor {Bi}$_2${Sr}$_2${CaCu}$_2${O}$_{8+x}$},\ }\href
  {https://doi.org/10.1038/35095012} {\bibfield  {journal} {\bibinfo  {journal}
  {Nature}\ }\textbf {\bibinfo {volume} {413}},\ \bibinfo {pages} {282}
  (\bibinfo {year} {2001})}\BibitemShut {NoStop}%
\bibitem [{\citenamefont {Kohsaka}\ \emph {et~al.}(2007)\citenamefont
  {Kohsaka}, \citenamefont {Taylor}, \citenamefont {Fujita}, \citenamefont
  {Schmidt}, \citenamefont {Lupien}, \citenamefont {Hanaguri}, \citenamefont
  {Azuma}, \citenamefont {Takano}, \citenamefont {Eisaki}, \citenamefont
  {Takagi}, \citenamefont {S.},\ and\ \citenamefont {Davis}}]{kohsaka07}%
  \BibitemOpen
  \bibfield  {author} {\bibinfo {author} {\bibfnamefont {Y.}~\bibnamefont
  {Kohsaka}}, \bibinfo {author} {\bibfnamefont {C.}~\bibnamefont {Taylor}},
  \bibinfo {author} {\bibfnamefont {K.}~\bibnamefont {Fujita}}, \bibinfo
  {author} {\bibfnamefont {A.}~\bibnamefont {Schmidt}}, \bibinfo {author}
  {\bibfnamefont {C.}~\bibnamefont {Lupien}}, \bibinfo {author} {\bibfnamefont
  {T.}~\bibnamefont {Hanaguri}}, \bibinfo {author} {\bibfnamefont
  {M.}~\bibnamefont {Azuma}}, \bibinfo {author} {\bibfnamefont
  {M.}~\bibnamefont {Takano}}, \bibinfo {author} {\bibfnamefont
  {H.}~\bibnamefont {Eisaki}}, \bibinfo {author} {\bibfnamefont
  {H.}~\bibnamefont {Takagi}}, \bibinfo {author} {\bibfnamefont
  {U.}~\bibnamefont {S.}},\ and\ \bibinfo {author} {\bibfnamefont {J.~C.}\
  \bibnamefont {Davis}},\ }\bibfield  {title} {\bibinfo {title} {An intrinsic
  bond-centered electronic glass with unidirectional domains in underdoped
  cuprates},\ }\href {https://doi.org/10.1126/science.1138584} {\bibfield
  {journal} {\bibinfo  {journal} {Science}\ }\textbf {\bibinfo {volume}
  {315}},\ \bibinfo {pages} {1380} (\bibinfo {year} {2007})}\BibitemShut
  {NoStop}%
\bibitem [{\citenamefont {Markiewicz}(2004)}]{markiewicz04}%
  \BibitemOpen
  \bibfield  {author} {\bibinfo {author} {\bibfnamefont {R.~S.}\ \bibnamefont
  {Markiewicz}},\ }\bibfield  {title} {\bibinfo {title} {Bridging $k$ and $q$
  space in the cuprates: Comparing angle-resolved photoemission and {STM}
  results},\ }\href {https://doi.org/10.1103/PhysRevB.69.214517} {\bibfield
  {journal} {\bibinfo  {journal} {Phys. Rev. B}\ }\textbf {\bibinfo {volume}
  {69}},\ \bibinfo {pages} {214517} (\bibinfo {year} {2004})}\BibitemShut
  {NoStop}%
\bibitem [{\citenamefont {Cheng}\ and\ \citenamefont {Su}(2005)}]{cheng05}%
  \BibitemOpen
  \bibfield  {author} {\bibinfo {author} {\bibfnamefont {M.}~\bibnamefont
  {Cheng}}\ and\ \bibinfo {author} {\bibfnamefont {W.~P.}\ \bibnamefont {Su}},\
  }\bibfield  {title} {\bibinfo {title} {Local density of states and
  angle-resolved photoemission spectral function of an inhomogeneous $d$-wave
  superconductor},\ }\href {https://doi.org/10.1103/PhysRevB.72.094512}
  {\bibfield  {journal} {\bibinfo  {journal} {Phys. Rev. B}\ }\textbf {\bibinfo
  {volume} {72}},\ \bibinfo {pages} {094512} (\bibinfo {year}
  {2005})}\BibitemShut {NoStop}%
\bibitem [{\citenamefont {McElroy}\ \emph {et~al.}(2006)\citenamefont
  {McElroy}, \citenamefont {Gweon}, \citenamefont {Zhou}, \citenamefont {Graf},
  \citenamefont {Uchida}, \citenamefont {Eisaki}, \citenamefont {Takagi},
  \citenamefont {Sasagawa}, \citenamefont {Lee},\ and\ \citenamefont
  {Lanzara}}]{mcelroy06}%
  \BibitemOpen
  \bibfield  {author} {\bibinfo {author} {\bibfnamefont {K.}~\bibnamefont
  {McElroy}}, \bibinfo {author} {\bibfnamefont {G.-H.}\ \bibnamefont {Gweon}},
  \bibinfo {author} {\bibfnamefont {S.~Y.}\ \bibnamefont {Zhou}}, \bibinfo
  {author} {\bibfnamefont {J.}~\bibnamefont {Graf}}, \bibinfo {author}
  {\bibfnamefont {S.}~\bibnamefont {Uchida}}, \bibinfo {author} {\bibfnamefont
  {H.}~\bibnamefont {Eisaki}}, \bibinfo {author} {\bibfnamefont
  {H.}~\bibnamefont {Takagi}}, \bibinfo {author} {\bibfnamefont
  {T.}~\bibnamefont {Sasagawa}}, \bibinfo {author} {\bibfnamefont {D.-H.}\
  \bibnamefont {Lee}},\ and\ \bibinfo {author} {\bibfnamefont {A.}~\bibnamefont
  {Lanzara}},\ }\bibfield  {title} {\bibinfo {title} {Elastic scattering
  susceptibility of the high temperature superconductor
  {Bi}$_{2}${Sr}$_{2}${CaCu}$_{2}${O}$_{8+\delta}$: A comparison between real
  and momentum space photoemission spectroscopies},\ }\href
  {https://doi.org/10.1103/PhysRevLett.96.067005} {\bibfield  {journal}
  {\bibinfo  {journal} {Phys. Rev. Lett.}\ }\textbf {\bibinfo {volume} {96}},\
  \bibinfo {pages} {067005} (\bibinfo {year} {2006})}\BibitemShut {NoStop}%
\bibitem [{\citenamefont {Chatterjee}\ \emph {et~al.}(2006)\citenamefont
  {Chatterjee}, \citenamefont {Shi}, \citenamefont {Kaminski}, \citenamefont
  {Kanigel}, \citenamefont {Fretwell}, \citenamefont {Terashima}, \citenamefont
  {Takahashi}, \citenamefont {Rosenkranz}, \citenamefont {Li}, \citenamefont
  {Raffy}, \citenamefont {Santander-Syro}, \citenamefont {Kadowaki},
  \citenamefont {Norman}, \citenamefont {Randeria},\ and\ \citenamefont
  {Campuzano}}]{chatterjee06}%
  \BibitemOpen
  \bibfield  {author} {\bibinfo {author} {\bibfnamefont {U.}~\bibnamefont
  {Chatterjee}}, \bibinfo {author} {\bibfnamefont {M.}~\bibnamefont {Shi}},
  \bibinfo {author} {\bibfnamefont {A.}~\bibnamefont {Kaminski}}, \bibinfo
  {author} {\bibfnamefont {A.}~\bibnamefont {Kanigel}}, \bibinfo {author}
  {\bibfnamefont {H.~M.}\ \bibnamefont {Fretwell}}, \bibinfo {author}
  {\bibfnamefont {K.}~\bibnamefont {Terashima}}, \bibinfo {author}
  {\bibfnamefont {T.}~\bibnamefont {Takahashi}}, \bibinfo {author}
  {\bibfnamefont {S.}~\bibnamefont {Rosenkranz}}, \bibinfo {author}
  {\bibfnamefont {Z.~Z.}\ \bibnamefont {Li}}, \bibinfo {author} {\bibfnamefont
  {H.}~\bibnamefont {Raffy}}, \bibinfo {author} {\bibfnamefont
  {A.}~\bibnamefont {Santander-Syro}}, \bibinfo {author} {\bibfnamefont
  {K.}~\bibnamefont {Kadowaki}}, \bibinfo {author} {\bibfnamefont {M.~R.}\
  \bibnamefont {Norman}}, \bibinfo {author} {\bibfnamefont {M.}~\bibnamefont
  {Randeria}},\ and\ \bibinfo {author} {\bibfnamefont {J.~C.}\ \bibnamefont
  {Campuzano}},\ }\bibfield  {title} {\bibinfo {title} {Nondispersive {Fermi}
  arcs and the absence of charge ordering in the pseudogap phase of
  {Bi}$_{2}${Sr}$_{2}${CaCu}$_{2}${O}$_{8+\delta}$},\ }\href
  {https://doi.org/10.1103/PhysRevLett.96.107006} {\bibfield  {journal}
  {\bibinfo  {journal} {Phys. Rev. Lett.}\ }\textbf {\bibinfo {volume} {96}},\
  \bibinfo {pages} {107006} (\bibinfo {year} {2006})}\BibitemShut {NoStop}%
\bibitem [{\citenamefont {Fujita}\ \emph {et~al.}(2020)\citenamefont {Fujita},
  \citenamefont {Drozdov}, \citenamefont {Du}, \citenamefont {Li},
  \citenamefont {Joo}, \citenamefont {Lee}, \citenamefont {Gu}, \citenamefont
  {Johnson},\ and\ \citenamefont {Valla}}]{fujita20}%
  \BibitemOpen
  \bibfield  {author} {\bibinfo {author} {\bibfnamefont {K.}~\bibnamefont
  {Fujita}}, \bibinfo {author} {\bibfnamefont {I.}~\bibnamefont {Drozdov}},
  \bibinfo {author} {\bibfnamefont {Z.}~\bibnamefont {Du}}, \bibinfo {author}
  {\bibfnamefont {H.}~\bibnamefont {Li}}, \bibinfo {author} {\bibfnamefont
  {S.-H.}\ \bibnamefont {Joo}}, \bibinfo {author} {\bibfnamefont
  {J.}~\bibnamefont {Lee}}, \bibinfo {author} {\bibfnamefont {G.}~\bibnamefont
  {Gu}}, \bibinfo {author} {\bibfnamefont {P.~D.}\ \bibnamefont {Johnson}},\
  and\ \bibinfo {author} {\bibfnamefont {T.}~\bibnamefont {Valla}},\ }\bibfield
   {title} {\bibinfo {title} {Combined spectroscopic imaging {STM} and {ARPES}
  study of different gaps measured in the cuprate phase diagram},\ }\href
  {https://doi.org/10.1103/PhysRevB.101.045136} {\bibfield  {journal} {\bibinfo
   {journal} {Phys. Rev. B}\ }\textbf {\bibinfo {volume} {101}},\ \bibinfo
  {pages} {045136} (\bibinfo {year} {2020})}\BibitemShut {NoStop}%
\bibitem [{\citenamefont {Vishik}\ \emph {et~al.}(2009)\citenamefont {Vishik},
  \citenamefont {Nowadnick}, \citenamefont {Lee}, \citenamefont {Shen},
  \citenamefont {Moritz}, \citenamefont {Devereaux}, \citenamefont {Tanaka},
  \citenamefont {Sasagawa},\ and\ \citenamefont {Fujii}}]{vishik09}%
  \BibitemOpen
  \bibfield  {author} {\bibinfo {author} {\bibfnamefont {I.~M.}\ \bibnamefont
  {Vishik}}, \bibinfo {author} {\bibfnamefont {E.~A.}\ \bibnamefont
  {Nowadnick}}, \bibinfo {author} {\bibfnamefont {W.~S.}\ \bibnamefont {Lee}},
  \bibinfo {author} {\bibfnamefont {Z.~X.}\ \bibnamefont {Shen}}, \bibinfo
  {author} {\bibfnamefont {B.}~\bibnamefont {Moritz}}, \bibinfo {author}
  {\bibfnamefont {T.~P.}\ \bibnamefont {Devereaux}}, \bibinfo {author}
  {\bibfnamefont {K.}~\bibnamefont {Tanaka}}, \bibinfo {author} {\bibfnamefont
  {T.}~\bibnamefont {Sasagawa}},\ and\ \bibinfo {author} {\bibfnamefont
  {T.}~\bibnamefont {Fujii}},\ }\bibfield  {title} {\bibinfo {title} {A
  momentum-dependent perspective on quasiparticle interference in
  {Bi}$_2${Sr}$_2${CaCu}$_2${O}$_{8+\delta}$},\ }\href
  {https://doi.org/10.1038/nphys1375} {\bibfield  {journal} {\bibinfo
  {journal} {Nature Physics}\ }\textbf {\bibinfo {volume} {5}},\ \bibinfo
  {pages} {718} (\bibinfo {year} {2009})}\BibitemShut {NoStop}%
\bibitem [{\citenamefont {Sakai}\ \emph
  {et~al.}(2016{\natexlab{a}})\citenamefont {Sakai}, \citenamefont {Civelli},\
  and\ \citenamefont {Imada}}]{sakai16}%
  \BibitemOpen
  \bibfield  {author} {\bibinfo {author} {\bibfnamefont {S.}~\bibnamefont
  {Sakai}}, \bibinfo {author} {\bibfnamefont {M.}~\bibnamefont {Civelli}},\
  and\ \bibinfo {author} {\bibfnamefont {M.}~\bibnamefont {Imada}},\ }\bibfield
   {title} {\bibinfo {title} {Hidden fermionic excitation boosting
  high-temperature superconductivity in cuprates},\ }\href
  {https://doi.org/10.1103/PhysRevLett.116.057003} {\bibfield  {journal}
  {\bibinfo  {journal} {Phys. Rev. Lett.}\ }\textbf {\bibinfo {volume} {116}},\
  \bibinfo {pages} {057003} (\bibinfo {year} {2016}{\natexlab{a}})}\BibitemShut
  {NoStop}%
\bibitem [{\citenamefont {Sakai}\ \emph
  {et~al.}(2016{\natexlab{b}})\citenamefont {Sakai}, \citenamefont {Civelli},\
  and\ \citenamefont {Imada}}]{sakai16PRB}%
  \BibitemOpen
  \bibfield  {author} {\bibinfo {author} {\bibfnamefont {S.}~\bibnamefont
  {Sakai}}, \bibinfo {author} {\bibfnamefont {M.}~\bibnamefont {Civelli}},\
  and\ \bibinfo {author} {\bibfnamefont {M.}~\bibnamefont {Imada}},\ }\bibfield
   {title} {\bibinfo {title} {Hidden-fermion representation of self-energy in
  pseudogap and superconducting states of the two-dimensional {Hubbard}
  model},\ }\href {https://doi.org/10.1103/PhysRevB.94.115130} {\bibfield
  {journal} {\bibinfo  {journal} {Phys. Rev. B}\ }\textbf {\bibinfo {volume}
  {94}},\ \bibinfo {pages} {115130} (\bibinfo {year}
  {2016}{\natexlab{b}})}\BibitemShut {NoStop}%
\bibitem [{\citenamefont {Tomonaga}(1950)}]{tomonaga50}%
  \BibitemOpen
  \bibfield  {author} {\bibinfo {author} {\bibfnamefont {S.-i.}\ \bibnamefont
  {Tomonaga}},\ }\bibfield  {title} {\bibinfo {title} {Remarks on {Bloch's}
  method of sound waves applied to many-fermion problems},\ }\href
  {https://doi.org/10.1143/ptp/5.4.544} {\bibfield  {journal} {\bibinfo
  {journal} {Progress of Theoretical Physics}\ }\textbf {\bibinfo {volume}
  {5}},\ \bibinfo {pages} {544} (\bibinfo {year} {1950})}\BibitemShut {NoStop}%
\bibitem [{\citenamefont {Luttinger}(1963)}]{luttinger63}%
  \BibitemOpen
  \bibfield  {author} {\bibinfo {author} {\bibfnamefont {J.~M.}\ \bibnamefont
  {Luttinger}},\ }\bibfield  {title} {\bibinfo {title} {An exactly soluble
  model of a many‐fermion system},\ }\href {https://doi.org/10.1063/1.1704046}
  {\bibfield  {journal} {\bibinfo  {journal} {Journal of Mathematical Physics}\
  }\textbf {\bibinfo {volume} {4}},\ \bibinfo {pages} {1154} (\bibinfo {year}
  {1963})}\BibitemShut {NoStop}%
\bibitem [{\citenamefont {Heeger}\ \emph {et~al.}(1988)\citenamefont {Heeger},
  \citenamefont {Kivelson}, \citenamefont {Schrieffer},\ and\ \citenamefont
  {Su}}]{heeger88}%
  \BibitemOpen
  \bibfield  {author} {\bibinfo {author} {\bibfnamefont {A.~J.}\ \bibnamefont
  {Heeger}}, \bibinfo {author} {\bibfnamefont {S.}~\bibnamefont {Kivelson}},
  \bibinfo {author} {\bibfnamefont {J.~R.}\ \bibnamefont {Schrieffer}},\ and\
  \bibinfo {author} {\bibfnamefont {W.~P.}\ \bibnamefont {Su}},\ }\bibfield
  {title} {\bibinfo {title} {Solitons in conducting polymers},\ }\href
  {https://doi.org/10.1103/RevModPhys.60.781} {\bibfield  {journal} {\bibinfo
  {journal} {Rev. Mod. Phys.}\ }\textbf {\bibinfo {volume} {60}},\ \bibinfo
  {pages} {781} (\bibinfo {year} {1988})}\BibitemShut {NoStop}%
\bibitem [{\citenamefont {Laughlin}(1983)}]{laughlin83}%
  \BibitemOpen
  \bibfield  {author} {\bibinfo {author} {\bibfnamefont {R.~B.}\ \bibnamefont
  {Laughlin}},\ }\bibfield  {title} {\bibinfo {title} {Anomalous quantum {Hall}
  effect: An incompressible quantum fluid with fractionally charged
  excitations},\ }\href {https://doi.org/10.1103/PhysRevLett.50.1395}
  {\bibfield  {journal} {\bibinfo  {journal} {Phys. Rev. Lett.}\ }\textbf
  {\bibinfo {volume} {50}},\ \bibinfo {pages} {1395} (\bibinfo {year}
  {1983})}\BibitemShut {NoStop}%
\bibitem [{\citenamefont {Sachdev}(2003)}]{sachdev03}%
  \BibitemOpen
  \bibfield  {author} {\bibinfo {author} {\bibfnamefont {S.}~\bibnamefont
  {Sachdev}},\ }\bibfield  {title} {\bibinfo {title} {Understanding correlated
  electron systems by a classification of {Mott} insulators},\ }\href
  {https://doi.org/https://doi.org/10.1016/S0003-4916(02)00024-6} {\bibfield
  {journal} {\bibinfo  {journal} {Annals of Physics}\ }\textbf {\bibinfo
  {volume} {303}},\ \bibinfo {pages} {226} (\bibinfo {year}
  {2003})}\BibitemShut {NoStop}%
\bibitem [{\citenamefont {Kotliar}\ and\ \citenamefont
  {Ruckenstein}(1986)}]{kotliar86}%
  \BibitemOpen
  \bibfield  {author} {\bibinfo {author} {\bibfnamefont {G.}~\bibnamefont
  {Kotliar}}\ and\ \bibinfo {author} {\bibfnamefont {A.~E.}\ \bibnamefont
  {Ruckenstein}},\ }\bibfield  {title} {\bibinfo {title} {New functional
  integral approach to strongly correlated {Fermi} systems: The {Gutzwiller}
  approximation as a saddle point},\ }\href
  {https://doi.org/10.1103/PhysRevLett.57.1362} {\bibfield  {journal} {\bibinfo
   {journal} {Phys. Rev. Lett.}\ }\textbf {\bibinfo {volume} {57}},\ \bibinfo
  {pages} {1362} (\bibinfo {year} {1986})}\BibitemShut {NoStop}%
\bibitem [{\citenamefont {Kotliar}\ \emph {et~al.}(2001)\citenamefont
  {Kotliar}, \citenamefont {Savrasov}, \citenamefont {P\'alsson},\ and\
  \citenamefont {Biroli}}]{kotliar01}%
  \BibitemOpen
  \bibfield  {author} {\bibinfo {author} {\bibfnamefont {G.}~\bibnamefont
  {Kotliar}}, \bibinfo {author} {\bibfnamefont {S.~Y.}\ \bibnamefont
  {Savrasov}}, \bibinfo {author} {\bibfnamefont {G.}~\bibnamefont
  {P\'alsson}},\ and\ \bibinfo {author} {\bibfnamefont {G.}~\bibnamefont
  {Biroli}},\ }\bibfield  {title} {\bibinfo {title} {Cellular dynamical mean
  field approach to strongly correlated systems},\ }\href
  {https://doi.org/10.1103/PhysRevLett.87.186401} {\bibfield  {journal}
  {\bibinfo  {journal} {Phys. Rev. Lett.}\ }\textbf {\bibinfo {volume} {87}},\
  \bibinfo {pages} {186401} (\bibinfo {year} {2001})}\BibitemShut {NoStop}%
\bibitem [{\citenamefont {Maier}\ \emph {et~al.}(2005)\citenamefont {Maier},
  \citenamefont {Jarrell}, \citenamefont {Pruschke},\ and\ \citenamefont
  {Hettler}}]{maier05RMP}%
  \BibitemOpen
  \bibfield  {author} {\bibinfo {author} {\bibfnamefont {T.}~\bibnamefont
  {Maier}}, \bibinfo {author} {\bibfnamefont {M.}~\bibnamefont {Jarrell}},
  \bibinfo {author} {\bibfnamefont {T.}~\bibnamefont {Pruschke}},\ and\
  \bibinfo {author} {\bibfnamefont {M.~H.}\ \bibnamefont {Hettler}},\
  }\bibfield  {title} {\bibinfo {title} {Quantum cluster theories},\ }\href
  {https://doi.org/10.1103/RevModPhys.77.1027} {\bibfield  {journal} {\bibinfo
  {journal} {Rev. Mod. Phys.}\ }\textbf {\bibinfo {volume} {77}},\ \bibinfo
  {pages} {1027} (\bibinfo {year} {2005})}\BibitemShut {NoStop}%
\bibitem [{\citenamefont {Metzner}\ and\ \citenamefont
  {Vollhardt}(1989)}]{metzner89}%
  \BibitemOpen
  \bibfield  {author} {\bibinfo {author} {\bibfnamefont {W.}~\bibnamefont
  {Metzner}}\ and\ \bibinfo {author} {\bibfnamefont {D.}~\bibnamefont
  {Vollhardt}},\ }\bibfield  {title} {\bibinfo {title} {Correlated lattice
  fermions in $d=\ensuremath{\infty}$ dimensions},\ }\href
  {https://doi.org/10.1103/PhysRevLett.62.324} {\bibfield  {journal} {\bibinfo
  {journal} {Phys. Rev. Lett.}\ }\textbf {\bibinfo {volume} {62}},\ \bibinfo
  {pages} {324} (\bibinfo {year} {1989})}\BibitemShut {NoStop}%
\bibitem [{\citenamefont {Georges}\ \emph {et~al.}(1996)\citenamefont
  {Georges}, \citenamefont {Kotliar}, \citenamefont {Krauth},\ and\
  \citenamefont {Rozenberg}}]{georges96}%
  \BibitemOpen
  \bibfield  {author} {\bibinfo {author} {\bibfnamefont {A.}~\bibnamefont
  {Georges}}, \bibinfo {author} {\bibfnamefont {G.}~\bibnamefont {Kotliar}},
  \bibinfo {author} {\bibfnamefont {W.}~\bibnamefont {Krauth}},\ and\ \bibinfo
  {author} {\bibfnamefont {M.~J.}\ \bibnamefont {Rozenberg}},\ }\bibfield
  {title} {\bibinfo {title} {Dynamical mean-field theory of strongly correlated
  fermion systems and the limit of infinite dimensions},\ }\href
  {https://doi.org/10.1103/RevModPhys.68.13} {\bibfield  {journal} {\bibinfo
  {journal} {Rev. Mod. Phys.}\ }\textbf {\bibinfo {volume} {68}},\ \bibinfo
  {pages} {13} (\bibinfo {year} {1996})}\BibitemShut {NoStop}%
\bibitem [{\citenamefont {Huscroft}\ \emph {et~al.}(2001)\citenamefont
  {Huscroft}, \citenamefont {Jarrell}, \citenamefont {Maier}, \citenamefont
  {Moukouri},\ and\ \citenamefont {Tahvildarzadeh}}]{huscroft01}%
  \BibitemOpen
  \bibfield  {author} {\bibinfo {author} {\bibfnamefont {C.}~\bibnamefont
  {Huscroft}}, \bibinfo {author} {\bibfnamefont {M.}~\bibnamefont {Jarrell}},
  \bibinfo {author} {\bibfnamefont {T.}~\bibnamefont {Maier}}, \bibinfo
  {author} {\bibfnamefont {S.}~\bibnamefont {Moukouri}},\ and\ \bibinfo
  {author} {\bibfnamefont {A.~N.}\ \bibnamefont {Tahvildarzadeh}},\ }\bibfield
  {title} {\bibinfo {title} {Pseudogaps in the 2d {Hubbard} model},\ }\href
  {https://doi.org/10.1103/PhysRevLett.86.139} {\bibfield  {journal} {\bibinfo
  {journal} {Phys. Rev. Lett.}\ }\textbf {\bibinfo {volume} {86}},\ \bibinfo
  {pages} {139} (\bibinfo {year} {2001})}\BibitemShut {NoStop}%
\bibitem [{\citenamefont {Maier}\ \emph {et~al.}(2002)\citenamefont {Maier},
  \citenamefont {Pruschke},\ and\ \citenamefont {Jarrell}}]{maier02}%
  \BibitemOpen
  \bibfield  {author} {\bibinfo {author} {\bibfnamefont {T.~A.}\ \bibnamefont
  {Maier}}, \bibinfo {author} {\bibfnamefont {T.}~\bibnamefont {Pruschke}},\
  and\ \bibinfo {author} {\bibfnamefont {M.}~\bibnamefont {Jarrell}},\
  }\bibfield  {title} {\bibinfo {title} {Angle-resolved photoemission spectra
  of the {Hubbard} model},\ }\href {https://doi.org/10.1103/PhysRevB.66.075102}
  {\bibfield  {journal} {\bibinfo  {journal} {Phys. Rev. B}\ }\textbf {\bibinfo
  {volume} {66}},\ \bibinfo {pages} {075102} (\bibinfo {year}
  {2002})}\BibitemShut {NoStop}%
\bibitem [{\citenamefont {S\'en\'echal}\ and\ \citenamefont
  {Tremblay}(2004)}]{senechal04}%
  \BibitemOpen
  \bibfield  {author} {\bibinfo {author} {\bibfnamefont {D.}~\bibnamefont
  {S\'en\'echal}}\ and\ \bibinfo {author} {\bibfnamefont {A.-M.~S.}\
  \bibnamefont {Tremblay}},\ }\bibfield  {title} {\bibinfo {title} {Hot spots
  and pseudogaps for hole- and electron-doped high-temperature
  superconductors},\ }\href {https://doi.org/10.1103/PhysRevLett.92.126401}
  {\bibfield  {journal} {\bibinfo  {journal} {Phys. Rev. Lett.}\ }\textbf
  {\bibinfo {volume} {92}},\ \bibinfo {pages} {126401} (\bibinfo {year}
  {2004})}\BibitemShut {NoStop}%
\bibitem [{\citenamefont {Civelli}\ \emph {et~al.}(2005)\citenamefont
  {Civelli}, \citenamefont {Capone}, \citenamefont {Kancharla}, \citenamefont
  {Parcollet},\ and\ \citenamefont {Kotliar}}]{civelli05}%
  \BibitemOpen
  \bibfield  {author} {\bibinfo {author} {\bibfnamefont {M.}~\bibnamefont
  {Civelli}}, \bibinfo {author} {\bibfnamefont {M.}~\bibnamefont {Capone}},
  \bibinfo {author} {\bibfnamefont {S.~S.}\ \bibnamefont {Kancharla}}, \bibinfo
  {author} {\bibfnamefont {O.}~\bibnamefont {Parcollet}},\ and\ \bibinfo
  {author} {\bibfnamefont {G.}~\bibnamefont {Kotliar}},\ }\bibfield  {title}
  {\bibinfo {title} {Dynamical breakup of the {Fermi} surface in a doped {Mott}
  insulator},\ }\href {https://doi.org/10.1103/PhysRevLett.95.106402}
  {\bibfield  {journal} {\bibinfo  {journal} {Phys. Rev. Lett.}\ }\textbf
  {\bibinfo {volume} {95}},\ \bibinfo {pages} {106402} (\bibinfo {year}
  {2005})}\BibitemShut {NoStop}%
\bibitem [{\citenamefont {S\'en\'echal}\ \emph {et~al.}(2005)\citenamefont
  {S\'en\'echal}, \citenamefont {Lavertu}, \citenamefont {Marois},\ and\
  \citenamefont {Tremblay}}]{senechal05}%
  \BibitemOpen
  \bibfield  {author} {\bibinfo {author} {\bibfnamefont {D.}~\bibnamefont
  {S\'en\'echal}}, \bibinfo {author} {\bibfnamefont {P.-L.}\ \bibnamefont
  {Lavertu}}, \bibinfo {author} {\bibfnamefont {M.-A.}\ \bibnamefont
  {Marois}},\ and\ \bibinfo {author} {\bibfnamefont {A.-M.~S.}\ \bibnamefont
  {Tremblay}},\ }\bibfield  {title} {\bibinfo {title} {Competition between
  antiferromagnetism and superconductivity in high-${T}_{c}$ cuprates},\ }\href
  {https://doi.org/10.1103/PhysRevLett.94.156404} {\bibfield  {journal}
  {\bibinfo  {journal} {Phys. Rev. Lett.}\ }\textbf {\bibinfo {volume} {94}},\
  \bibinfo {pages} {156404} (\bibinfo {year} {2005})}\BibitemShut {NoStop}%
\bibitem [{\citenamefont {Kyung}\ \emph {et~al.}(2006)\citenamefont {Kyung},
  \citenamefont {Kancharla}, \citenamefont {S\'en\'echal}, \citenamefont
  {Tremblay}, \citenamefont {Civelli},\ and\ \citenamefont
  {Kotliar}}]{kyung06}%
  \BibitemOpen
  \bibfield  {author} {\bibinfo {author} {\bibfnamefont {B.}~\bibnamefont
  {Kyung}}, \bibinfo {author} {\bibfnamefont {S.~S.}\ \bibnamefont
  {Kancharla}}, \bibinfo {author} {\bibfnamefont {D.}~\bibnamefont
  {S\'en\'echal}}, \bibinfo {author} {\bibfnamefont {A.-M.~S.}\ \bibnamefont
  {Tremblay}}, \bibinfo {author} {\bibfnamefont {M.}~\bibnamefont {Civelli}},\
  and\ \bibinfo {author} {\bibfnamefont {G.}~\bibnamefont {Kotliar}},\
  }\bibfield  {title} {\bibinfo {title} {Pseudogap induced by short-range spin
  correlations in a doped {Mott} insulator},\ }\href
  {https://doi.org/10.1103/PhysRevB.73.165114} {\bibfield  {journal} {\bibinfo
  {journal} {Phys. Rev. B}\ }\textbf {\bibinfo {volume} {73}},\ \bibinfo
  {pages} {165114} (\bibinfo {year} {2006})}\BibitemShut {NoStop}%
\bibitem [{\citenamefont {Aichhorn}\ \emph {et~al.}(2007)\citenamefont
  {Aichhorn}, \citenamefont {Arrigoni}, \citenamefont {Huang},\ and\
  \citenamefont {Hanke}}]{aichhorn07}%
  \BibitemOpen
  \bibfield  {author} {\bibinfo {author} {\bibfnamefont {M.}~\bibnamefont
  {Aichhorn}}, \bibinfo {author} {\bibfnamefont {E.}~\bibnamefont {Arrigoni}},
  \bibinfo {author} {\bibfnamefont {Z.~B.}\ \bibnamefont {Huang}},\ and\
  \bibinfo {author} {\bibfnamefont {W.}~\bibnamefont {Hanke}},\ }\bibfield
  {title} {\bibinfo {title} {Superconducting gap in the hubbard model and the
  two-gap energy scales of high-${T}_{c}$ cuprate superconductors},\ }\href
  {https://doi.org/10.1103/PhysRevLett.99.257002} {\bibfield  {journal}
  {\bibinfo  {journal} {Phys. Rev. Lett.}\ }\textbf {\bibinfo {volume} {99}},\
  \bibinfo {pages} {257002} (\bibinfo {year} {2007})}\BibitemShut {NoStop}%
\bibitem [{\citenamefont {Kancharla}\ \emph {et~al.}(2008)\citenamefont
  {Kancharla}, \citenamefont {Kyung}, \citenamefont {S\'en\'echal},
  \citenamefont {Civelli}, \citenamefont {Capone}, \citenamefont {Kotliar},\
  and\ \citenamefont {Tremblay}}]{kancharla08}%
  \BibitemOpen
  \bibfield  {author} {\bibinfo {author} {\bibfnamefont {S.~S.}\ \bibnamefont
  {Kancharla}}, \bibinfo {author} {\bibfnamefont {B.}~\bibnamefont {Kyung}},
  \bibinfo {author} {\bibfnamefont {D.}~\bibnamefont {S\'en\'echal}}, \bibinfo
  {author} {\bibfnamefont {M.}~\bibnamefont {Civelli}}, \bibinfo {author}
  {\bibfnamefont {M.}~\bibnamefont {Capone}}, \bibinfo {author} {\bibfnamefont
  {G.}~\bibnamefont {Kotliar}},\ and\ \bibinfo {author} {\bibfnamefont
  {A.-M.~S.}\ \bibnamefont {Tremblay}},\ }\bibfield  {title} {\bibinfo {title}
  {Anomalous superconductivity and its competition with antiferromagnetism in
  doped {Mott} insulators},\ }\href
  {https://doi.org/10.1103/PhysRevB.77.184516} {\bibfield  {journal} {\bibinfo
  {journal} {Phys. Rev. B}\ }\textbf {\bibinfo {volume} {77}},\ \bibinfo
  {pages} {184516} (\bibinfo {year} {2008})}\BibitemShut {NoStop}%
\bibitem [{\citenamefont {Civelli}\ \emph {et~al.}(2008)\citenamefont
  {Civelli}, \citenamefont {Capone}, \citenamefont {Georges}, \citenamefont
  {Haule}, \citenamefont {Parcollet}, \citenamefont {Stanescu},\ and\
  \citenamefont {Kotliar}}]{civelli08}%
  \BibitemOpen
  \bibfield  {author} {\bibinfo {author} {\bibfnamefont {M.}~\bibnamefont
  {Civelli}}, \bibinfo {author} {\bibfnamefont {M.}~\bibnamefont {Capone}},
  \bibinfo {author} {\bibfnamefont {A.}~\bibnamefont {Georges}}, \bibinfo
  {author} {\bibfnamefont {K.}~\bibnamefont {Haule}}, \bibinfo {author}
  {\bibfnamefont {O.}~\bibnamefont {Parcollet}}, \bibinfo {author}
  {\bibfnamefont {T.~D.}\ \bibnamefont {Stanescu}},\ and\ \bibinfo {author}
  {\bibfnamefont {G.}~\bibnamefont {Kotliar}},\ }\bibfield  {title} {\bibinfo
  {title} {Nodal-antinodal dichotomy and the two gaps of a superconducting
  doped mott insulator},\ }\href
  {https://doi.org/10.1103/PhysRevLett.100.046402} {\bibfield  {journal}
  {\bibinfo  {journal} {Phys. Rev. Lett.}\ }\textbf {\bibinfo {volume} {100}},\
  \bibinfo {pages} {046402} (\bibinfo {year} {2008})}\BibitemShut {NoStop}%
\bibitem [{\citenamefont {Liebsch}\ and\ \citenamefont
  {Tong}(2009)}]{liebsch09}%
  \BibitemOpen
  \bibfield  {author} {\bibinfo {author} {\bibfnamefont {A.}~\bibnamefont
  {Liebsch}}\ and\ \bibinfo {author} {\bibfnamefont {N.-H.}\ \bibnamefont
  {Tong}},\ }\bibfield  {title} {\bibinfo {title} {Finite-temperature exact
  diagonalization cluster dynamical mean-field study of the two-dimensional
  {Hubbard} model: Pseudogap, non-{Fermi}-liquid behavior, and particle-hole
  asymmetry},\ }\href {https://doi.org/10.1103/PhysRevB.80.165126} {\bibfield
  {journal} {\bibinfo  {journal} {Phys. Rev. B}\ }\textbf {\bibinfo {volume}
  {80}},\ \bibinfo {pages} {165126} (\bibinfo {year} {2009})}\BibitemShut
  {NoStop}%
\bibitem [{\citenamefont {Okamoto}\ \emph {et~al.}(2010)\citenamefont
  {Okamoto}, \citenamefont {S\'en\'echal}, \citenamefont {Civelli},\ and\
  \citenamefont {Tremblay}}]{okamoto10PRB82}%
  \BibitemOpen
  \bibfield  {author} {\bibinfo {author} {\bibfnamefont {S.}~\bibnamefont
  {Okamoto}}, \bibinfo {author} {\bibfnamefont {D.}~\bibnamefont
  {S\'en\'echal}}, \bibinfo {author} {\bibfnamefont {M.}~\bibnamefont
  {Civelli}},\ and\ \bibinfo {author} {\bibfnamefont {A.-M.~S.}\ \bibnamefont
  {Tremblay}},\ }\bibfield  {title} {\bibinfo {title} {Dynamical electronic
  nematicity from mott physics},\ }\href
  {https://doi.org/10.1103/PhysRevB.82.180511} {\bibfield  {journal} {\bibinfo
  {journal} {Phys. Rev. B}\ }\textbf {\bibinfo {volume} {82}},\ \bibinfo
  {pages} {180511} (\bibinfo {year} {2010})}\BibitemShut {NoStop}%
\bibitem [{\citenamefont {Chen}\ \emph {et~al.}(2012)\citenamefont {Chen},
  \citenamefont {Meng}, \citenamefont {Pruschke}, \citenamefont {Moreno},\ and\
  \citenamefont {Jarrell}}]{chen12}%
  \BibitemOpen
  \bibfield  {author} {\bibinfo {author} {\bibfnamefont {K.-S.}\ \bibnamefont
  {Chen}}, \bibinfo {author} {\bibfnamefont {Z.~Y.}\ \bibnamefont {Meng}},
  \bibinfo {author} {\bibfnamefont {T.}~\bibnamefont {Pruschke}}, \bibinfo
  {author} {\bibfnamefont {J.}~\bibnamefont {Moreno}},\ and\ \bibinfo {author}
  {\bibfnamefont {M.}~\bibnamefont {Jarrell}},\ }\bibfield  {title} {\bibinfo
  {title} {Lifshitz transition in the two-dimensional {Hubbard} model},\ }\href
  {https://doi.org/10.1103/PhysRevB.86.165136} {\bibfield  {journal} {\bibinfo
  {journal} {Phys. Rev. B}\ }\textbf {\bibinfo {volume} {86}},\ \bibinfo
  {pages} {165136} (\bibinfo {year} {2012})}\BibitemShut {NoStop}%
\bibitem [{\citenamefont {Gull}\ \emph {et~al.}(2013)\citenamefont {Gull},
  \citenamefont {Parcollet},\ and\ \citenamefont {Millis}}]{gull13PRL}%
  \BibitemOpen
  \bibfield  {author} {\bibinfo {author} {\bibfnamefont {E.}~\bibnamefont
  {Gull}}, \bibinfo {author} {\bibfnamefont {O.}~\bibnamefont {Parcollet}},\
  and\ \bibinfo {author} {\bibfnamefont {A.~J.}\ \bibnamefont {Millis}},\
  }\bibfield  {title} {\bibinfo {title} {Superconductivity and the pseudogap in
  the two-dimensional {Hubbard} model},\ }\href
  {https://doi.org/10.1103/PhysRevLett.110.216405} {\bibfield  {journal}
  {\bibinfo  {journal} {Phys. Rev. Lett.}\ }\textbf {\bibinfo {volume} {110}},\
  \bibinfo {pages} {216405} (\bibinfo {year} {2013})}\BibitemShut {NoStop}%
\bibitem [{\citenamefont {Kohno}(2014)}]{kohno14}%
  \BibitemOpen
  \bibfield  {author} {\bibinfo {author} {\bibfnamefont {M.}~\bibnamefont
  {Kohno}},\ }\bibfield  {title} {\bibinfo {title} {Spectral properties near
  the {Mott} transition in the two-dimensional {Hubbard} model with
  next-nearest-neighbor hopping},\ }\href
  {https://doi.org/10.1103/PhysRevB.90.035111} {\bibfield  {journal} {\bibinfo
  {journal} {Phys. Rev. B}\ }\textbf {\bibinfo {volume} {90}},\ \bibinfo
  {pages} {035111} (\bibinfo {year} {2014})}\BibitemShut {NoStop}%
\bibitem [{\citenamefont {Lin}\ \emph {et~al.}(2012)\citenamefont {Lin},
  \citenamefont {Gull},\ and\ \citenamefont {Millis}}]{lin12}%
  \BibitemOpen
  \bibfield  {author} {\bibinfo {author} {\bibfnamefont {N.}~\bibnamefont
  {Lin}}, \bibinfo {author} {\bibfnamefont {E.}~\bibnamefont {Gull}},\ and\
  \bibinfo {author} {\bibfnamefont {A.~J.}\ \bibnamefont {Millis}},\ }\bibfield
   {title} {\bibinfo {title} {Two-particle response in cluster dynamical
  mean-field theory: Formalism and application to the raman response of
  high-temperature superconductors},\ }\href
  {https://doi.org/10.1103/PhysRevLett.109.106401} {\bibfield  {journal}
  {\bibinfo  {journal} {Phys. Rev. Lett.}\ }\textbf {\bibinfo {volume} {109}},\
  \bibinfo {pages} {106401} (\bibinfo {year} {2012})}\BibitemShut {NoStop}%
\bibitem [{\citenamefont {Sakai}\ \emph {et~al.}(2013)\citenamefont {Sakai},
  \citenamefont {Blanc}, \citenamefont {Civelli}, \citenamefont {Gallais},
  \citenamefont {Cazayous}, \citenamefont {M\'easson}, \citenamefont {Wen},
  \citenamefont {Xu}, \citenamefont {Gu}, \citenamefont {Sangiovanni},
  \citenamefont {Motome}, \citenamefont {Held}, \citenamefont {Sacuto},
  \citenamefont {Georges},\ and\ \citenamefont {Imada}}]{sakai13}%
  \BibitemOpen
  \bibfield  {author} {\bibinfo {author} {\bibfnamefont {S.}~\bibnamefont
  {Sakai}}, \bibinfo {author} {\bibfnamefont {S.}~\bibnamefont {Blanc}},
  \bibinfo {author} {\bibfnamefont {M.}~\bibnamefont {Civelli}}, \bibinfo
  {author} {\bibfnamefont {Y.}~\bibnamefont {Gallais}}, \bibinfo {author}
  {\bibfnamefont {M.}~\bibnamefont {Cazayous}}, \bibinfo {author}
  {\bibfnamefont {M.-A.}\ \bibnamefont {M\'easson}}, \bibinfo {author}
  {\bibfnamefont {J.~S.}\ \bibnamefont {Wen}}, \bibinfo {author} {\bibfnamefont
  {Z.~J.}\ \bibnamefont {Xu}}, \bibinfo {author} {\bibfnamefont {G.~D.}\
  \bibnamefont {Gu}}, \bibinfo {author} {\bibfnamefont {G.}~\bibnamefont
  {Sangiovanni}}, \bibinfo {author} {\bibfnamefont {Y.}~\bibnamefont {Motome}},
  \bibinfo {author} {\bibfnamefont {K.}~\bibnamefont {Held}}, \bibinfo {author}
  {\bibfnamefont {A.}~\bibnamefont {Sacuto}}, \bibinfo {author} {\bibfnamefont
  {A.}~\bibnamefont {Georges}},\ and\ \bibinfo {author} {\bibfnamefont
  {M.}~\bibnamefont {Imada}},\ }\bibfield  {title} {\bibinfo {title}
  {Raman-scattering measurements and theory of the energy-momentum spectrum for
  underdoped {Bi}$_{2}${Sr}$_{2}${CaCuO}$_{8+\delta}$ superconductors: Evidence
  of an $s$-wave structure for the pseudogap},\ }\href
  {https://doi.org/10.1103/PhysRevLett.111.107001} {\bibfield  {journal}
  {\bibinfo  {journal} {Phys. Rev. Lett.}\ }\textbf {\bibinfo {volume} {111}},\
  \bibinfo {pages} {107001} (\bibinfo {year} {2013})}\BibitemShut {NoStop}%
\bibitem [{\citenamefont {Loret}\ \emph {et~al.}(2016)\citenamefont {Loret},
  \citenamefont {Sakai}, \citenamefont {Gallais}, \citenamefont {Cazayous},
  \citenamefont {M\'easson}, \citenamefont {Forget}, \citenamefont {Colson},
  \citenamefont {Civelli},\ and\ \citenamefont {Sacuto}}]{loret16}%
  \BibitemOpen
  \bibfield  {author} {\bibinfo {author} {\bibfnamefont {B.}~\bibnamefont
  {Loret}}, \bibinfo {author} {\bibfnamefont {S.}~\bibnamefont {Sakai}},
  \bibinfo {author} {\bibfnamefont {Y.}~\bibnamefont {Gallais}}, \bibinfo
  {author} {\bibfnamefont {M.}~\bibnamefont {Cazayous}}, \bibinfo {author}
  {\bibfnamefont {M.-A.}\ \bibnamefont {M\'easson}}, \bibinfo {author}
  {\bibfnamefont {A.}~\bibnamefont {Forget}}, \bibinfo {author} {\bibfnamefont
  {D.}~\bibnamefont {Colson}}, \bibinfo {author} {\bibfnamefont
  {M.}~\bibnamefont {Civelli}},\ and\ \bibinfo {author} {\bibfnamefont
  {A.}~\bibnamefont {Sacuto}},\ }\bibfield  {title} {\bibinfo {title}
  {Unconventional high-energy-state contribution to the cooper pairing in the
  underdoped copper-oxide superconductor
  {HgBa}$_{2}${Ca}$_{2}${Cu}$_{3}${O}$_{8+\delta}$},\ }\href
  {https://doi.org/10.1103/PhysRevLett.116.197001} {\bibfield  {journal}
  {\bibinfo  {journal} {Phys. Rev. Lett.}\ }\textbf {\bibinfo {volume} {116}},\
  \bibinfo {pages} {197001} (\bibinfo {year} {2016})}\BibitemShut {NoStop}%
\bibitem [{\citenamefont {Loret}\ \emph {et~al.}(2017)\citenamefont {Loret},
  \citenamefont {Sakai}, \citenamefont {Benhabib}, \citenamefont {Gallais},
  \citenamefont {Cazayous}, \citenamefont {M\'easson}, \citenamefont {Zhong},
  \citenamefont {Schneeloch}, \citenamefont {Gu}, \citenamefont {Forget},
  \citenamefont {Colson}, \citenamefont {Paul}, \citenamefont {Civelli},\ and\
  \citenamefont {Sacuto}}]{loret17}%
  \BibitemOpen
  \bibfield  {author} {\bibinfo {author} {\bibfnamefont {B.}~\bibnamefont
  {Loret}}, \bibinfo {author} {\bibfnamefont {S.}~\bibnamefont {Sakai}},
  \bibinfo {author} {\bibfnamefont {S.}~\bibnamefont {Benhabib}}, \bibinfo
  {author} {\bibfnamefont {Y.}~\bibnamefont {Gallais}}, \bibinfo {author}
  {\bibfnamefont {M.}~\bibnamefont {Cazayous}}, \bibinfo {author}
  {\bibfnamefont {M.~A.}\ \bibnamefont {M\'easson}}, \bibinfo {author}
  {\bibfnamefont {R.~D.}\ \bibnamefont {Zhong}}, \bibinfo {author}
  {\bibfnamefont {J.}~\bibnamefont {Schneeloch}}, \bibinfo {author}
  {\bibfnamefont {G.~D.}\ \bibnamefont {Gu}}, \bibinfo {author} {\bibfnamefont
  {A.}~\bibnamefont {Forget}}, \bibinfo {author} {\bibfnamefont
  {D.}~\bibnamefont {Colson}}, \bibinfo {author} {\bibfnamefont
  {I.}~\bibnamefont {Paul}}, \bibinfo {author} {\bibfnamefont {M.}~\bibnamefont
  {Civelli}},\ and\ \bibinfo {author} {\bibfnamefont {A.}~\bibnamefont
  {Sacuto}},\ }\bibfield  {title} {\bibinfo {title} {Vertical temperature
  boundary of the pseudogap under the superconducting dome in the phase diagram
  of {Bi}$_{2}${Sr}$_{2}${CaCu}$_{2}${O}$_{8+\delta}$},\ }\href
  {https://doi.org/10.1103/PhysRevB.96.094525} {\bibfield  {journal} {\bibinfo
  {journal} {Phys. Rev. B}\ }\textbf {\bibinfo {volume} {96}},\ \bibinfo
  {pages} {094525} (\bibinfo {year} {2017})}\BibitemShut {NoStop}%
\bibitem [{\citenamefont {Lin}\ \emph {et~al.}(2009)\citenamefont {Lin},
  \citenamefont {Gull},\ and\ \citenamefont {Millis}}]{lin09}%
  \BibitemOpen
  \bibfield  {author} {\bibinfo {author} {\bibfnamefont {N.}~\bibnamefont
  {Lin}}, \bibinfo {author} {\bibfnamefont {E.}~\bibnamefont {Gull}},\ and\
  \bibinfo {author} {\bibfnamefont {A.~J.}\ \bibnamefont {Millis}},\ }\bibfield
   {title} {\bibinfo {title} {Optical conductivity from cluster dynamical
  mean-field theory: Formalism and application to high-temperature
  superconductors},\ }\href {https://doi.org/10.1103/PhysRevB.80.161105}
  {\bibfield  {journal} {\bibinfo  {journal} {Phys. Rev. B}\ }\textbf {\bibinfo
  {volume} {80}},\ \bibinfo {pages} {161105} (\bibinfo {year}
  {2009})}\BibitemShut {NoStop}%
\bibitem [{\citenamefont {Ferrero}\ \emph {et~al.}(2010)\citenamefont
  {Ferrero}, \citenamefont {Parcollet}, \citenamefont {Georges}, \citenamefont
  {Kotliar},\ and\ \citenamefont {Basov}}]{ferrero10}%
  \BibitemOpen
  \bibfield  {author} {\bibinfo {author} {\bibfnamefont {M.}~\bibnamefont
  {Ferrero}}, \bibinfo {author} {\bibfnamefont {O.}~\bibnamefont {Parcollet}},
  \bibinfo {author} {\bibfnamefont {A.}~\bibnamefont {Georges}}, \bibinfo
  {author} {\bibfnamefont {G.}~\bibnamefont {Kotliar}},\ and\ \bibinfo {author}
  {\bibfnamefont {D.~N.}\ \bibnamefont {Basov}},\ }\bibfield  {title} {\bibinfo
  {title} {Interplane charge dynamics in a valence-bond dynamical mean-field
  theory of cuprate superconductors},\ }\href
  {https://doi.org/10.1103/PhysRevB.82.054502} {\bibfield  {journal} {\bibinfo
  {journal} {Phys. Rev. B}\ }\textbf {\bibinfo {volume} {82}},\ \bibinfo
  {pages} {054502} (\bibinfo {year} {2010})}\BibitemShut {NoStop}%
\bibitem [{\citenamefont {Ferrero}\ \emph {et~al.}(2009)\citenamefont
  {Ferrero}, \citenamefont {Cornaglia}, \citenamefont {De~Leo}, \citenamefont
  {Parcollet}, \citenamefont {Kotliar},\ and\ \citenamefont
  {Georges}}]{ferrero09}%
  \BibitemOpen
  \bibfield  {author} {\bibinfo {author} {\bibfnamefont {M.}~\bibnamefont
  {Ferrero}}, \bibinfo {author} {\bibfnamefont {P.~S.}\ \bibnamefont
  {Cornaglia}}, \bibinfo {author} {\bibfnamefont {L.}~\bibnamefont {De~Leo}},
  \bibinfo {author} {\bibfnamefont {O.}~\bibnamefont {Parcollet}}, \bibinfo
  {author} {\bibfnamefont {G.}~\bibnamefont {Kotliar}},\ and\ \bibinfo {author}
  {\bibfnamefont {A.}~\bibnamefont {Georges}},\ }\bibfield  {title} {\bibinfo
  {title} {Pseudogap opening and formation of fermi arcs as an
  orbital-selective {Mott} transition in momentum space},\ }\href
  {https://doi.org/10.1103/PhysRevB.80.064501} {\bibfield  {journal} {\bibinfo
  {journal} {Phys. Rev. B}\ }\textbf {\bibinfo {volume} {80}},\ \bibinfo
  {pages} {064501} (\bibinfo {year} {2009})}\BibitemShut {NoStop}%
\bibitem [{\citenamefont {Sakai}\ \emph {et~al.}(2010)\citenamefont {Sakai},
  \citenamefont {Motome},\ and\ \citenamefont {Imada}}]{sakai10}%
  \BibitemOpen
  \bibfield  {author} {\bibinfo {author} {\bibfnamefont {S.}~\bibnamefont
  {Sakai}}, \bibinfo {author} {\bibfnamefont {Y.}~\bibnamefont {Motome}},\ and\
  \bibinfo {author} {\bibfnamefont {M.}~\bibnamefont {Imada}},\ }\bibfield
  {title} {\bibinfo {title} {Doped high-${T}_{c}$ cuprate superconductors
  elucidated in the light of zeros and poles of the electronic {Green's}
  function},\ }\href {https://doi.org/10.1103/PhysRevB.82.134505} {\bibfield
  {journal} {\bibinfo  {journal} {Phys. Rev. B}\ }\textbf {\bibinfo {volume}
  {82}},\ \bibinfo {pages} {134505} (\bibinfo {year} {2010})}\BibitemShut
  {NoStop}%
\bibitem [{\citenamefont {Sordi}\ \emph {et~al.}(2012)\citenamefont {Sordi},
  \citenamefont {S\'emon}, \citenamefont {Haule},\ and\ \citenamefont
  {Tremblay}}]{sordi12PRL}%
  \BibitemOpen
  \bibfield  {author} {\bibinfo {author} {\bibfnamefont {G.}~\bibnamefont
  {Sordi}}, \bibinfo {author} {\bibfnamefont {P.}~\bibnamefont {S\'emon}},
  \bibinfo {author} {\bibfnamefont {K.}~\bibnamefont {Haule}},\ and\ \bibinfo
  {author} {\bibfnamefont {A.-M.~S.}\ \bibnamefont {Tremblay}},\ }\bibfield
  {title} {\bibinfo {title} {Strong coupling superconductivity, pseudogap, and
  {Mott} transition},\ }\href {https://doi.org/10.1103/PhysRevLett.108.216401}
  {\bibfield  {journal} {\bibinfo  {journal} {Phys. Rev. Lett.}\ }\textbf
  {\bibinfo {volume} {108}},\ \bibinfo {pages} {216401} (\bibinfo {year}
  {2012})}\BibitemShut {NoStop}%
\bibitem [{\citenamefont {Stanescu}\ and\ \citenamefont
  {Kotliar}(2006)}]{stanescu06}%
  \BibitemOpen
  \bibfield  {author} {\bibinfo {author} {\bibfnamefont {T.~D.}\ \bibnamefont
  {Stanescu}}\ and\ \bibinfo {author} {\bibfnamefont {G.}~\bibnamefont
  {Kotliar}},\ }\bibfield  {title} {\bibinfo {title} {Fermi arcs and hidden
  zeros of the {Green} function in the pseudogap state},\ }\href
  {https://doi.org/10.1103/PhysRevB.74.125110} {\bibfield  {journal} {\bibinfo
  {journal} {Phys. Rev. B}\ }\textbf {\bibinfo {volume} {74}},\ \bibinfo
  {pages} {125110} (\bibinfo {year} {2006})}\BibitemShut {NoStop}%
\bibitem [{\citenamefont {Civelli}(2009)}]{civelli09PRL}%
  \BibitemOpen
  \bibfield  {author} {\bibinfo {author} {\bibfnamefont {M.}~\bibnamefont
  {Civelli}},\ }\bibfield  {title} {\bibinfo {title} {Evolution of the
  dynamical pairing across the phase diagram of a strongly correlated
  high-temperature superconductor},\ }\href
  {https://doi.org/10.1103/PhysRevLett.103.136402} {\bibfield  {journal}
  {\bibinfo  {journal} {Phys. Rev. Lett.}\ }\textbf {\bibinfo {volume} {103}},\
  \bibinfo {pages} {136402} (\bibinfo {year} {2009})}\BibitemShut {NoStop}%
\bibitem [{\citenamefont {Kyung}\ \emph {et~al.}(2009)\citenamefont {Kyung},
  \citenamefont {S\'en\'echal},\ and\ \citenamefont {Tremblay}}]{kyung09}%
  \BibitemOpen
  \bibfield  {author} {\bibinfo {author} {\bibfnamefont {B.}~\bibnamefont
  {Kyung}}, \bibinfo {author} {\bibfnamefont {D.}~\bibnamefont
  {S\'en\'echal}},\ and\ \bibinfo {author} {\bibfnamefont {A.-M.~S.}\
  \bibnamefont {Tremblay}},\ }\bibfield  {title} {\bibinfo {title} {Pairing
  dynamics in strongly correlated superconductivity},\ }\href
  {https://doi.org/10.1103/PhysRevB.80.205109} {\bibfield  {journal} {\bibinfo
  {journal} {Phys. Rev. B}\ }\textbf {\bibinfo {volume} {80}},\ \bibinfo
  {pages} {205109} (\bibinfo {year} {2009})}\BibitemShut {NoStop}%
\bibitem [{\citenamefont {Sakai}\ \emph {et~al.}(2009)\citenamefont {Sakai},
  \citenamefont {Motome},\ and\ \citenamefont {Imada}}]{sakai09}%
  \BibitemOpen
  \bibfield  {author} {\bibinfo {author} {\bibfnamefont {S.}~\bibnamefont
  {Sakai}}, \bibinfo {author} {\bibfnamefont {Y.}~\bibnamefont {Motome}},\ and\
  \bibinfo {author} {\bibfnamefont {M.}~\bibnamefont {Imada}},\ }\bibfield
  {title} {\bibinfo {title} {Evolution of electronic structure of doped {Mott}
  insulators: Reconstruction of poles and zeros of {Green's} function},\ }\href
  {https://doi.org/10.1103/PhysRevLett.102.056404} {\bibfield  {journal}
  {\bibinfo  {journal} {Phys. Rev. Lett.}\ }\textbf {\bibinfo {volume} {102}},\
  \bibinfo {pages} {056404} (\bibinfo {year} {2009})}\BibitemShut {NoStop}%
\bibitem [{\citenamefont {Gull}\ and\ \citenamefont {Millis}(2015)}]{gull15}%
  \BibitemOpen
  \bibfield  {author} {\bibinfo {author} {\bibfnamefont {E.}~\bibnamefont
  {Gull}}\ and\ \bibinfo {author} {\bibfnamefont {A.~J.}\ \bibnamefont
  {Millis}},\ }\bibfield  {title} {\bibinfo {title} {Quasiparticle properties
  of the superconducting state of the two-dimensional hubbard model},\ }\href
  {https://doi.org/10.1103/PhysRevB.91.085116} {\bibfield  {journal} {\bibinfo
  {journal} {Phys. Rev. B}\ }\textbf {\bibinfo {volume} {91}},\ \bibinfo
  {pages} {085116} (\bibinfo {year} {2015})}\BibitemShut {NoStop}%
\bibitem [{\citenamefont {Sakai}\ \emph {et~al.}(2018)\citenamefont {Sakai},
  \citenamefont {Civelli},\ and\ \citenamefont {Imada}}]{sakai18}%
  \BibitemOpen
  \bibfield  {author} {\bibinfo {author} {\bibfnamefont {S.}~\bibnamefont
  {Sakai}}, \bibinfo {author} {\bibfnamefont {M.}~\bibnamefont {Civelli}},\
  and\ \bibinfo {author} {\bibfnamefont {M.}~\bibnamefont {Imada}},\ }\bibfield
   {title} {\bibinfo {title} {Direct connection between mott insulators and
  $d$-wave high-temperature superconductors revealed by continuous evolution of
  self-energy poles},\ }\href {https://doi.org/10.1103/PhysRevB.98.195109}
  {\bibfield  {journal} {\bibinfo  {journal} {Phys. Rev. B}\ }\textbf {\bibinfo
  {volume} {98}},\ \bibinfo {pages} {195109} (\bibinfo {year}
  {2018})}\BibitemShut {NoStop}%
\bibitem [{\citenamefont {Imada}(2024)}]{imada24}%
  \BibitemOpen
  \bibfield  {author} {\bibinfo {author} {\bibfnamefont {M.}~\bibnamefont
  {Imada}},\ }\bibfield  {title} {\bibinfo {title} {Fermi machine - quantum
  many-body solver derived from correspondence between noninteracting and
  strongly correlated fermions},\ }\href
  {https://doi.org/10.7566/JPSJ.93.104002} {\bibfield  {journal} {\bibinfo
  {journal} {J. Physical. Soc. Jpn.}\ }\textbf {\bibinfo {volume} {93}},\
  \bibinfo {pages} {104002} (\bibinfo {year} {2024})}\BibitemShut {NoStop}%
\bibitem [{\citenamefont {Sakai}(2023)}]{sakai23}%
  \BibitemOpen
  \bibfield  {author} {\bibinfo {author} {\bibfnamefont {S.}~\bibnamefont
  {Sakai}},\ }\bibfield  {title} {\bibinfo {title} {Nonperturbative
  calculations for spectroscopic properties of cuprate high-temperature
  superconductors},\ }\href {https://doi.org/10.7566/JPSJ.92.092001} {\bibfield
   {journal} {\bibinfo  {journal} {Journal of the Physical Society of Japan}\
  }\textbf {\bibinfo {volume} {92}},\ \bibinfo {pages} {092001} (\bibinfo
  {year} {2023})}\BibitemShut {NoStop}%
\bibitem [{\citenamefont {Yamaji}\ \emph {et~al.}(2021)\citenamefont {Yamaji},
  \citenamefont {Yoshida}, \citenamefont {Fujimori},\ and\ \citenamefont
  {Imada}}]{yamaji21}%
  \BibitemOpen
  \bibfield  {author} {\bibinfo {author} {\bibfnamefont {Y.}~\bibnamefont
  {Yamaji}}, \bibinfo {author} {\bibfnamefont {T.}~\bibnamefont {Yoshida}},
  \bibinfo {author} {\bibfnamefont {A.}~\bibnamefont {Fujimori}},\ and\
  \bibinfo {author} {\bibfnamefont {M.}~\bibnamefont {Imada}},\ }\bibfield
  {title} {\bibinfo {title} {Hidden self-energies as origin of cuprate
  superconductivity revealed by machine learning},\ }\href
  {https://doi.org/10.1103/PhysRevResearch.3.043099} {\bibfield  {journal}
  {\bibinfo  {journal} {Phys. Rev. Research}\ }\textbf {\bibinfo {volume}
  {3}},\ \bibinfo {pages} {043099} (\bibinfo {year} {2021})}\BibitemShut
  {NoStop}%
\bibitem [{\citenamefont {Imada}(2021)}]{imada21}%
  \BibitemOpen
  \bibfield  {author} {\bibinfo {author} {\bibfnamefont {M.}~\bibnamefont
  {Imada}},\ }\bibfield  {title} {\bibinfo {title} {Resonant inelastic x-ray
  scattering spectra of cuprate superconductors predicted by model of
  fractionalized fermions},\ }\href {https://doi.org/10.7566/JPSJ.90.074702}
  {\bibfield  {journal} {\bibinfo  {journal} {Journal of the Physical Society
  of Japan}\ }\textbf {\bibinfo {volume} {90}},\ \bibinfo {pages} {074702}
  (\bibinfo {year} {2021})},\ \Eprint
  {https://arxiv.org/abs/https://doi.org/10.7566/JPSJ.90.074702}
  {https://doi.org/10.7566/JPSJ.90.074702} \BibitemShut {NoStop}%
\bibitem [{\citenamefont {Singh}\ \emph {et~al.}(2022)\citenamefont {Singh},
  \citenamefont {Huang}, \citenamefont {Xie}, \citenamefont {Okamoto},
  \citenamefont {Chen}, \citenamefont {Watanabe}, \citenamefont {Fujimori},
  \citenamefont {Imada},\ and\ \citenamefont {Huang}}]{singh22}%
  \BibitemOpen
  \bibfield  {author} {\bibinfo {author} {\bibfnamefont {A.}~\bibnamefont
  {Singh}}, \bibinfo {author} {\bibfnamefont {H.~Y.}\ \bibnamefont {Huang}},
  \bibinfo {author} {\bibfnamefont {J.~D.}\ \bibnamefont {Xie}}, \bibinfo
  {author} {\bibfnamefont {J.}~\bibnamefont {Okamoto}}, \bibinfo {author}
  {\bibfnamefont {C.~T.}\ \bibnamefont {Chen}}, \bibinfo {author}
  {\bibfnamefont {T.}~\bibnamefont {Watanabe}}, \bibinfo {author}
  {\bibfnamefont {A.}~\bibnamefont {Fujimori}}, \bibinfo {author}
  {\bibfnamefont {M.}~\bibnamefont {Imada}},\ and\ \bibinfo {author}
  {\bibfnamefont {D.~J.}\ \bibnamefont {Huang}},\ }\bibfield  {title} {\bibinfo
  {title} {Unconventional exciton evolution from the pseudogap to
  superconducting phases in cuprates},\ }\href
  {https://doi.org/10.1038/s41467-022-35210-8} {\bibfield  {journal} {\bibinfo
  {journal} {Nature Communications}\ }\textbf {\bibinfo {volume} {13}},\
  \bibinfo {pages} {7906} (\bibinfo {year} {2022})}\BibitemShut {NoStop}%
\bibitem [{\citenamefont {Lanat\`a}\ \emph {et~al.}(2017)\citenamefont
  {Lanat\`a}, \citenamefont {Lee}, \citenamefont {Yao},\ and\ \citenamefont
  {Dobrosavljevi\ifmmode~\acute{c}\else \'{c}\fi{}}}]{lanata17}%
  \BibitemOpen
  \bibfield  {author} {\bibinfo {author} {\bibfnamefont {N.}~\bibnamefont
  {Lanat\`a}}, \bibinfo {author} {\bibfnamefont {T.-H.}\ \bibnamefont {Lee}},
  \bibinfo {author} {\bibfnamefont {Y.-X.}\ \bibnamefont {Yao}},\ and\ \bibinfo
  {author} {\bibfnamefont {V.}~\bibnamefont
  {Dobrosavljevi\ifmmode~\acute{c}\else \'{c}\fi{}}},\ }\bibfield  {title}
  {\bibinfo {title} {Emergent bloch excitations in mott matter},\ }\href
  {https://doi.org/10.1103/PhysRevB.96.195126} {\bibfield  {journal} {\bibinfo
  {journal} {Phys. Rev. B}\ }\textbf {\bibinfo {volume} {96}},\ \bibinfo
  {pages} {195126} (\bibinfo {year} {2017})}\BibitemShut {NoStop}%
\bibitem [{\citenamefont {Zhang}\ and\ \citenamefont
  {Sachdev}(2020)}]{zhang20}%
  \BibitemOpen
  \bibfield  {author} {\bibinfo {author} {\bibfnamefont {Y.-H.}\ \bibnamefont
  {Zhang}}\ and\ \bibinfo {author} {\bibfnamefont {S.}~\bibnamefont
  {Sachdev}},\ }\bibfield  {title} {\bibinfo {title} {From the pseudogap metal
  to the fermi liquid using ancilla qubits},\ }\href
  {https://doi.org/10.1103/PhysRevResearch.2.023172} {\bibfield  {journal}
  {\bibinfo  {journal} {Phys. Rev. Res.}\ }\textbf {\bibinfo {volume} {2}},\
  \bibinfo {pages} {023172} (\bibinfo {year} {2020})}\BibitemShut {NoStop}%
\bibitem [{\citenamefont {Imada}\ and\ \citenamefont {Suzuki}(2019)}]{imada19}%
  \BibitemOpen
  \bibfield  {author} {\bibinfo {author} {\bibfnamefont {M.}~\bibnamefont
  {Imada}}\ and\ \bibinfo {author} {\bibfnamefont {T.~J.}\ \bibnamefont
  {Suzuki}},\ }\bibfield  {title} {\bibinfo {title} {Excitons and dark fermions
  as origins of {Mott} gap, pseudogap and superconductivity in cuprate
  superconductors ― general concept and basic formalism based on gap
  physics},\ }\href {https://doi.org/10.7566/JPSJ.88.024701} {\bibfield
  {journal} {\bibinfo  {journal} {Journal of the Physical Society of Japan}\
  }\textbf {\bibinfo {volume} {88}},\ \bibinfo {pages} {024701} (\bibinfo
  {year} {2019})}\BibitemShut {NoStop}%
\bibitem [{\citenamefont {Schmid}\ \emph {et~al.}(2023)\citenamefont {Schmid},
  \citenamefont {Mor\'ee}, \citenamefont {Kaneko}, \citenamefont {Yamaji},\
  and\ \citenamefont {Imada}}]{schmid23}%
  \BibitemOpen
  \bibfield  {author} {\bibinfo {author} {\bibfnamefont {M.~T.}\ \bibnamefont
  {Schmid}}, \bibinfo {author} {\bibfnamefont {J.-B.}\ \bibnamefont {Mor\'ee}},
  \bibinfo {author} {\bibfnamefont {R.}~\bibnamefont {Kaneko}}, \bibinfo
  {author} {\bibfnamefont {Y.}~\bibnamefont {Yamaji}},\ and\ \bibinfo {author}
  {\bibfnamefont {M.}~\bibnamefont {Imada}},\ }\bibfield  {title} {\bibinfo
  {title} {Superconductivity studied by solving \textit{Ab Initio} low-energy
  effective hamiltonians for carrier doped {CaCuO$_2$}, {Bi$_2$Sr$_2$CuO$_6$},
  {Bi$_2$Sr$_2$CaCu$_2$O$_8$}, and {HgBa$_2$CuO$_4$}},\ }\href
  {https://doi.org/10.1103/PhysRevX.13.041036} {\bibfield  {journal} {\bibinfo
  {journal} {Phys. Rev. X}\ }\textbf {\bibinfo {volume} {13}},\ \bibinfo
  {pages} {041036} (\bibinfo {year} {2023})}\BibitemShut {NoStop}%
\bibitem [{\citenamefont {Palczewski}\ \emph {et~al.}(2010)\citenamefont
  {Palczewski}, \citenamefont {Kondo}, \citenamefont {Wen}, \citenamefont {Xu},
  \citenamefont {Gu},\ and\ \citenamefont {Kaminski}}]{palczewski10}%
  \BibitemOpen
  \bibfield  {author} {\bibinfo {author} {\bibfnamefont {A.~D.}\ \bibnamefont
  {Palczewski}}, \bibinfo {author} {\bibfnamefont {T.}~\bibnamefont {Kondo}},
  \bibinfo {author} {\bibfnamefont {J.~S.}\ \bibnamefont {Wen}}, \bibinfo
  {author} {\bibfnamefont {G.~Z.~J.}\ \bibnamefont {Xu}}, \bibinfo {author}
  {\bibfnamefont {G.}~\bibnamefont {Gu}},\ and\ \bibinfo {author}
  {\bibfnamefont {A.}~\bibnamefont {Kaminski}},\ }\bibfield  {title} {\bibinfo
  {title} {Controlling the carrier concentration of the high-temperature
  superconductor
  {${\text{Bi}}_{2}{\text{Sr}}_{2}{\text{CaCu}}_{2}{\text{O}}_{8+\ensuremath{\delta}}$}
  in angle-resolved photoemission spectroscopy experiments},\ }\href
  {https://doi.org/10.1103/PhysRevB.81.104521} {\bibfield  {journal} {\bibinfo
  {journal} {Phys. Rev. B}\ }\textbf {\bibinfo {volume} {81}},\ \bibinfo
  {pages} {104521} (\bibinfo {year} {2010})}\BibitemShut {NoStop}%
\bibitem [{\citenamefont {Yamaji}\ and\ \citenamefont
  {Imada}(2011{\natexlab{a}})}]{yamaji11PRB}%
  \BibitemOpen
  \bibfield  {author} {\bibinfo {author} {\bibfnamefont {Y.}~\bibnamefont
  {Yamaji}}\ and\ \bibinfo {author} {\bibfnamefont {M.}~\bibnamefont {Imada}},\
  }\bibfield  {title} {\bibinfo {title} {Composite fermion theory for pseudogap
  phenomena and superconductivity in underdoped cuprate superconductors},\
  }\href {https://doi.org/10.1103/PhysRevB.83.214522} {\bibfield  {journal}
  {\bibinfo  {journal} {Phys. Rev. B}\ }\textbf {\bibinfo {volume} {83}},\
  \bibinfo {pages} {214522} (\bibinfo {year} {2011}{\natexlab{a}})}\BibitemShut
  {NoStop}%
\bibitem [{\citenamefont {Yamaji}\ and\ \citenamefont
  {Imada}(2011{\natexlab{b}})}]{yamaji11PRL}%
  \BibitemOpen
  \bibfield  {author} {\bibinfo {author} {\bibfnamefont {Y.}~\bibnamefont
  {Yamaji}}\ and\ \bibinfo {author} {\bibfnamefont {M.}~\bibnamefont {Imada}},\
  }\bibfield  {title} {\bibinfo {title} {Composite-fermion theory for
  pseudogap, {Fermi} arc, hole pocket, and non-{Fermi} liquid of underdoped
  cuprate superconductors},\ }\href
  {https://doi.org/10.1103/PhysRevLett.106.016404} {\bibfield  {journal}
  {\bibinfo  {journal} {Phys. Rev. Lett.}\ }\textbf {\bibinfo {volume} {106}},\
  \bibinfo {pages} {016404} (\bibinfo {year} {2011}{\natexlab{b}})}\BibitemShut
  {NoStop}%
\bibitem [{\citenamefont {Mor\'ee}\ \emph {et~al.}(2022)\citenamefont
  {Mor\'ee}, \citenamefont {Hirayama}, \citenamefont {Schmid}, \citenamefont
  {Yamaji},\ and\ \citenamefont {Imada}}]{moree22}%
  \BibitemOpen
  \bibfield  {author} {\bibinfo {author} {\bibfnamefont {J.-B.}\ \bibnamefont
  {Mor\'ee}}, \bibinfo {author} {\bibfnamefont {M.}~\bibnamefont {Hirayama}},
  \bibinfo {author} {\bibfnamefont {M.~T.}\ \bibnamefont {Schmid}}, \bibinfo
  {author} {\bibfnamefont {Y.}~\bibnamefont {Yamaji}},\ and\ \bibinfo {author}
  {\bibfnamefont {M.}~\bibnamefont {Imada}},\ }\bibfield  {title} {\bibinfo
  {title} {Ab initio low-energy effective hamiltonians for the high-temperature
  superconducting cuprates {Bi}$_{2}${Sr}$_{2}${CuO}$_{6}$,
  {Bi}$_{2}${Sr}$_{2}${CaCu}$_{2}${O}$_{8}$,{HgBa}$_{2}${CuO}$_{4}$, and
  {CaCuO}$_{2}$},\ }\href {https://doi.org/10.1103/PhysRevB.106.235150}
  {\bibfield  {journal} {\bibinfo  {journal} {Phys. Rev. B}\ }\textbf {\bibinfo
  {volume} {106}},\ \bibinfo {pages} {235150} (\bibinfo {year}
  {2022})}\BibitemShut {NoStop}%
\bibitem [{\citenamefont {Varma}\ \emph {et~al.}(1989)\citenamefont {Varma},
  \citenamefont {Littlewood}, \citenamefont {Schmitt-Rink}, \citenamefont
  {Abrahams},\ and\ \citenamefont {Ruckenstein}}]{varma89}%
  \BibitemOpen
  \bibfield  {author} {\bibinfo {author} {\bibfnamefont {C.~M.}\ \bibnamefont
  {Varma}}, \bibinfo {author} {\bibfnamefont {P.~B.}\ \bibnamefont
  {Littlewood}}, \bibinfo {author} {\bibfnamefont {S.}~\bibnamefont
  {Schmitt-Rink}}, \bibinfo {author} {\bibfnamefont {E.}~\bibnamefont
  {Abrahams}},\ and\ \bibinfo {author} {\bibfnamefont {A.~E.}\ \bibnamefont
  {Ruckenstein}},\ }\bibfield  {title} {\bibinfo {title} {Phenomenology of the
  normal state of {Cu-O} high-temperature superconductors},\ }\href
  {https://doi.org/10.1103/PhysRevLett.63.1996} {\bibfield  {journal} {\bibinfo
   {journal} {Phys. Rev. Lett.}\ }\textbf {\bibinfo {volume} {63}},\ \bibinfo
  {pages} {1996} (\bibinfo {year} {1989})}\BibitemShut {NoStop}%
\bibitem [{\citenamefont {Sachdev}(2010)}]{sachdev10}%
  \BibitemOpen
  \bibfield  {author} {\bibinfo {author} {\bibfnamefont {S.}~\bibnamefont
  {Sachdev}},\ }\bibfield  {title} {\bibinfo {title} {Holographic metals and
  the fractionalized {Fermi} liquid},\ }\href
  {https://doi.org/10.1103/PhysRevLett.105.151602} {\bibfield  {journal}
  {\bibinfo  {journal} {Phys. Rev. Lett.}\ }\textbf {\bibinfo {volume} {105}},\
  \bibinfo {pages} {151602} (\bibinfo {year} {2010})}\BibitemShut {NoStop}%
\bibitem [{\citenamefont {Zaanen}(2019)}]{zaanen19}%
  \BibitemOpen
  \bibfield  {author} {\bibinfo {author} {\bibfnamefont {J.}~\bibnamefont
  {Zaanen}},\ }\bibfield  {title} {\bibinfo {title} {{Planckian dissipation,
  minimal viscosity and the transport in cuprate strange metals}},\ }\href
  {https://doi.org/10.21468/SciPostPhys.6.5.061} {\bibfield  {journal}
  {\bibinfo  {journal} {SciPost Phys.}\ }\textbf {\bibinfo {volume} {6}},\
  \bibinfo {pages} {061} (\bibinfo {year} {2019})}\BibitemShut {NoStop}%
\bibitem [{\citenamefont {Carbotte}\ \emph {et~al.}(2011)\citenamefont
  {Carbotte}, \citenamefont {Timusk},\ and\ \citenamefont
  {Hwang}}]{carbotte11}%
  \BibitemOpen
  \bibfield  {author} {\bibinfo {author} {\bibfnamefont {J.~P.}\ \bibnamefont
  {Carbotte}}, \bibinfo {author} {\bibfnamefont {T.}~\bibnamefont {Timusk}},\
  and\ \bibinfo {author} {\bibfnamefont {J.}~\bibnamefont {Hwang}},\ }\bibfield
   {title} {\bibinfo {title} {Bosons in high-temperature superconductors: an
  experimental survey},\ }\href {https://doi.org/10.1088/0034-4885/74/6/066501}
  {\bibfield  {journal} {\bibinfo  {journal} {Reports on Progress in Physics}\
  }\textbf {\bibinfo {volume} {74}},\ \bibinfo {pages} {066501} (\bibinfo
  {year} {2011})}\BibitemShut {NoStop}%
\bibitem [{\citenamefont {Norman}\ \emph {et~al.}(1995)\citenamefont {Norman},
  \citenamefont {Randeria}, \citenamefont {Ding},\ and\ \citenamefont
  {Campuzano}}]{norman95}%
  \BibitemOpen
  \bibfield  {author} {\bibinfo {author} {\bibfnamefont {M.~R.}\ \bibnamefont
  {Norman}}, \bibinfo {author} {\bibfnamefont {M.}~\bibnamefont {Randeria}},
  \bibinfo {author} {\bibfnamefont {H.}~\bibnamefont {Ding}},\ and\ \bibinfo
  {author} {\bibfnamefont {J.~C.}\ \bibnamefont {Campuzano}},\ }\bibfield
  {title} {\bibinfo {title} {Phenomenological models for the gap anisotropy of
  {Bi}$_{2}${Sr}$_{2}${CaCu}$_{2}${O}$_{8}$ as measured by angle-resolved
  photoemission spectroscopy},\ }\href
  {https://doi.org/10.1103/PhysRevB.52.615} {\bibfield  {journal} {\bibinfo
  {journal} {Phys. Rev. B}\ }\textbf {\bibinfo {volume} {52}},\ \bibinfo
  {pages} {615} (\bibinfo {year} {1995})}\BibitemShut {NoStop}%
\bibitem [{\citenamefont {Li}\ \emph {et~al.}(2018)\citenamefont {Li},
  \citenamefont {Zhou}, \citenamefont {Parham}, \citenamefont {Reber},
  \citenamefont {Berger}, \citenamefont {Arnold},\ and\ \citenamefont
  {Dessau}}]{li2018coherent}%
  \BibitemOpen
  \bibfield  {author} {\bibinfo {author} {\bibfnamefont {H.}~\bibnamefont
  {Li}}, \bibinfo {author} {\bibfnamefont {X.}~\bibnamefont {Zhou}}, \bibinfo
  {author} {\bibfnamefont {S.}~\bibnamefont {Parham}}, \bibinfo {author}
  {\bibfnamefont {T.~J.}\ \bibnamefont {Reber}}, \bibinfo {author}
  {\bibfnamefont {H.}~\bibnamefont {Berger}}, \bibinfo {author} {\bibfnamefont
  {G.~B.}\ \bibnamefont {Arnold}},\ and\ \bibinfo {author} {\bibfnamefont
  {D.~S.}\ \bibnamefont {Dessau}},\ }\bibfield  {title} {\bibinfo {title}
  {Coherent organization of electronic correlations as a mechanism to enhance
  and stabilize high-${T}_c$ cuprate superconductivity},\ }\href
  {https://doi.org/10.1038/s41467-017-02422-2} {\bibfield  {journal} {\bibinfo
  {journal} {Nature Communications}\ }\textbf {\bibinfo {volume} {9}},\
  \bibinfo {pages} {26} (\bibinfo {year} {2018})}\BibitemShut {NoStop}%
\bibitem [{\citenamefont {Kaminski}\ \emph {et~al.}(2001)\citenamefont
  {Kaminski}, \citenamefont {Randeria}, \citenamefont {Campuzano},
  \citenamefont {Norman}, \citenamefont {Fretwell}, \citenamefont {Mesot},
  \citenamefont {Sato}, \citenamefont {Takahashi},\ and\ \citenamefont
  {Kadowaki}}]{kaminski01}%
  \BibitemOpen
  \bibfield  {author} {\bibinfo {author} {\bibfnamefont {A.}~\bibnamefont
  {Kaminski}}, \bibinfo {author} {\bibfnamefont {M.}~\bibnamefont {Randeria}},
  \bibinfo {author} {\bibfnamefont {J.~C.}\ \bibnamefont {Campuzano}}, \bibinfo
  {author} {\bibfnamefont {M.~R.}\ \bibnamefont {Norman}}, \bibinfo {author}
  {\bibfnamefont {H.}~\bibnamefont {Fretwell}}, \bibinfo {author}
  {\bibfnamefont {J.}~\bibnamefont {Mesot}}, \bibinfo {author} {\bibfnamefont
  {T.}~\bibnamefont {Sato}}, \bibinfo {author} {\bibfnamefont {T.}~\bibnamefont
  {Takahashi}},\ and\ \bibinfo {author} {\bibfnamefont {K.}~\bibnamefont
  {Kadowaki}},\ }\bibfield  {title} {\bibinfo {title} {Renormalization of
  spectral line shape and dispersion below ${T}_{c}$ in
  {Bi}$_{2}${Sr}$_{2}${CaCu}$_{2}${O}$_{8+\delta}$},\ }\href
  {https://doi.org/10.1103/PhysRevLett.86.1070} {\bibfield  {journal} {\bibinfo
   {journal} {Phys. Rev. Lett.}\ }\textbf {\bibinfo {volume} {86}},\ \bibinfo
  {pages} {1070} (\bibinfo {year} {2001})}\BibitemShut {NoStop}%
\bibitem [{\citenamefont {Liu}\ \emph {et~al.}(2012)\citenamefont {Liu},
  \citenamefont {Qi},\ and\ \citenamefont {Zhang}}]{liu12}%
  \BibitemOpen
  \bibfield  {author} {\bibinfo {author} {\bibfnamefont {Q.}~\bibnamefont
  {Liu}}, \bibinfo {author} {\bibfnamefont {X.-L.}\ \bibnamefont {Qi}},\ and\
  \bibinfo {author} {\bibfnamefont {S.-C.}\ \bibnamefont {Zhang}},\ }\bibfield
  {title} {\bibinfo {title} {Stationary phase approximation approach to the
  quasiparticle interference on the surface of a strong topological
  insulator},\ }\href {https://doi.org/10.1103/PhysRevB.85.125314} {\bibfield
  {journal} {\bibinfo  {journal} {Phys. Rev. B}\ }\textbf {\bibinfo {volume}
  {85}},\ \bibinfo {pages} {125314} (\bibinfo {year} {2012})}\BibitemShut
  {NoStop}%
\bibitem [{\citenamefont {Kohsaka}()}]{kohsaka_unpub}%
  \BibitemOpen
  \bibfield  {author} {\bibinfo {author} {\bibfnamefont {Y.}~\bibnamefont
  {Kohsaka}},\ }\href@noop {} {}\Eprint {https://arxiv.org/abs/unpublished}
  {unpublished} \BibitemShut {NoStop}%
\bibitem [{\citenamefont {Dell'Anna}\ \emph {et~al.}(2005)\citenamefont
  {Dell'Anna}, \citenamefont {Lorenzana}, \citenamefont {Capone}, \citenamefont
  {Castellani},\ and\ \citenamefont {Grilli}}]{dellanna05}%
  \BibitemOpen
  \bibfield  {author} {\bibinfo {author} {\bibfnamefont {L.}~\bibnamefont
  {Dell'Anna}}, \bibinfo {author} {\bibfnamefont {J.}~\bibnamefont
  {Lorenzana}}, \bibinfo {author} {\bibfnamefont {M.}~\bibnamefont {Capone}},
  \bibinfo {author} {\bibfnamefont {C.}~\bibnamefont {Castellani}},\ and\
  \bibinfo {author} {\bibfnamefont {M.}~\bibnamefont {Grilli}},\ }\bibfield
  {title} {\bibinfo {title} {Effect of mesoscopic inhomogeneities on local
  tunneling density of states in cuprates},\ }\href
  {https://doi.org/10.1103/PhysRevB.71.064518} {\bibfield  {journal} {\bibinfo
  {journal} {Phys. Rev. B}\ }\textbf {\bibinfo {volume} {71}},\ \bibinfo
  {pages} {064518} (\bibinfo {year} {2005})}\BibitemShut {NoStop}%
\bibitem [{\citenamefont {Tohyama}(2004)}]{tohyama04}%
  \BibitemOpen
  \bibfield  {author} {\bibinfo {author} {\bibfnamefont {T.}~\bibnamefont
  {Tohyama}},\ }\bibfield  {title} {\bibinfo {title} {Asymmetry of the
  electronic states in hole- and electron-doped cuprates: Exact diagonalization
  study of the $t-t'-t''-{J}$ model},\ }\href
  {https://doi.org/10.1103/PhysRevB.70.174517} {\bibfield  {journal} {\bibinfo
  {journal} {Phys. Rev. B}\ }\textbf {\bibinfo {volume} {70}},\ \bibinfo
  {pages} {174517} (\bibinfo {year} {2004})}\BibitemShut {NoStop}%
\bibitem [{\citenamefont {Chi}\ \emph {et~al.}(2017{\natexlab{a}})\citenamefont
  {Chi}, \citenamefont {Hardy}, \citenamefont {Liang}, \citenamefont {Dosanjh},
  \citenamefont {Wahl}, \citenamefont {Burke},\ and\ \citenamefont
  {Bonn}}]{chi_2017_1}%
  \BibitemOpen
  \bibfield  {author} {\bibinfo {author} {\bibfnamefont {S.}~\bibnamefont
  {Chi}}, \bibinfo {author} {\bibfnamefont {W.~N.}\ \bibnamefont {Hardy}},
  \bibinfo {author} {\bibfnamefont {R.}~\bibnamefont {Liang}}, \bibinfo
  {author} {\bibfnamefont {P.}~\bibnamefont {Dosanjh}}, \bibinfo {author}
  {\bibfnamefont {P.}~\bibnamefont {Wahl}}, \bibinfo {author} {\bibfnamefont
  {S.~A.}\ \bibnamefont {Burke}},\ and\ \bibinfo {author} {\bibfnamefont
  {D.~A.}\ \bibnamefont {Bonn}},\ }\bibfield  {title} {\bibinfo {title}
  {{Extracting phase information about the superconducting order parameter from
  defect bound states}},\ }\href {https://arxiv.org/abs/1710.09088} {\bibfield
  {journal} {\bibinfo  {journal} {arXiv:1710.09088}\ } (\bibinfo {year}
  {2017}{\natexlab{a}})}\BibitemShut {NoStop}%
\bibitem [{\citenamefont {Chi}\ \emph {et~al.}(2017{\natexlab{b}})\citenamefont
  {Chi}, \citenamefont {Hardy}, \citenamefont {Liang}, \citenamefont {Dosanjh},
  \citenamefont {Wahl}, \citenamefont {Burke},\ and\ \citenamefont
  {Bonn}}]{chi_2017_2}%
  \BibitemOpen
  \bibfield  {author} {\bibinfo {author} {\bibfnamefont {S.}~\bibnamefont
  {Chi}}, \bibinfo {author} {\bibfnamefont {W.~N.}\ \bibnamefont {Hardy}},
  \bibinfo {author} {\bibfnamefont {R.}~\bibnamefont {Liang}}, \bibinfo
  {author} {\bibfnamefont {P.}~\bibnamefont {Dosanjh}}, \bibinfo {author}
  {\bibfnamefont {P.}~\bibnamefont {Wahl}}, \bibinfo {author} {\bibfnamefont
  {S.~A.}\ \bibnamefont {Burke}},\ and\ \bibinfo {author} {\bibfnamefont
  {D.~A.}\ \bibnamefont {Bonn}},\ }\bibfield  {title} {\bibinfo {title}
  {{Determination of the Superconducting Order Parameter from Defect Bound
  State Quasiparticle Interference}},\ }\href
  {https://arxiv.org/abs/1710.09089} {\bibfield  {journal} {\bibinfo  {journal}
  {arXiv:1710.09089}\ } (\bibinfo {year} {2017}{\natexlab{b}})}\BibitemShut
  {NoStop}%
\bibitem [{sup()}]{suppl}%
  \BibitemOpen
  \href@noop {} {}\bibinfo {note} {See Supplemental Material at
  [URL].}\BibitemShut {Stop}%
\bibitem [{\citenamefont {Smit}\ \emph {et~al.}(2024)\citenamefont {Smit},
  \citenamefont {Mauri}, \citenamefont {Bawden}, \citenamefont {Heringa},
  \citenamefont {Gerritsen}, \citenamefont {van Heumen}, \citenamefont {Huang},
  \citenamefont {Kondo}, \citenamefont {Takeuchi}, \citenamefont {Hussey},
  \citenamefont {Allan}, \citenamefont {Kim}, \citenamefont {Cacho},
  \citenamefont {Krikun}, \citenamefont {Schalm}, \citenamefont {Stoof},\ and\
  \citenamefont {Golden}}]{smit2024momentum}%
  \BibitemOpen
  \bibfield  {author} {\bibinfo {author} {\bibfnamefont {S.}~\bibnamefont
  {Smit}}, \bibinfo {author} {\bibfnamefont {E.}~\bibnamefont {Mauri}},
  \bibinfo {author} {\bibfnamefont {L.}~\bibnamefont {Bawden}}, \bibinfo
  {author} {\bibfnamefont {F.}~\bibnamefont {Heringa}}, \bibinfo {author}
  {\bibfnamefont {F.}~\bibnamefont {Gerritsen}}, \bibinfo {author}
  {\bibfnamefont {E.}~\bibnamefont {van Heumen}}, \bibinfo {author}
  {\bibfnamefont {Y.~K.}\ \bibnamefont {Huang}}, \bibinfo {author}
  {\bibfnamefont {T.}~\bibnamefont {Kondo}}, \bibinfo {author} {\bibfnamefont
  {T.}~\bibnamefont {Takeuchi}}, \bibinfo {author} {\bibfnamefont {N.~E.}\
  \bibnamefont {Hussey}}, \bibinfo {author} {\bibfnamefont {M.}~\bibnamefont
  {Allan}}, \bibinfo {author} {\bibfnamefont {T.~K.}\ \bibnamefont {Kim}},
  \bibinfo {author} {\bibfnamefont {C.}~\bibnamefont {Cacho}}, \bibinfo
  {author} {\bibfnamefont {A.}~\bibnamefont {Krikun}}, \bibinfo {author}
  {\bibfnamefont {K.}~\bibnamefont {Schalm}}, \bibinfo {author} {\bibfnamefont
  {H.~T.~C.}\ \bibnamefont {Stoof}},\ and\ \bibinfo {author} {\bibfnamefont
  {M.~S.}\ \bibnamefont {Golden}},\ }\bibfield  {title} {\bibinfo {title}
  {Momentum-dependent scaling exponents of nodal self-energies measured in
  strange metal cuprates and modelled using semi-holography},\ }\href
  {https://doi.org/10.1038/s41467-024-48594-6} {\bibfield  {journal} {\bibinfo
  {journal} {Nature Communications}\ }\textbf {\bibinfo {volume} {15}},\
  \bibinfo {pages} {4581} (\bibinfo {year} {2024})}\BibitemShut {NoStop}%
\bibitem [{\citenamefont {Bok}\ \emph {et~al.}(2010)\citenamefont {Bok},
  \citenamefont {Yun}, \citenamefont {Choi}, \citenamefont {Zhang},
  \citenamefont {Zhou},\ and\ \citenamefont {Varma}}]{bok10}%
  \BibitemOpen
  \bibfield  {author} {\bibinfo {author} {\bibfnamefont {J.~M.}\ \bibnamefont
  {Bok}}, \bibinfo {author} {\bibfnamefont {J.~H.}\ \bibnamefont {Yun}},
  \bibinfo {author} {\bibfnamefont {H.-Y.}\ \bibnamefont {Choi}}, \bibinfo
  {author} {\bibfnamefont {W.}~\bibnamefont {Zhang}}, \bibinfo {author}
  {\bibfnamefont {X.~J.}\ \bibnamefont {Zhou}},\ and\ \bibinfo {author}
  {\bibfnamefont {C.~M.}\ \bibnamefont {Varma}},\ }\bibfield  {title} {\bibinfo
  {title} {Momentum dependence of the single-particle self-energy and
  fluctuation spectrum of slightly underdoped
  {Bi}$_{2}${Sr}$_{2}${CaCu}$_{2}${O}$_{8+\delta}$ from high-resolution laser
  angle-resolved photoemission},\ }\href
  {https://doi.org/10.1103/PhysRevB.81.174516} {\bibfield  {journal} {\bibinfo
  {journal} {Phys. Rev. B}\ }\textbf {\bibinfo {volume} {81}},\ \bibinfo
  {pages} {174516} (\bibinfo {year} {2010})}\BibitemShut {NoStop}%
\bibitem [{\citenamefont {Schrieffer}\ \emph {et~al.}(1963)\citenamefont
  {Schrieffer}, \citenamefont {Scalapino},\ and\ \citenamefont
  {Wilkins}}]{schrieffer63}%
  \BibitemOpen
  \bibfield  {author} {\bibinfo {author} {\bibfnamefont {J.~R.}\ \bibnamefont
  {Schrieffer}}, \bibinfo {author} {\bibfnamefont {D.~J.}\ \bibnamefont
  {Scalapino}},\ and\ \bibinfo {author} {\bibfnamefont {J.~W.}\ \bibnamefont
  {Wilkins}},\ }\bibfield  {title} {\bibinfo {title} {Effective tunneling
  density of states in superconductors},\ }\href
  {https://doi.org/10.1103/PhysRevLett.10.336} {\bibfield  {journal} {\bibinfo
  {journal} {Phys. Rev. Lett.}\ }\textbf {\bibinfo {volume} {10}},\ \bibinfo
  {pages} {336} (\bibinfo {year} {1963})}\BibitemShut {NoStop}%
\bibitem [{\citenamefont {Scalapino}\ \emph {et~al.}(1966)\citenamefont
  {Scalapino}, \citenamefont {Schrieffer},\ and\ \citenamefont
  {Wilkins}}]{scalapino66}%
  \BibitemOpen
  \bibfield  {author} {\bibinfo {author} {\bibfnamefont {D.~J.}\ \bibnamefont
  {Scalapino}}, \bibinfo {author} {\bibfnamefont {J.~R.}\ \bibnamefont
  {Schrieffer}},\ and\ \bibinfo {author} {\bibfnamefont {J.~W.}\ \bibnamefont
  {Wilkins}},\ }\bibfield  {title} {\bibinfo {title} {Strong-coupling
  superconductivity. {I}},\ }\href {https://doi.org/10.1103/PhysRev.148.263}
  {\bibfield  {journal} {\bibinfo  {journal} {Phys. Rev.}\ }\textbf {\bibinfo
  {volume} {148}},\ \bibinfo {pages} {263} (\bibinfo {year}
  {1966})}\BibitemShut {NoStop}%
\bibitem [{\citenamefont {Chubukov}\ and\ \citenamefont
  {Schmalian}(2020)}]{chubukov20}%
  \BibitemOpen
  \bibfield  {author} {\bibinfo {author} {\bibfnamefont {A.~V.}\ \bibnamefont
  {Chubukov}}\ and\ \bibinfo {author} {\bibfnamefont {J.}~\bibnamefont
  {Schmalian}},\ }\bibfield  {title} {\bibinfo {title} {Pairing glue in cuprate
  superconductors from the self-energy revealed via machine learning},\ }\href
  {https://doi.org/10.1103/PhysRevB.101.180510} {\bibfield  {journal} {\bibinfo
   {journal} {Phys. Rev. B}\ }\textbf {\bibinfo {volume} {101}},\ \bibinfo
  {pages} {180510} (\bibinfo {year} {2020})}\BibitemShut {NoStop}%
\bibitem [{\citenamefont {Pavlyukh}\ \emph {et~al.}(2016)\citenamefont
  {Pavlyukh}, \citenamefont {Uimonen}, \citenamefont {Stefanucci},\ and\
  \citenamefont {van Leeuwen}}]{PhysRevLett.117.206402}%
  \BibitemOpen
  \bibfield  {author} {\bibinfo {author} {\bibfnamefont {Y.}~\bibnamefont
  {Pavlyukh}}, \bibinfo {author} {\bibfnamefont {A.-M.}\ \bibnamefont
  {Uimonen}}, \bibinfo {author} {\bibfnamefont {G.}~\bibnamefont
  {Stefanucci}},\ and\ \bibinfo {author} {\bibfnamefont {R.}~\bibnamefont {van
  Leeuwen}},\ }\bibfield  {title} {\bibinfo {title} {Vertex corrections for
  positive-definite spectral functions of simple metals},\ }\href
  {https://doi.org/10.1103/PhysRevLett.117.206402} {\bibfield  {journal}
  {\bibinfo  {journal} {Phys. Rev. Lett.}\ }\textbf {\bibinfo {volume} {117}},\
  \bibinfo {pages} {206402} (\bibinfo {year} {2016})}\BibitemShut {NoStop}%
\bibitem [{\citenamefont {Macridin}\ \emph {et~al.}(2007)\citenamefont
  {Macridin}, \citenamefont {Jarrell}, \citenamefont {Maier},\ and\
  \citenamefont {Scalapino}}]{macridin07}%
  \BibitemOpen
  \bibfield  {author} {\bibinfo {author} {\bibfnamefont {A.}~\bibnamefont
  {Macridin}}, \bibinfo {author} {\bibfnamefont {M.}~\bibnamefont {Jarrell}},
  \bibinfo {author} {\bibfnamefont {T.}~\bibnamefont {Maier}},\ and\ \bibinfo
  {author} {\bibfnamefont {D.~J.}\ \bibnamefont {Scalapino}},\ }\bibfield
  {title} {\bibinfo {title} {{High-Energy Kink in the Single-Particle Spectra
  of the Two-Dimensional Hubbard Model}},\ }\href
  {https://doi.org/10.1103/PhysRevLett.99.237001} {\bibfield  {journal}
  {\bibinfo  {journal} {Phys. Rev. Lett.}\ }\textbf {\bibinfo {volume} {99}},\
  \bibinfo {pages} {237001} (\bibinfo {year} {2007})}\BibitemShut {NoStop}%
\bibitem [{\citenamefont {Charlebois}\ and\ \citenamefont
  {Imada}(2020)}]{charlebois20}%
  \BibitemOpen
  \bibfield  {author} {\bibinfo {author} {\bibfnamefont {M.}~\bibnamefont
  {Charlebois}}\ and\ \bibinfo {author} {\bibfnamefont {M.}~\bibnamefont
  {Imada}},\ }\bibfield  {title} {\bibinfo {title} {Single-particle spectral
  function formulated and calculated by variational {Monte} {Carlo} method with
  application to $d$-wave superconducting state},\ }\href
  {https://doi.org/10.1103/PhysRevX.10.041023} {\bibfield  {journal} {\bibinfo
  {journal} {Phys. Rev. X}\ }\textbf {\bibinfo {volume} {10}},\ \bibinfo
  {pages} {041023} (\bibinfo {year} {2020})}\BibitemShut {NoStop}%
\bibitem [{\citenamefont {Li}\ \emph {et~al.}(2021)\citenamefont {Li},
  \citenamefont {Wu}, \citenamefont {Chan},\ and\ \citenamefont
  {Louie}}]{PhysRevLett.126.146401}%
  \BibitemOpen
  \bibfield  {author} {\bibinfo {author} {\bibfnamefont {Z.}~\bibnamefont
  {Li}}, \bibinfo {author} {\bibfnamefont {M.}~\bibnamefont {Wu}}, \bibinfo
  {author} {\bibfnamefont {Y.-H.}\ \bibnamefont {Chan}},\ and\ \bibinfo
  {author} {\bibfnamefont {S.~G.}\ \bibnamefont {Louie}},\ }\bibfield  {title}
  {\bibinfo {title} {Unmasking the origin of kinks in the photoemission spectra
  of cuprate superconductors},\ }\href
  {https://doi.org/10.1103/PhysRevLett.126.146401} {\bibfield  {journal}
  {\bibinfo  {journal} {Phys. Rev. Lett.}\ }\textbf {\bibinfo {volume} {126}},\
  \bibinfo {pages} {146401} (\bibinfo {year} {2021})}\BibitemShut {NoStop}%
\bibitem [{\citenamefont {Wang}\ and\ \citenamefont {Lee}(2003)}]{wang03}%
  \BibitemOpen
  \bibfield  {author} {\bibinfo {author} {\bibfnamefont {Q.-H.}\ \bibnamefont
  {Wang}}\ and\ \bibinfo {author} {\bibfnamefont {D.-H.}\ \bibnamefont {Lee}},\
  }\bibfield  {title} {\bibinfo {title} {Quasiparticle scattering interference
  in high-temperature superconductors},\ }\href
  {https://doi.org/10.1103/PhysRevB.67.020511} {\bibfield  {journal} {\bibinfo
  {journal} {Phys. Rev. B}\ }\textbf {\bibinfo {volume} {67}},\ \bibinfo
  {pages} {020511} (\bibinfo {year} {2003})}\BibitemShut {NoStop}%
\bibitem [{\citenamefont {Choubey}\ \emph {et~al.}(2014)\citenamefont
  {Choubey}, \citenamefont {Berlijn}, \citenamefont {Kreisel}, \citenamefont
  {Cao},\ and\ \citenamefont {Hirschfeld}}]{choubey14}%
  \BibitemOpen
  \bibfield  {author} {\bibinfo {author} {\bibfnamefont {P.}~\bibnamefont
  {Choubey}}, \bibinfo {author} {\bibfnamefont {T.}~\bibnamefont {Berlijn}},
  \bibinfo {author} {\bibfnamefont {A.}~\bibnamefont {Kreisel}}, \bibinfo
  {author} {\bibfnamefont {C.}~\bibnamefont {Cao}},\ and\ \bibinfo {author}
  {\bibfnamefont {P.~J.}\ \bibnamefont {Hirschfeld}},\ }\bibfield  {title}
  {\bibinfo {title} {Visualization of atomic-scale phenomena in
  superconductors: Application to {FeSe}},\ }\href
  {https://doi.org/10.1103/PhysRevB.90.134520} {\bibfield  {journal} {\bibinfo
  {journal} {Phys. Rev. B}\ }\textbf {\bibinfo {volume} {90}},\ \bibinfo
  {pages} {134520} (\bibinfo {year} {2014})}\BibitemShut {NoStop}%
\bibitem [{\citenamefont {Kreisel}\ \emph {et~al.}(2015)\citenamefont
  {Kreisel}, \citenamefont {Choubey}, \citenamefont {Berlijn}, \citenamefont
  {Ku}, \citenamefont {Andersen},\ and\ \citenamefont
  {Hirschfeld}}]{kreisel15}%
  \BibitemOpen
  \bibfield  {author} {\bibinfo {author} {\bibfnamefont {A.}~\bibnamefont
  {Kreisel}}, \bibinfo {author} {\bibfnamefont {P.}~\bibnamefont {Choubey}},
  \bibinfo {author} {\bibfnamefont {T.}~\bibnamefont {Berlijn}}, \bibinfo
  {author} {\bibfnamefont {W.}~\bibnamefont {Ku}}, \bibinfo {author}
  {\bibfnamefont {B.~M.}\ \bibnamefont {Andersen}},\ and\ \bibinfo {author}
  {\bibfnamefont {P.~J.}\ \bibnamefont {Hirschfeld}},\ }\bibfield  {title}
  {\bibinfo {title} {Interpretation of scanning tunneling quasiparticle
  interference and impurity states in cuprates},\ }\href
  {https://doi.org/10.1103/PhysRevLett.114.217002} {\bibfield  {journal}
  {\bibinfo  {journal} {Phys. Rev. Lett.}\ }\textbf {\bibinfo {volume} {114}},\
  \bibinfo {pages} {217002} (\bibinfo {year} {2015})}\BibitemShut {NoStop}%
\bibitem [{\citenamefont {Dalla~Torre}\ \emph {et~al.}(2016)\citenamefont
  {Dalla~Torre}, \citenamefont {He},\ and\ \citenamefont {Demler}}]{torre16}%
  \BibitemOpen
  \bibfield  {author} {\bibinfo {author} {\bibfnamefont {E.}~\bibnamefont
  {Dalla~Torre}}, \bibinfo {author} {\bibfnamefont {Y.}~\bibnamefont {He}},\
  and\ \bibinfo {author} {\bibfnamefont {E.}~\bibnamefont {Demler}},\
  }\bibfield  {title} {\bibinfo {title} {Holographic maps of quasiparticle
  interference},\ }\href {https://doi.org/10.1038/NPHYS3829} {\bibfield
  {journal} {\bibinfo  {journal} {Nature Phys.}\ }\textbf {\bibinfo {volume}
  {12}},\ \bibinfo {pages} {1052} (\bibinfo {year} {2016})}\BibitemShut
  {NoStop}%
\bibitem [{\citenamefont {Choubey}\ \emph {et~al.}(2017)\citenamefont
  {Choubey}, \citenamefont {Kreisel}, \citenamefont {Berlijn}, \citenamefont
  {Andersen},\ and\ \citenamefont {Hirschfeld}}]{choubey17}%
  \BibitemOpen
  \bibfield  {author} {\bibinfo {author} {\bibfnamefont {P.}~\bibnamefont
  {Choubey}}, \bibinfo {author} {\bibfnamefont {A.}~\bibnamefont {Kreisel}},
  \bibinfo {author} {\bibfnamefont {T.}~\bibnamefont {Berlijn}}, \bibinfo
  {author} {\bibfnamefont {B.~M.}\ \bibnamefont {Andersen}},\ and\ \bibinfo
  {author} {\bibfnamefont {P.~J.}\ \bibnamefont {Hirschfeld}},\ }\bibfield
  {title} {\bibinfo {title} {Universality of scanning tunneling microscopy in
  cuprate superconductors},\ }\href
  {https://doi.org/10.1103/PhysRevB.96.174523} {\bibfield  {journal} {\bibinfo
  {journal} {Phys. Rev. B}\ }\textbf {\bibinfo {volume} {96}},\ \bibinfo
  {pages} {174523} (\bibinfo {year} {2017})}\BibitemShut {NoStop}%
\bibitem [{\citenamefont {Andersen}\ and\ \citenamefont
  {Hedeg$\aa$rd}(2003)}]{andersen03}%
  \BibitemOpen
  \bibfield  {author} {\bibinfo {author} {\bibfnamefont {B.~M.}\ \bibnamefont
  {Andersen}}\ and\ \bibinfo {author} {\bibfnamefont {P.}~\bibnamefont
  {Hedeg$\aa$rd}},\ }\bibfield  {title} {\bibinfo {title} {Quantum interference
  between multiple impurities in anisotropic superconductors},\ }\href
  {https://doi.org/10.1103/PhysRevB.67.172505} {\bibfield  {journal} {\bibinfo
  {journal} {Phys. Rev. B}\ }\textbf {\bibinfo {volume} {67}},\ \bibinfo
  {pages} {172505} (\bibinfo {year} {2003})}\BibitemShut {NoStop}%
\bibitem [{\citenamefont {Capriotti}\ \emph {et~al.}(2003)\citenamefont
  {Capriotti}, \citenamefont {Scalapino},\ and\ \citenamefont
  {Sedgewick}}]{capriotti03}%
  \BibitemOpen
  \bibfield  {author} {\bibinfo {author} {\bibfnamefont {L.}~\bibnamefont
  {Capriotti}}, \bibinfo {author} {\bibfnamefont {D.~J.}\ \bibnamefont
  {Scalapino}},\ and\ \bibinfo {author} {\bibfnamefont {R.~D.}\ \bibnamefont
  {Sedgewick}},\ }\bibfield  {title} {\bibinfo {title} {Wave-vector power
  spectrum of the local tunneling density of states: Ripples in a $d$-wave
  sea},\ }\href {https://doi.org/10.1103/PhysRevB.68.014508} {\bibfield
  {journal} {\bibinfo  {journal} {Phys. Rev. B}\ }\textbf {\bibinfo {volume}
  {68}},\ \bibinfo {pages} {014508} (\bibinfo {year} {2003})}\BibitemShut
  {NoStop}%
\bibitem [{\citenamefont {Pereg-Barnea}\ and\ \citenamefont
  {Franz}(2003)}]{pereg-barnea03}%
  \BibitemOpen
  \bibfield  {author} {\bibinfo {author} {\bibfnamefont {T.}~\bibnamefont
  {Pereg-Barnea}}\ and\ \bibinfo {author} {\bibfnamefont {M.}~\bibnamefont
  {Franz}},\ }\bibfield  {title} {\bibinfo {title} {Theory of quasiparticle
  interference patterns in the pseudogap phase of the cuprate
  superconductors},\ }\href {https://doi.org/10.1103/PhysRevB.68.180506}
  {\bibfield  {journal} {\bibinfo  {journal} {Phys. Rev. B}\ }\textbf {\bibinfo
  {volume} {68}},\ \bibinfo {pages} {180506} (\bibinfo {year}
  {2003})}\BibitemShut {NoStop}%
\bibitem [{\citenamefont {Zhang}\ and\ \citenamefont {Ting}(2003)}]{zhang03}%
  \BibitemOpen
  \bibfield  {author} {\bibinfo {author} {\bibfnamefont {D.}~\bibnamefont
  {Zhang}}\ and\ \bibinfo {author} {\bibfnamefont {C.~S.}\ \bibnamefont
  {Ting}},\ }\bibfield  {title} {\bibinfo {title} {Energy-dependent modulations
  in the local density of states of the cuprate superconductors},\ }\href
  {https://doi.org/10.1103/PhysRevB.67.100506} {\bibfield  {journal} {\bibinfo
  {journal} {Phys. Rev. B}\ }\textbf {\bibinfo {volume} {67}},\ \bibinfo
  {pages} {100506} (\bibinfo {year} {2003})}\BibitemShut {NoStop}%
\bibitem [{\citenamefont {B\"oker}\ \emph {et~al.}(2020)\citenamefont
  {B\"oker}, \citenamefont {Sulangi}, \citenamefont {Akbari}, \citenamefont
  {Davis}, \citenamefont {Hirschfeld},\ and\ \citenamefont
  {Eremin}}]{boeker20}%
  \BibitemOpen
  \bibfield  {author} {\bibinfo {author} {\bibfnamefont {J.}~\bibnamefont
  {B\"oker}}, \bibinfo {author} {\bibfnamefont {M.~A.}\ \bibnamefont
  {Sulangi}}, \bibinfo {author} {\bibfnamefont {A.}~\bibnamefont {Akbari}},
  \bibinfo {author} {\bibfnamefont {J.~C.~S.}\ \bibnamefont {Davis}}, \bibinfo
  {author} {\bibfnamefont {P.~J.}\ \bibnamefont {Hirschfeld}},\ and\ \bibinfo
  {author} {\bibfnamefont {I.~M.}\ \bibnamefont {Eremin}},\ }\bibfield  {title}
  {\bibinfo {title} {Phase-sensitive determination of nodal $d$-wave order
  parameter in single-band and multiband superconductors},\ }\href
  {https://doi.org/10.1103/PhysRevB.101.214505} {\bibfield  {journal} {\bibinfo
   {journal} {Phys. Rev. B}\ }\textbf {\bibinfo {volume} {101}},\ \bibinfo
  {pages} {214505} (\bibinfo {year} {2020})}\BibitemShut {NoStop}%
\bibitem [{\citenamefont {Zhu}\ \emph {et~al.}(2004)\citenamefont {Zhu},
  \citenamefont {Atkinson},\ and\ \citenamefont {Hirschfeld}}]{zhu04}%
  \BibitemOpen
  \bibfield  {author} {\bibinfo {author} {\bibfnamefont {L.}~\bibnamefont
  {Zhu}}, \bibinfo {author} {\bibfnamefont {W.~A.}\ \bibnamefont {Atkinson}},\
  and\ \bibinfo {author} {\bibfnamefont {P.~J.}\ \bibnamefont {Hirschfeld}},\
  }\bibfield  {title} {\bibinfo {title} {Power spectrum of many impurities in a
  $d$-wave superconductor},\ }\href
  {https://doi.org/10.1103/PhysRevB.69.060503} {\bibfield  {journal} {\bibinfo
  {journal} {Phys. Rev. B}\ }\textbf {\bibinfo {volume} {69}},\ \bibinfo
  {pages} {060503} (\bibinfo {year} {2004})}\BibitemShut {NoStop}%
\bibitem [{\citenamefont {Nunner}\ \emph {et~al.}(2006)\citenamefont {Nunner},
  \citenamefont {Chen}, \citenamefont {Andersen}, \citenamefont {Melikyan},\
  and\ \citenamefont {Hirschfeld}}]{nunner06}%
  \BibitemOpen
  \bibfield  {author} {\bibinfo {author} {\bibfnamefont {T.~S.}\ \bibnamefont
  {Nunner}}, \bibinfo {author} {\bibfnamefont {W.}~\bibnamefont {Chen}},
  \bibinfo {author} {\bibfnamefont {B.~M.}\ \bibnamefont {Andersen}}, \bibinfo
  {author} {\bibfnamefont {A.}~\bibnamefont {Melikyan}},\ and\ \bibinfo
  {author} {\bibfnamefont {P.~J.}\ \bibnamefont {Hirschfeld}},\ }\bibfield
  {title} {\bibinfo {title} {Fourier transform spectroscopy of $d$-wave
  quasiparticles in the presence of atomic scale pairing disorder},\ }\href
  {https://doi.org/10.1103/PhysRevB.73.104511} {\bibfield  {journal} {\bibinfo
  {journal} {Phys. Rev. B}\ }\textbf {\bibinfo {volume} {73}},\ \bibinfo
  {pages} {104511} (\bibinfo {year} {2006})}\BibitemShut {NoStop}%
\bibitem [{\citenamefont {Nunner}\ \emph {et~al.}(2005)\citenamefont {Nunner},
  \citenamefont {Andersen}, \citenamefont {Melikyan},\ and\ \citenamefont
  {Hirschfeld}}]{nunner05}%
  \BibitemOpen
  \bibfield  {author} {\bibinfo {author} {\bibfnamefont {T.~S.}\ \bibnamefont
  {Nunner}}, \bibinfo {author} {\bibfnamefont {B.~M.}\ \bibnamefont
  {Andersen}}, \bibinfo {author} {\bibfnamefont {A.}~\bibnamefont {Melikyan}},\
  and\ \bibinfo {author} {\bibfnamefont {P.~J.}\ \bibnamefont {Hirschfeld}},\
  }\bibfield  {title} {\bibinfo {title} {Dopant-modulated pair interaction in
  cuprate superconductors},\ }\href
  {https://doi.org/10.1103/PhysRevLett.95.177003} {\bibfield  {journal}
  {\bibinfo  {journal} {Phys. Rev. Lett.}\ }\textbf {\bibinfo {volume} {95}},\
  \bibinfo {pages} {177003} (\bibinfo {year} {2005})}\BibitemShut {NoStop}%
\bibitem [{\citenamefont {Sulangi}\ \emph {et~al.}(2017)\citenamefont
  {Sulangi}, \citenamefont {Allan},\ and\ \citenamefont {Zaanen}}]{sulangi17}%
  \BibitemOpen
  \bibfield  {author} {\bibinfo {author} {\bibfnamefont {M.~A.}\ \bibnamefont
  {Sulangi}}, \bibinfo {author} {\bibfnamefont {M.~P.}\ \bibnamefont {Allan}},\
  and\ \bibinfo {author} {\bibfnamefont {J.}~\bibnamefont {Zaanen}},\
  }\bibfield  {title} {\bibinfo {title} {Revisiting quasiparticle scattering
  interference in high-temperature superconductors: The problem of narrow
  peaks},\ }\href {https://doi.org/10.1103/PhysRevB.96.134507} {\bibfield
  {journal} {\bibinfo  {journal} {Phys. Rev. B}\ }\textbf {\bibinfo {volume}
  {96}},\ \bibinfo {pages} {134507} (\bibinfo {year} {2017})}\BibitemShut
  {NoStop}%
\bibitem [{\citenamefont {Kampf}\ and\ \citenamefont
  {Devereaux}(1997)}]{kampf97}%
  \BibitemOpen
  \bibfield  {author} {\bibinfo {author} {\bibfnamefont {A.~P.}\ \bibnamefont
  {Kampf}}\ and\ \bibinfo {author} {\bibfnamefont {T.~P.}\ \bibnamefont
  {Devereaux}},\ }\bibfield  {title} {\bibinfo {title} {Extended impurity
  potential in a ${\mathit{d}}_{{x}^{2}\ensuremath{-}{y}^{2}}$
  superconductor},\ }\href {https://doi.org/10.1103/PhysRevB.56.2360}
  {\bibfield  {journal} {\bibinfo  {journal} {Phys. Rev. B}\ }\textbf {\bibinfo
  {volume} {56}},\ \bibinfo {pages} {2360} (\bibinfo {year}
  {1997})}\BibitemShut {NoStop}%
\bibitem [{\citenamefont {Vekhter}\ and\ \citenamefont
  {Varma}(2003)}]{vekhter03}%
  \BibitemOpen
  \bibfield  {author} {\bibinfo {author} {\bibfnamefont {I.}~\bibnamefont
  {Vekhter}}\ and\ \bibinfo {author} {\bibfnamefont {C.~M.}\ \bibnamefont
  {Varma}},\ }\bibfield  {title} {\bibinfo {title} {Proposal to determine the
  spectrum of pairing glue in high-temperature superconductors},\ }\href
  {https://doi.org/10.1103/PhysRevLett.90.237003} {\bibfield  {journal}
  {\bibinfo  {journal} {Phys. Rev. Lett.}\ }\textbf {\bibinfo {volume} {90}},\
  \bibinfo {pages} {237003} (\bibinfo {year} {2003})}\BibitemShut {NoStop}%
\end{thebibliography}%


\begin{thebibliography}{0}%
\makeatletter
\providecommand \@ifxundefined [1]{%
 \@ifx{#1\undefined}
}%
\providecommand \@ifnum [1]{%
 \ifnum #1\expandafter \@firstoftwo
 \else \expandafter \@secondoftwo
 \fi
}%
\providecommand \@ifx [1]{%
 \ifx #1\expandafter \@firstoftwo
 \else \expandafter \@secondoftwo
 \fi
}%
\providecommand \natexlab [1]{#1}%
\providecommand \enquote  [1]{``#1''}%
\providecommand \bibnamefont  [1]{#1}%
\providecommand \bibfnamefont [1]{#1}%
\providecommand \citenamefont [1]{#1}%
\providecommand \href@noop [0]{\@secondoftwo}%
\providecommand \href [0]{\begingroup \@sanitize@url \@href}%
\providecommand \@href[1]{\@@startlink{#1}\@@href}%
\providecommand \@@href[1]{\endgroup#1\@@endlink}%
\providecommand \@sanitize@url [0]{\catcode `\\12\catcode `\$12\catcode
  `\&12\catcode `\#12\catcode `\^12\catcode `\_12\catcode `\%12\relax}%
\providecommand \@@startlink[1]{}%
\providecommand \@@endlink[0]{}%
\providecommand \url  [0]{\begingroup\@sanitize@url \@url }%
\providecommand \@url [1]{\endgroup\@href {#1}{\urlprefix }}%
\providecommand \urlprefix  [0]{URL }%
\providecommand \Eprint [0]{\href }%
\providecommand \doibase [0]{https://doi.org/}%
\providecommand \selectlanguage [0]{\@gobble}%
\providecommand \bibinfo  [0]{\@secondoftwo}%
\providecommand \bibfield  [0]{\@secondoftwo}%
\providecommand \translation [1]{[#1]}%
\providecommand \BibitemOpen [0]{}%
\providecommand \bibitemStop [0]{}%
\providecommand \bibitemNoStop [0]{.\EOS\space}%
\providecommand \EOS [0]{\spacefactor3000\relax}%
\providecommand \BibitemShut  [1]{\csname bibitem#1\endcsname}%
\let\auto@bib@innerbib\@empty
\end{thebibliography}%
\end{document}


\title{Supplemental Material for `Unified description of cuprate superconductors by fractionalized electrons emerging from integrated analyses of photoemission spectra and quasiparticle interference'}

\author{Shiro Sakai$^{1,2}$, Youhei Yamaji$^{3,4}$, Fumihiro Imoto$^5$, Tsuyoshi Tamegai$^6$, Adam Kaminski$^{7,8}$, Takeshi Kondo$^9$, Yuhki Kohsaka$^{2,10}$, Tetsuo Hanaguri$^2$, and Masatoshi Imada$^{1,6}$}
\affiliation{
$^1$ \mbox{Physics Division, Sophia University, Chiyoda, Tokyo 102-8554, Japan}\\
$^2$ \mbox{RIKEN Center for Emergent Matter Science, Wako, Saitama 351-0198, Japan} \\
$^3$ \mbox{Research Center for Materials Nanoarchitectonics, National Institute for Materials Science,} \\
\mbox{Tsukuba, Ibaraki 305-0044, Japan} \\
$^4$ \mbox{RIKEN Center for Computational Science, Kobe, 650-0047, Japan}\\
$^5$ \mbox{Schr\"odinger Inc., Tokyo 100-0005, Japan}\\
$^6$ \mbox{Department of Applied Physics, University of Tokyo, Tokyo 113-8656, Japan}\\
$^7$ \mbox{Ames National Laboratory, U.S. Department of Energy, Ames, IA 50011, USA}\\
$^8$ \mbox{Department of Physics and Astronomy, Iowa State University, Ames, IA 50011, USA}\\
$^9$ \mbox{ISSP, University of Tokyo, Kashiwa, Chiba 277-8581, Japan}\\
$^{10}$ \mbox{Department of Physics, Kyoto University, Kyoto 606-8502, Japan}
} 

\date{\today}

\maketitle

\section{Comparison of averaged EDCs}\label{sec:arpes_suppl}

\begin{figure}[tb]
\centering
\includegraphics[width=0.48\textwidth]{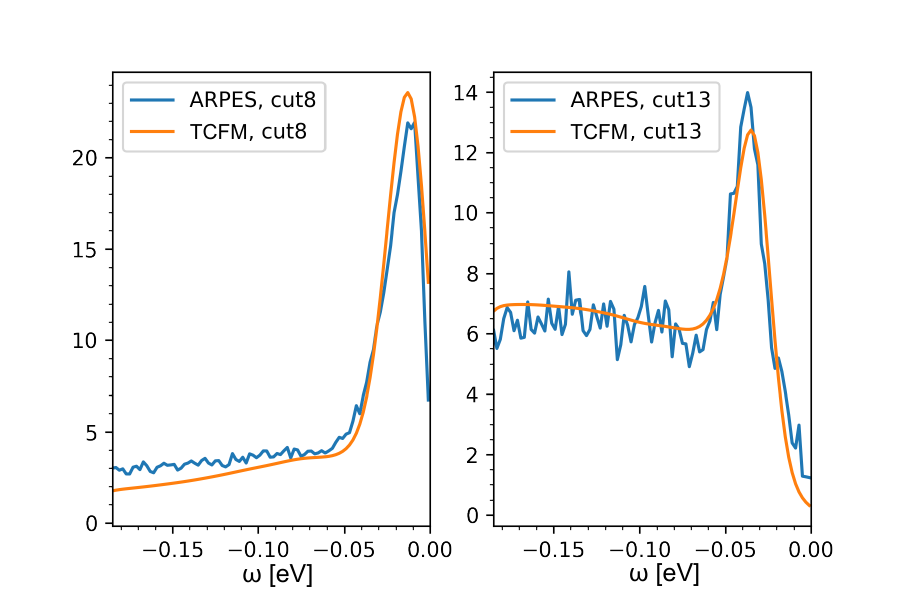}
\caption{Comparison of momentum-averaged EDCs between ARPES and TCFM around $\kk=\kk_{\rm max}$ along the cuts 8 and 13.} \label{fig:EDC_av}
\end{figure}
 
Figure \ref{fig:EDC_av} compares the EDCs of the normalized spectra for $\kk=\kk_{\rm max}$ along the cuts 8 (nodal direction) and 13 (antinodal direction).
Here, we average 
$\tilde{I}_{\rm ARPES}(\kk,\w)$ and $\tilde{I}_{\rm TCFM}(\kk,\w)$ over five momentum points around $\kk=\kk_{\rm max}$ along each cut, to suppress the noise, as has often been done in the analysis of the ARPES data. 
Note that the raw data at $\kk=\kk_{\rm max}$ are shown in Fig.~4.

Regarding the ARPES experimental data, the noise is significantly suppressed compared to Fig.~4.
In particular, for the cut 13 (antinode), this suppression of the noise allows us to identify a dip around $\w=-0.07$eV and a hump around $\w=-0.1$ to $-0.15$eV. 
Since we used all the raw data for fitting, the TCFM results succeeded in reproducing these intrinsic features of the ARPES spectra.

\section{$q-\w$ maps of QPI}\label{sec:qwmap_suppl}

\subsection{$L_r$ along $(0,0)-(2\pi,0)$}\label{ssec:Lr_suppl}

\begin{figure}[tb]
\centering
\includegraphics[width=0.48\textwidth]{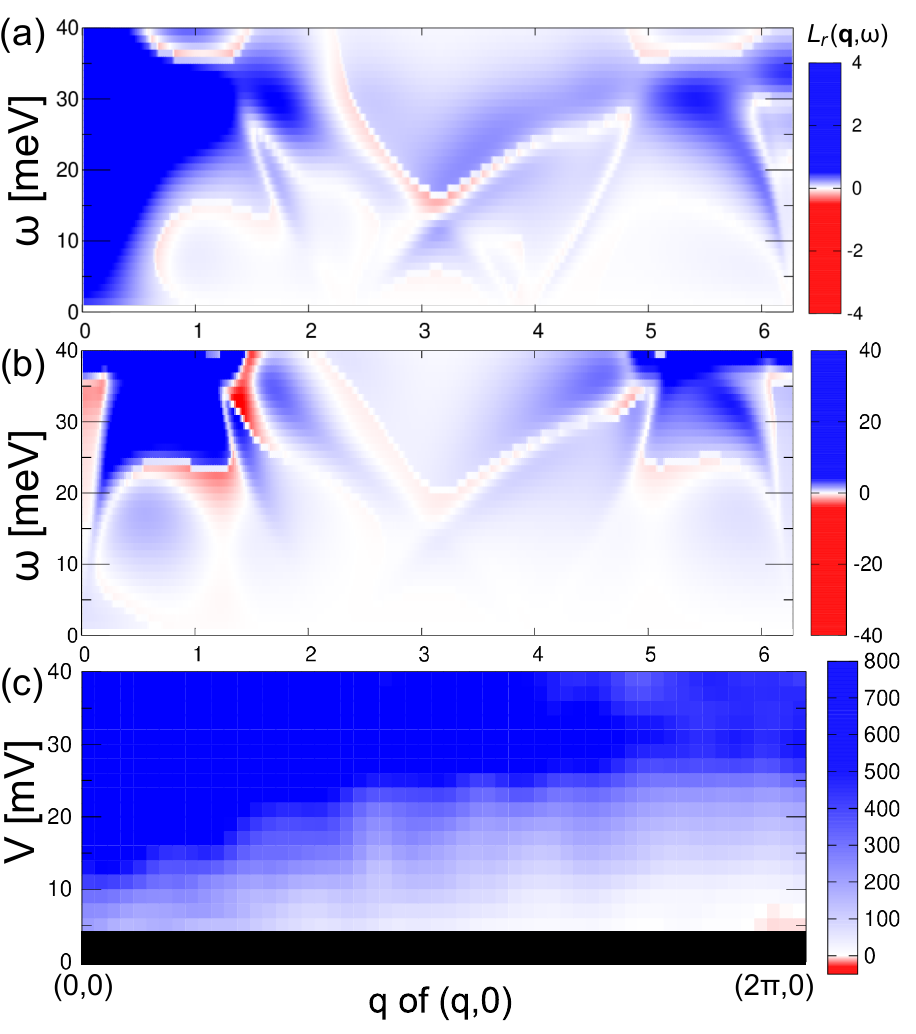}
\caption{Relative-sign structure, $L_r(\qq,\w>0)$,  along $(0,0)-(2\pi,0)$, obtained by (a) the TCFM, (b) the Hamiltonian, Eq.~(11),  
and (c) the experiment.
Black region in (c) indicates an energy range where accurate data is difficult to obtain because both $\dv{I}{V}$ and $\frac{I}{V}$ in $L=\frac{V}{I}\dv{I}{V}$ are not large enough compared to noise.} \label{fig:Lr_00-2P0}
\end{figure}

In Fig.~13 of the main text, we show $L_r(\qq,\w)$ along $(0,0)-(\pi,\pi)$, where we find a sign change characteristic of the TCFM, in consistency with the experimental result.
Figure \ref{fig:Lr_00-2P0} shows corresponding plots along $(0,0)-(2\pi,0)$.
The sign is overall positive in both (a) the TCFM and (b) the single-component Hamiltonian, Eq.~(11), as well as in (c) the experimental result.
The signature of the pseudogap in this direction becomes clearer at a higher energy ($\sim 0.1$ eV), as we see in Sec.~\ref{ssec:highE_L_suppl}.

\subsection{Sign structure of $g(\qq,\w)$}\label{ssec:gsign_suppl}

\begin{figure}[tb]
\centering
\includegraphics[width=0.48\textwidth]{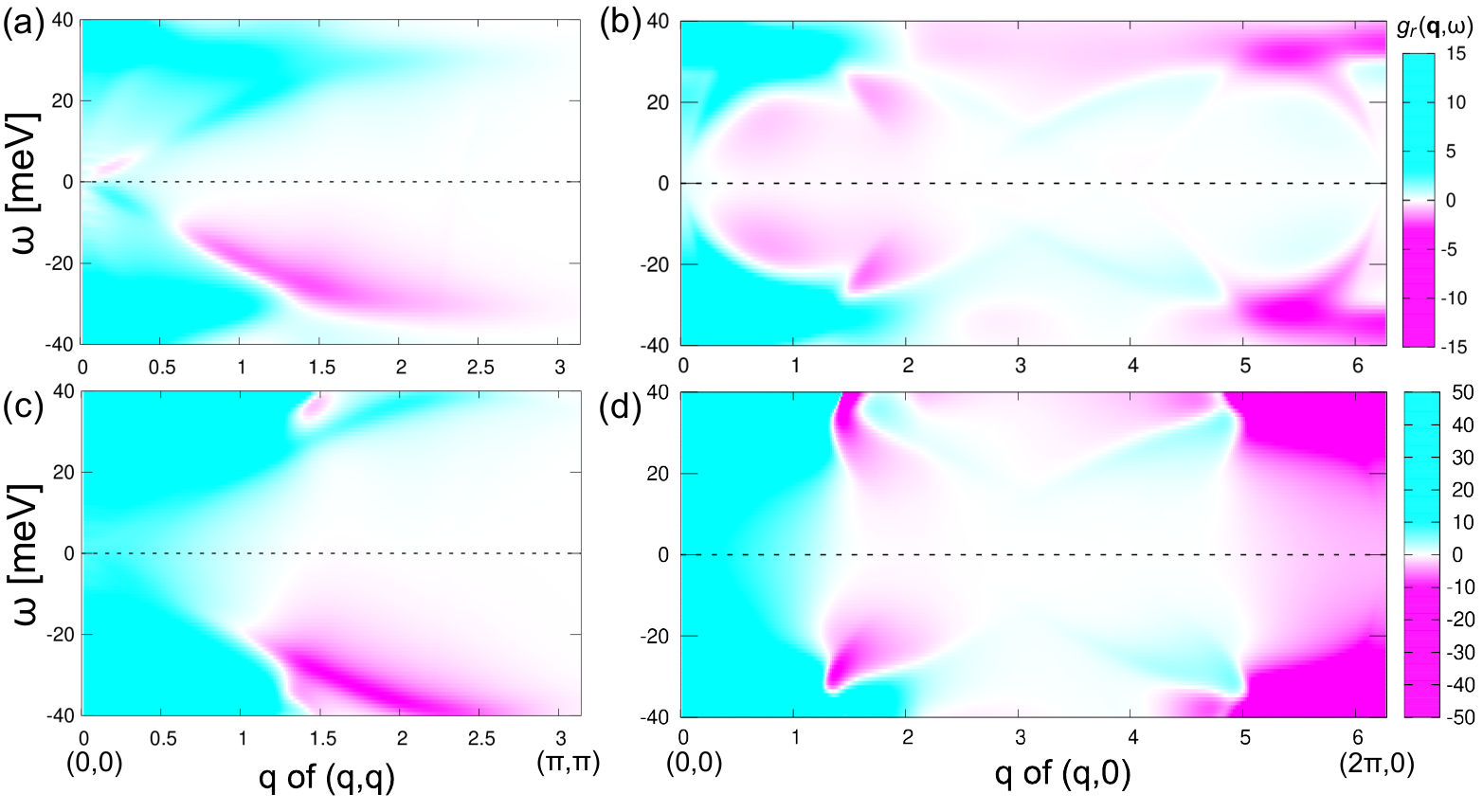}
\caption{Sign structure of $g(\qq,\w)$ along $(0,0)-(\pi,\pi)$ and $(0,0)-(2\pi,0)$ calculated with (a,b) the TCFM and (c,d) the Hamiltonian, Eq.~(11).} \label{fig:g_sign}
\end{figure}

\begin{figure}[tb]
\centering
\includegraphics[width=0.48\textwidth]{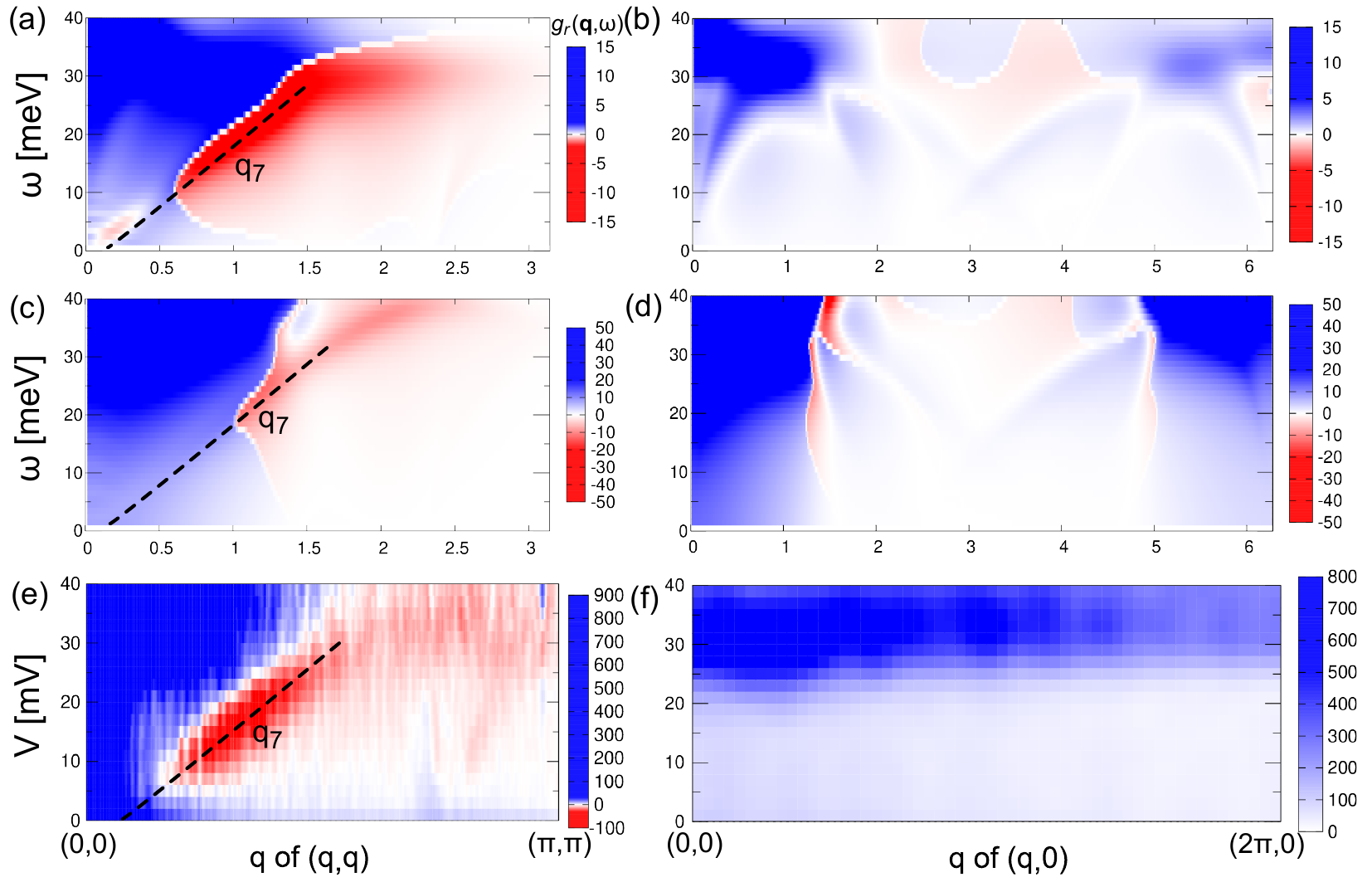}
\caption{Relative-sign structure, $g_r(\qq,\w>0)$, along $(0,0)-(\pi,\pi)$ and $(0,0)-(2\pi,0)$, obtained by (a,b) the TCFM, (c,d) the Hamiltonian, Eq.~(11), and (e,f) the experiment.
Note that the overall amplitude in the experimental results depends on the set-point current, which is not taken into account in the theory.}
\label{fig:gr_disp}
\end{figure}

Here, we present the results for $g$ and $g_r$, which correspond to Figs.~12, 13, and 22 of $L$ and $L_r$ in the main text. 
We used $\Vset=-0.1$ eV in both the calculation and the experiment.
Figure \ref{fig:g_sign} shows the sign structure of $g(\qq,\w)$ along $(0,0)-(\pi,\pi)$ and $(0,0)-(2\pi,0)$ calculated with (a,b) the TCFM and (c,d) the Hamiltonian, Eq.~(11).

A notable difference between the two models is in the low-$\w$ and small-$q$ region along $(0,0)-(2\pi,0)$ direction: In the TCFM [(a)], $g(\qq,\w)$ changes sign and becomes negative for $0\lesssim \qq\lesssim 1.5$ while it is always positive in the single-component model [(d)].
However, this difference is hard to detect in $g_r$, because the sign structure in \ref{fig:g_sign}(b) and \ref{fig:g_sign}(d) is almost symmetric around $\w=0$ and hence the relative-sign structure represented by $g_r$ is similar to each other [Fig.~\ref{fig:gr_disp}(b) and \ref{fig:gr_disp}(d)]. In fact, the experimental result [(f)] does not exhibit any conspicuous structure, either. 

Comparing the results along $(0,0)-(\pi,\pi)$ direction in Fig.~\ref{fig:gr_disp}, we find that the TCFM result [(a)] resembles to the experimental result [(e)] more than the single-component one [(c)], in particular regarding the size of the red region around $\qq_7$.

Unlike $L$ presented in Fig.~13(a), $g$ in the $(0,0)-(\pi,\pi)$ direction does not show the negative region on the smaller-$q$ side of the $\qq_7$ dispersion.
This can be understood based on Eq.~(A49), where the second term on the right-hand side is small for our choice of $e\Vset=-0.1$ eV because $\d\r$ changes the sign in the range $[e\Vset,0]$ as seen in Fig.~\ref{fig:e0.2}(a). 
Therefore, the sign structure of $\gqw$ is similar to that of $\d\r$ for $\qq\simeq (1,1)$.

\subsection{High-energy structures of $\d\r$ and $L$}\label{ssec:highE_L_suppl}

\begin{figure}[tb]
\centering
\includegraphics[width=0.48\textwidth]{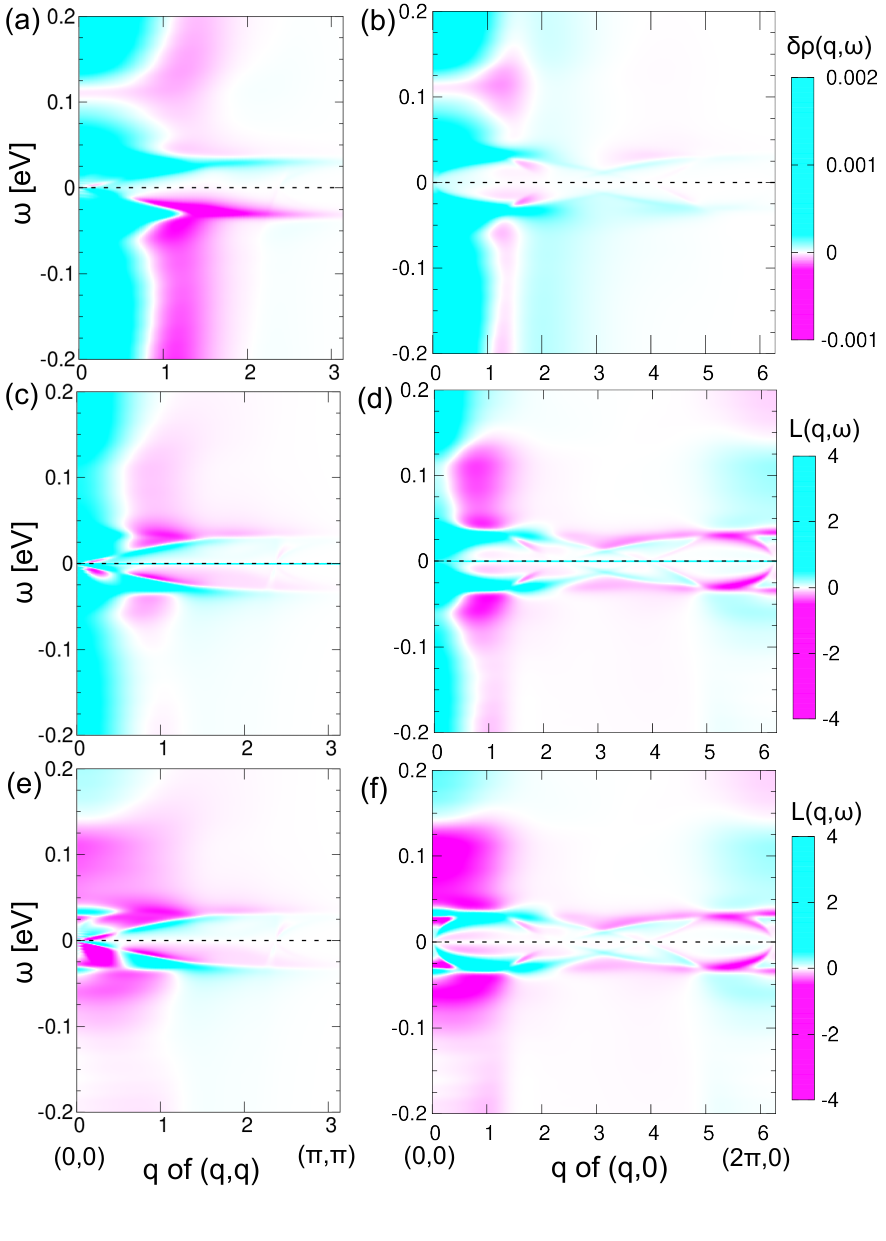}
\caption{High-energy structure of 
 (a,b) $\d\r(\qq,\w)$, (c,d) $L(\qq,\w)$ along $(0,0)-(\pi,\pi)$ and $(0,0)-(2\pi,0)$, calculated with the TCFM, and (e,f) the same as (c,d) but without the  Gaussian contribution, Eq.~(A47), around $\qq\simeq \Vec{0}$.} \label{fig:e0.2}
\end{figure}

\begin{figure}[tb]
\centering
\includegraphics[width=0.48\textwidth]{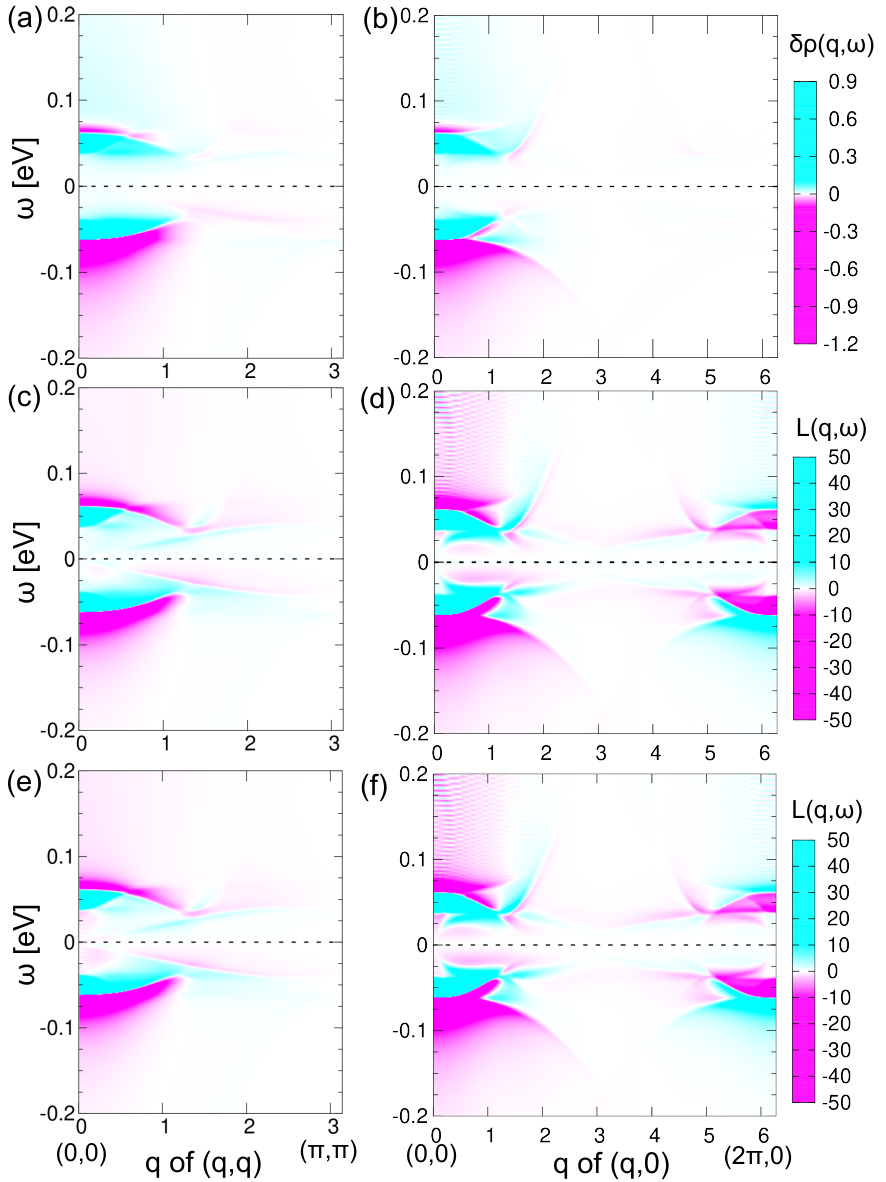}
\caption{High-energy structure of 
 (a,b) $\d\r(\qq,\w)$, (c,d) $L(\qq,\w)$ along $(0,0)-(\pi,\pi)$ and $(0,0)-(2\pi,0)$, calculated with the Hamiltonian, Eq.~(11), and (e,f) the same as (c,d) but without the  Gaussian contribution, Eq.~(A47), around $\qq\simeq \Vec{0}$.}
  \label{fig:1c_e0.2}
\end{figure}

Figure \ref{fig:e0.2} displays $\d\r$ and $L$ in the range of $[-0.2{\rm eV},0.2{\rm eV}]$ along $(0,0)-(\pi,\pi)$ and $(0,0)-(2\pi,0)$.
Enlarged views of the small-$q$ region ($|\qq|<1$) are presented in Fig.~14 of the main text and discussed in Sec.V.C.4.
For $|\qq|>1$, the QPI signals are faint except for small energies ($|\w|<40$ meV) because of the impurity potential $\hat{V}_{\rm imp}(\qq)$ rapidly decaying with $|\qq|$.
A relatively strong intensity around $(2\pi,0)$ is due to the Bragg reflection.
Similar features in the large-$|q|$ and high-$|\w|$ region are seen in the single-component results presented in 
Fig.~\ref{fig:1c_e0.2}, too.

\subsection{High-energy structure of $g$}\label{ssec:highE_g_suppl}
\begin{figure}[tb]
\centering
\includegraphics[width=0.48\textwidth]{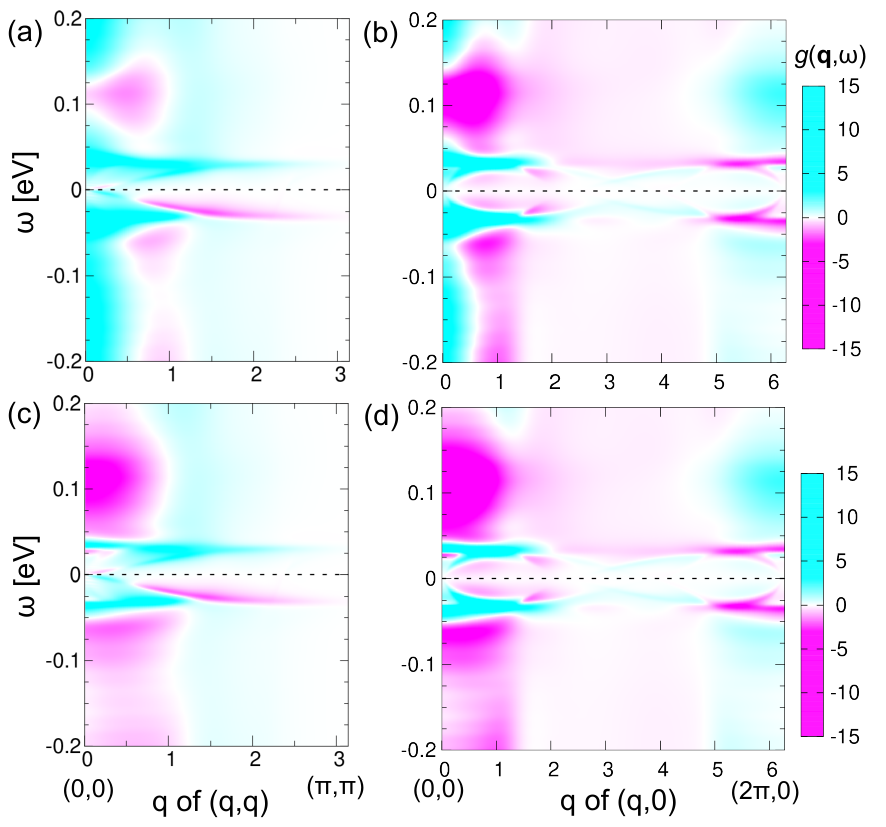}
\caption{Sign structure of $g(\qq,\w)$ along $(0,0)-(\pi,\pi)$ and $(0,0)-(2\pi,0)$ calculated with (a,b) the TCFM and (c,d) that without the Gaussian contribution, Eq.~(A46), around $\qq\simeq \Vec{0}$.} \label{fig:g_sign_E0.2}
\end{figure}

\begin{figure}[tb]
\centering
\includegraphics[width=0.48\textwidth]{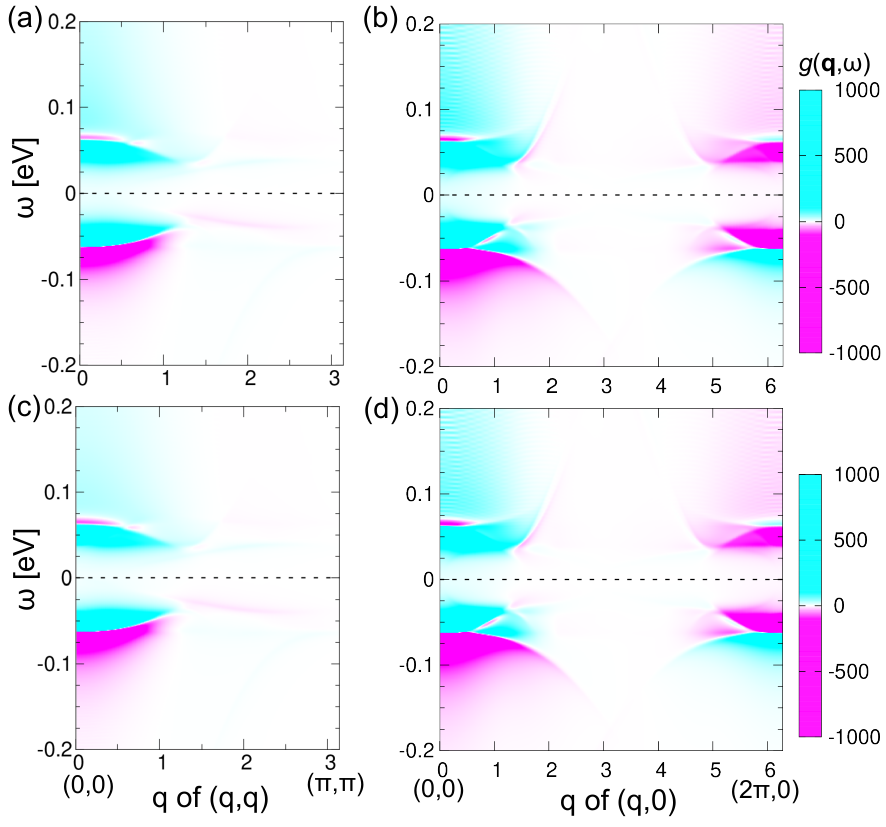}
\caption{Sign structure of $g(\qq,\w)$ along $(0,0)-(\pi,\pi)$ and $(0,0)-(2\pi,0)$ calculated with (a,b) the Hamiltonian, Eq.~(11), and (c,d) that without the Gaussian contribution, Eq.~(A46), around $\qq\simeq \Vec{0}$.} \label{fig:g_sign_E0.2_1c}
\end{figure}

Figure \ref{fig:g_sign_E0.2} shows the high-energy structure of $g$ along $(0,0)-(\pi,\pi)$ and $(0,0)-(2\pi,0)$.
Although sign changes around $\w=0.1$ eV, similar to that of $\d\r$ [Figs.~\ref{fig:e0.2}(a,b)], are seen in the TCFM results with the $\qq\simeq \Vec{0}$ Gaussian contribution [Figs.~\ref{fig:g_sign_E0.2}(a,b)], these are not seen in the results without the Gaussian contribution [Figs.~\ref{fig:g_sign_E0.2}(c,d)]. 
Namely, the sign changes around $\w=0.1$eV in Figs.~\ref{fig:g_sign_E0.2}(a,b) are not due to the presence of the pseudogap but the inhomogeneity producing the Gaussian contribution.
Such sign changes are not seen in the results for the single-component Hamiltonian, Eq.~(11) (Fig.~\ref{fig:g_sign_E0.2_1c}).

\section{$q$-space maps of QPI} \label{sec:qmap_suppl}
\subsection{$g$ and $L$ maps}\label{ssec:gLmap_suppl}
\begin{figure*}[tb]
\centering
\includegraphics[width=0.96\textwidth]{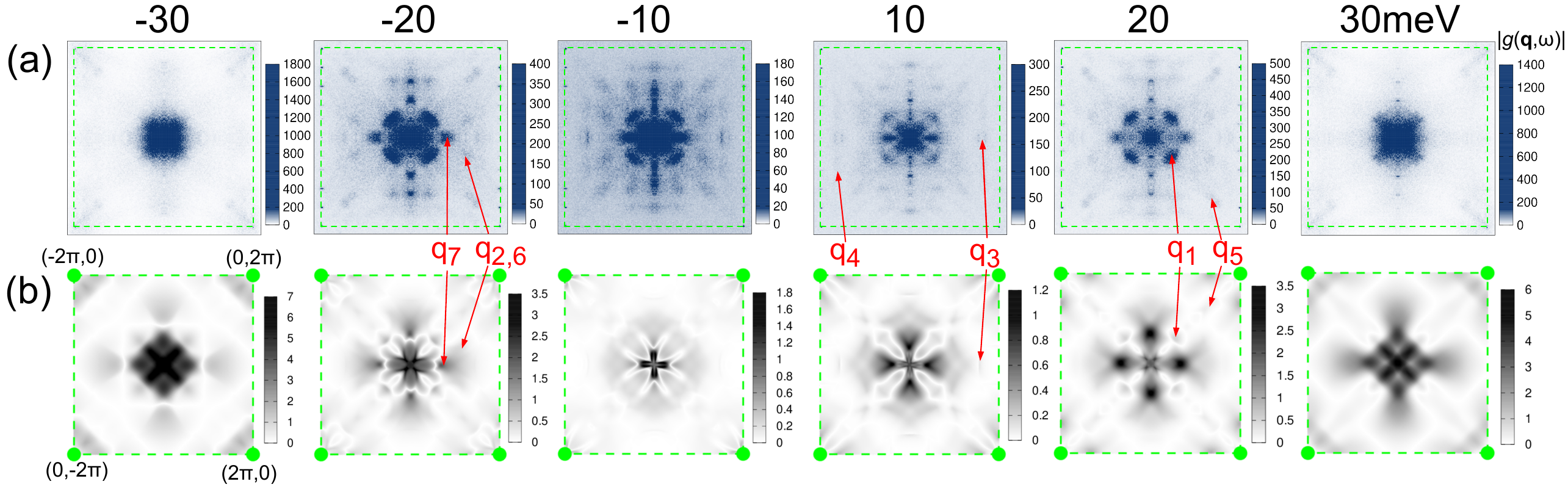}
\caption{Comparison of $|g(\qq,\w)|$ between (a) the experiment and (b) the TCFM.
The light-green square represents the Brillouin zone. The comparison should be made with the experimental data of the horizontal direction because additional signals due to the superstructure appear in the vertical direction. 
} \label{fig:gmap}
\end{figure*}

\begin{figure*}[tb]
\centering
\includegraphics[width=0.96\textwidth]{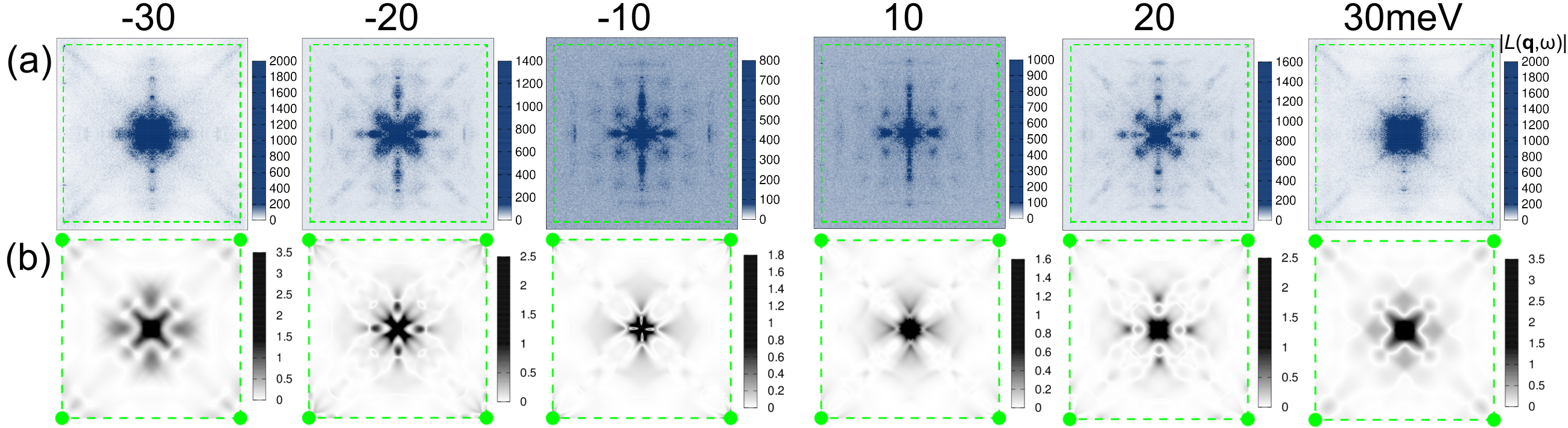}
\caption{Comparison of $|L(\qq,\w)|$ between (a) experiment and (b) the TCFM.
The light-green square represents the Brillouin zone.
The comparison should be made with the experimental data of the horizontal direction because additional signals due to the superstructure appear in the vertical direction. 
} \label{fig:Lmap}
\end{figure*}

Figures \ref{fig:gmap} and \ref{fig:Lmap} respectively display the $\qq$-space maps of $|g|$ and $|L|$ calculated at $\w=-30$ meV to $30$ meV.
Note that the calculated values of these data are real since the present theory places a single impurity at the origin and hence holds the inversion symmetry,
while experimental data are generally complex in $\qq$ space because multiple impurities contribute to the QPI. 
Since the phases of $g$, $L$ and $\d\r$ depend significantly on the positions of the impurities, which are not easily determined in experiments, we focus on the absolute values of these data.  

For $|\w|<30$ meV, we find $\qq_i (i=1,2,...,7)$ as denoted by red arrows in Fig.~\ref{fig:gmap} though $\qq_4$ is not very clear in the calculated results. The $\qq_i$ signals become vague around $|\w|=30$meV, and disappear for $|\w|\geq 40$ meV, consistently with the behavior known as extinction in QPI experiments.
In our theory, this extinction is attributed to the background self-energy, $\Sbg$, introduced to reproduce the flat spectra in ARPES (Appendix A.1.c).

\subsection{$Z$ map}\label{ssec:Zmap_suppl}

\begin{figure}[tb]
\centering
\includegraphics[width=0.48\textwidth]{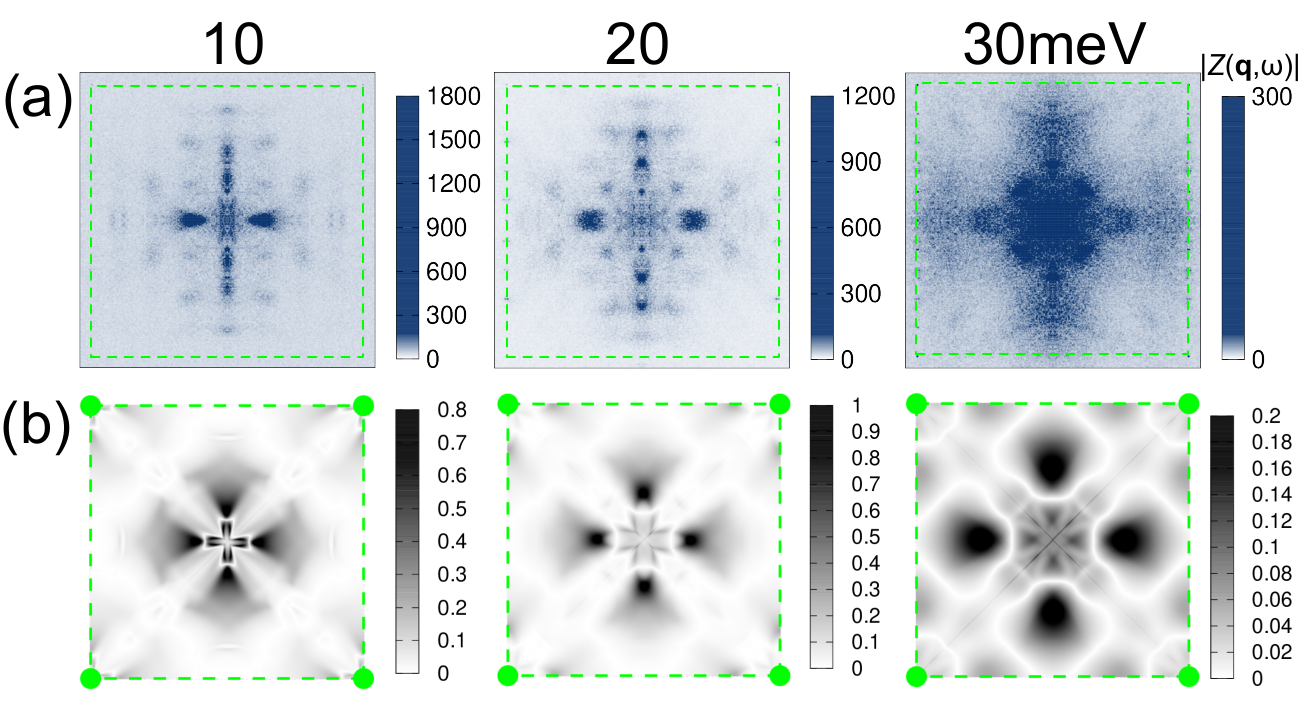}
\caption{Comparison of $|Z(\qq,\w)|$ between the experiment and the theory. (a) Experimental data of an optimally-doped Bi2212 bulk sample. The light-green square represents the Brillouin zone.
(b) TCFM results at the corresponding energies. The light-green circles denote the Bragg points, $(\pm 2\pi,0)$ and $(0,\pm 2\pi)$.
The comparison should be made with the experimental data of the horizontal direction because additional signals due to the superstructure appear in the vertical direction. } \label{fig:zmap}
\end{figure}

Figure \ref{fig:zmap} presents $Z(\qq,\w)$ defined by Eqs.~(A42) and (A45). 
At 10 meV, we find strong $\qq_7$ spots in the $(0,0)-(\pi,\pi)$ (horizontal) direction and $\qq_{2,6}$ spots around them in both the experiment and the theory.
As $\w$ increases, $\qq_1$ spots in the diagonal direction become relatively stronger while the horizontal direction becomes broader. This is also consistent between the experiment and the theory. 

\subsection{$g_r$ map}\label{ssec:grmap_suppl}
\begin{figure}[tb]
\centering
\includegraphics[width=0.48\textwidth]{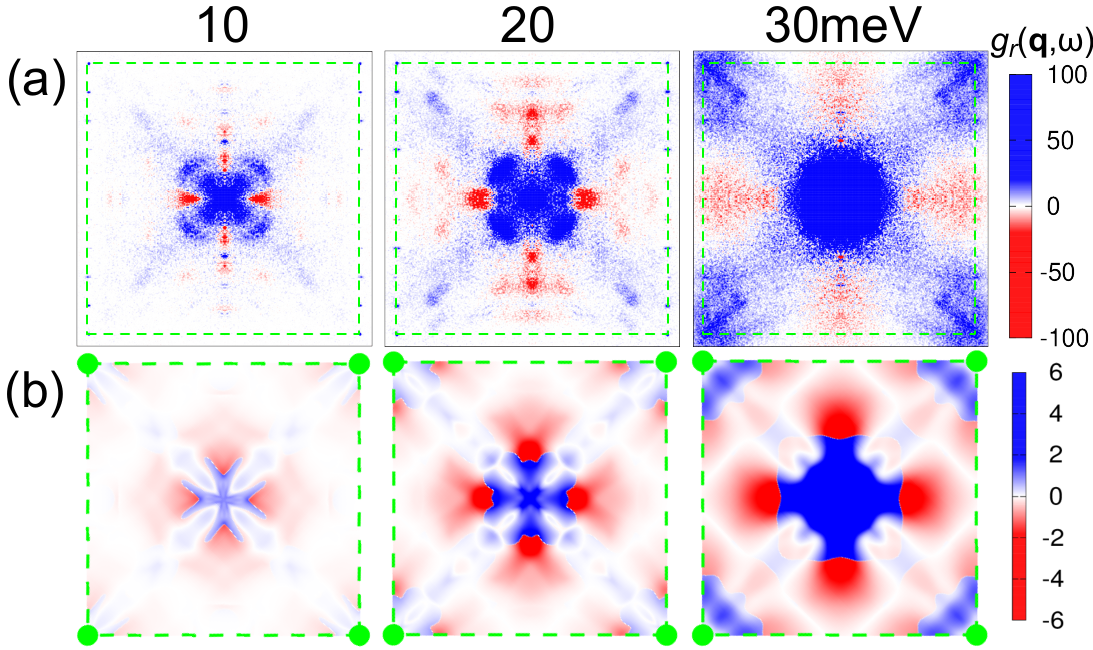}
\caption{Comparison of $g_r$ map between (a) the experiment and (b) the TCFM.} \label{fig:grmap}
\end{figure}

Figure \ref{fig:grmap} displays $g_r(\qq,\w)$ defined by Eq.~(11). 
The overall sign structure is similar to that of $L_r(\qq,\w)$ shown in Fig.~23.
A main difference from $L_r$ is the lack of the negative signal inside $\qq_7$, which corresponds to the difference between Fig.~13(a) and Fig.~\ref{fig:gr_disp}(a).